\documentclass[aippof,
 amsmath,amssymb,
 reprint,%
]{revtex4-1}

\usepackage{graphicx}
\usepackage{dcolumn}
\usepackage{bm}

\usepackage[utf8]{inputenc}
\usepackage[T1]{fontenc}
\usepackage{mathptmx}
\usepackage{color}

\begin{document}

\preprint{AIP/123-QED}

\title[Local dynamics of multiarm star polymer melts]{Dynamical heterogeneities in non-entangled polystyrene and poly(ethylene oxide) star melts}

\author{Petra Ba\v{c}ov\'a}
 \email{pbacova@iacm.forth.gr}
 \affiliation{Institute of Applied and Computational Mathematics (IACM), Foundation for Research and Technology Hellas (FORTH), GR-70013 Heraklion, Crete, Greece}
\affiliation{Computation-based Science and Technology Research Center, The Cyprus Institute, 20 Constantinou Kavafi Str., Nicosia 2121, Cyprus}
\author{Eirini Gkolfi}%
\affiliation{Institute of Applied and Computational Mathematics (IACM), Foundation for Research and Technology Hellas (FORTH), GR-70013 Heraklion, Crete, Greece}
\affiliation{Department of Mathematics and Applied Mathematics, University of Crete, GR-71409 Heraklion, Crete, Greece}

\author{Laurence G. D. Hawke}
\affiliation{
Institute of Condensed Matter and Nanosciences (IMCN), Bio and Soft Matter Division (BSMA), Universit\'e catholique de Louvain, Croix du Sud 1 \& Place L. Pasteur 1, B-1348 Louvain-la-Neuve, Belgium
}
\author{Vagelis Harmandaris}
\affiliation{Institute of Applied and Computational Mathematics (IACM), Foundation for Research and Technology Hellas (FORTH), GR-70013 Heraklion, Crete, Greece}
\affiliation{Department of Mathematics and Applied Mathematics, University of Crete, GR-71409 Heraklion, Crete, Greece}
\affiliation{Computation-based Science and Technology Research Center, The Cyprus Institute, 20 Constantinou Kavafi Str., Nicosia 2121, Cyprus}

\date{\today}

\begin{abstract}
The following article has been accepted by Physics of Fluids. After it is published, it will be found at: \url{https://publishing.aip.org/resources/librarians/products/journals/}.
\\
\\
Star polymers can exhibit a heterogeneous dynamical behavior due to their internal structure. In this work we employ atomistic molecular dynamics simulations to study translational motion in non-entangled polystyrene and poly(ethylene oxide) star-shaped melts.
We focus on the local heterogeneous dynamics originating from the multi-arm star-like architecture and quantify the intramolecular dynamical gradient.
By examining the translational motion at length scales of the order of the Kuhn length, we aim to find common features for both studied chemistries and to provide a critical and direct comparison with theoretical models of polymer dynamics. 
We discuss the observed tendencies with respect to the continuous Rouse model adjusted for the star-like architectures.
Two versions of the Rouse model are examined: one assuming uniform friction on every Rouse bead and another one considering larger branch point friction. Apart from chain connectivity between neighboring beads, both versions disregard interactions between the chains. Despite the tolerable description of the simulation data, neither model appears to reflect the mobility gradient accurately.
The detailed quantitative atomistic models employed here bridge the gap between the theoretical and general, coarse-granined models of star-like polymers which lack the indispensable chemical details.   
\end{abstract}

\maketitle

\section{\label{intro}Introduction}
Most of the industrially prepared polymer materials consist of polymers with branched or even hyperbranched architecture.~\cite{synth50} 
The ramification of the polymer structure due to the presence of one or more branch points is responsible for complex viscoelastic and dynamical properties as compared to linear chains.~\cite{Vlassopoulos2016,Daniel,Kanso} More specifically, the relaxation spectrum of branched polymer melts extends over various time decades, having a huge effect on the viscosity of these materials.~\cite{Mcleish} 
In the common attempt of the recent years to undercover hidden pieces of the complex picture concerning the dynamical behavior of macromolecular systems, the symmetric star-shaped polymers served as an excellent example of a model polymer with only one ramification point.~\cite{Milner_mcleish,Ball,Vlas_multiarm_dyn,Pakula2,Petra_branch,PetraStefan,petra_lau,stars,Grest,Christos_review,biomed_app}
Due to their well-defined architecture with $f$ equally-long arms connected to a one common branch point (in what follows we call it kernel), the main players in the structural-dynamical relationship are reduced to two principal characteristics: the number of the arms (or functionality) and the arm length. The latter determines whether the chain is classified as unentangled or entangled. 

From a theoretical standpoint, the Rouse model, initially developed for linear chains~\cite{Doi,Rouse}, is the standard model for studying the dynamics of unentangled chains. It approximates the polymer chain as a sequence of $N$ connected Kuhn segments or beads with each bead being characterized by its size $b$, the so-called Kuhn length, its friction $\zeta_0$, and its relaxation time, $\tau_0$.
It is to be noticed that the Rouse model is a coarse grained model owing to the fact that each Rouse segment comprises several actual monomers. The essential physics of the model are thermal (Brownian) motion and chain connectivity.~\cite{Doi,Rouse} Thermal motion gives rise to drag and random forces while chain connectivity gives rise to spring forces between adjacent beads. This physics is cast into a Langevin equation (see Eq.~\ref{Rouse_Equation} below), which describes the (translational and orientational) dynamics of the Rouse chain. 

This equation is solved by introducing normal modes.~\cite{Doi} Such modes represent chain reorientation within a subchain (blob) comprising $p$ segments, where $p$ denotes the mode index. Essentially, the chain is split into $N/p$ blobs with each
blob relaxing (reorienting) at time $\tau_p$, according to the expression:
\begin{equation}
\tau_p=\frac{\tau_R}{p^2} \quad\quad p=1,2,3...N.
\end{equation}
At the given time $\tau_p$, there are $p$ unrelaxed modes, that contribute to the stress relaxation modulus $G(\tau_p)$. 
The longest one ($p=1$) is called the Rouse time $\tau_R$. 
The mean square displacement $\langle r^2(t)\rangle$ of a Rouse segment can be expressed in terms of Rouse modes and this procedure leads to the following scaling regimes at short and long times:       
\begin{equation}
\langle r^2(t)\rangle\simeq \left\{ \begin{array}{ll}
	{2b^2}/{\pi^{1.5}} \tau_0^{-1/2} t^{1/2} & \mbox{$\tau_0\leq t\leq \tau_R$},\\
6D_{\rm{CM}}t & \mbox{$\tau_R\leq t$ }\end{array}\right.
\end{equation} 
where $T$ is the temperature and $D_{\rm{CM}}$ is the diffusion coefficient of the center-of-mass of the chain, i.e., $D_{\rm{CM}}=\tfrac{k_BT}{N\zeta_0}$ with $k_{\textrm{B}}$ being the Boltzmann constant (see Table~\ref{table:parameters} for the full list of variables and corresponding units). Apart from the segmental mean squared displacement (MSD), analytical predictions for several other physical quantities can be derived from the Rouse equation. We note for example, the decay of the autocorrelation function of the end-to-end vector~\cite{Mcleish}, the relaxation modulus $G(t)$ folowing a step shear strain~\cite{Doi}, and the dynamic (coherent and/or incoherent) structure factor~\cite{Doi}.  

When model predictions are compared to scattering data of specific linear polymers such as polyethylene or polybutadiene, various discrepancies have been found, especially at higher values of the scattering vector, i.e., at short length scales, where the simple coarse-grained picture cease to be valid.~\cite{VH_98,VH_reptation,Krushev,Richter_PRL,Bulacu,Smith_nongaus,Arantxa} 
Modifications of the Rouse model based on the internal viscosity or chain stiffness (semiflexible chain model SFCM) have been suggested in order to improve the agreement with the experimental data.~\cite{Allegra_IVM,Ganazzoli_IVM,Harnau} According to the findings of Refs.~\cite{VH_98,Krushev,Smith_nongaus}, the main drawback of the Rouse model is its inability to describe non-Gaussian monomeric motions.
These findings are further supported by those of Refs.~\cite{Diddens_PEO_2010,Diddens_PEO_2015}; there, the transalational dynamics of unentangled linear PEO chains, as obtained by atomistic molecular dynamics simulations, are better described by the SFCM model than the Rouse model. Recently, an alternative, microscopic approach explaining heterogeneous dynamics in unentangled polyethylene melts well above their glass transition temperature has been proposed, relating the anomalous dynamics to entropy fluctuations.~\cite{Borah} 

Extension of the Rouse model to polymers with star topology has been the subject of several works.~\cite{petra_lau,Colby_student,Watanabe_dist,Zimm_kilb} For example, Zimm and Kilb~\cite{Zimm_kilb} derived the expansion of the segmental position vector in terms of eigenfunctions for various model branched topologies, including symmetric stars. Further, using the derived expansions, they calculated the 
intrinsic viscosity of the chains both with and without hydrodynamic interactions. Following the eigenfunction expansions proposed by Zimm and Kilb, Watanabe {\it{et al.}}~\cite{Watanabe_dist} derived expressions for the relaxation modulus and the dielectric decay function of both symmetric and asymmetric unentangled stars. Nevertheless, theoretical predictions were not compared against experimental findings. 
Theoretical predictions for the viscoelastic properties of unentangled symmetric star polymers have been also obtained by means of a discrete Rouse model.~\cite{Colby_student}
In a more recent contribution, Ba\v{c}ov\'a {\it{et al.}}~\cite{petra_lau} derived analytical expressions for the MSD of both unentangled and entangled symmetric stars. Their expressions were derived from a 
continuous Rouse model for symmetric star polymers. Concerning entangled stars, each Rouse segment was confined by an additional localizing spring to represent entanglements. For such entangled chains, the authors~\cite{petra_lau} compared their theoretical segmental MSD against corresponding MSD data obtained from molecular dynamics simulations, which were based on a coarse-grained bead-spring description of polymer chains. A good comparison between theory and simulation findings was reported for chain section in the vicinity of the branch point. Further, the simulation results revealed a strong dispersion, over several decades, of the relaxation times after the local reptative (Rouse in tube) regime. Relaxation was dramatically slowed down by approaching the branch point from the outer segments.~\cite{petra_lau}
In other words, a significant mobility gradient along the star arm was observed, with the slowest components being placed close to the branch point (or kernel). This gradient was attributed to the entanglement constraints.~\cite{petra_lau,PetraStefan} Nevertheless, such a mobility gradient has been also observed in non-entangled mikto-arm stars.~\cite{Petra2} This finding indicates that mobility 
(or friction) gradient along a star arm is a general feature of star-like polymers. 

The second key characteristic of the stars, functionality $f$ or the number of arms, also heavily influences their dynamics. Depending on the functionality, the dynamic properties of the stars range from linear-like to colloid-like.~\cite{Vlas_multiarm_dyn,Pakula2,Vlassopoulos_colloid,Christos_review,Evelyne_multi,Leo} At low $f$, the theoretical assumption of an independent arm relaxation is
justified, however, with increasing functionality the presence of a second slower relaxation process is evident.~\cite{Vlas_multiarm_dyn} This process has a colloidal-like nature and is related to the nonuniform single-star monomer density distribution.~\cite{Christos_review} In other words, due to the high number of arms the central part of the star close to the kernel is impenetrable, with arms closely packed together, resembling a particle consisting of high-density core and penerable corona.  
As a consequence of the star-like architecture and factors contributing to the colloid-like nature of the multi-arm stars, their linear viscoelastic spectrum extends over various time decades and combines multiple relaxation mechanisms.~\cite{Glynos,Vlas_multiarm_dyn,Evelyne_multi} This feature renders them suitable as fillers in all-polymer nanocomposites.~\cite{Senses,Manos_review}
Simulations techniques allow for closer, separate inspection of each relaxation process, providing detailed information at studied time and length scales. Highly-coarse grained models in combination with rheological measurements have been used to address the shear-thinning behavior in the unentangled polystyrene star melts.~\cite{Fitzgerald} 
Probing the local dynamical properties, generic, bead-spring models and Monte Carlo techniques have been applied to study mostly the rotational segmental dynamics, analogical to the dynamics measured by dielectric spectroscopy.~\cite{Pakula,Pakula2,Vlas_multiarm_dyn,Chremos_Manos,Chremos_Tg}

In this work, we perform atomistic molecular dynamics simulations to study the local translational motion in star-shaped polymer melts comprising unentangled arms. The chosen method allows us to ``step back'' in the hierarchy of simulation methods and to probe the structural-dynamical relationship of specific star polymers by maintaning all chemistry-related details. 
By employing a realistic model of specific star polymers, we aim to provide a ``bottom-up'' description of the sub-diffusive (anomalous) dynamics and of the dynamical heterogeneities of star melts with very distinct flexibility and packing. We chose two representative polymers, poly(ethylene oxide), PEO, and polystyrene, PS.~\cite{Fetters_Me,PS_cinf,PEO_cinf}  
Both types of stars have been used in experimental studies, serving as model systems in non-entangled~\cite{Fitzgerald} and entangled regime\cite{Glynos,Coppola,Polgar_ketone,Clarke,Roovers_PSstars_f4_6,Roovers_PSstars_f4,Khasat_PSstars_f3_12}. 
We model non-entangled stars, having in mind the complications stemming from the topological constraints in entangled systems, as well as recent experimental data, which suggested that in the polystyrene stars with arm of low molecular weight the transition to colloidal-like behavior occurs at relatively low functionalities.~\cite{Glynos} 

The computational design of the model systems allows us to control the two main parameters of stars molecular architecture, i.e., the arm length and the functionality, ruling out the architectural dispersity as one of the main factors affecting the properties of synthetically prepared branched polymers.~\cite{Watanabe_dist,Frank1,Frank2} 
The monodisperse character of our samples also makes them an ideal candidate for testing the theoretically predicted behavior. Similar to Ref.~\cite{petra_lau} we compare our simulation results with theoretical predictions obtained from a continuous Rouse model for symmetric stars.  

\begin{table*}[t]
\begin{center}
\begin{tabular}{|l|l|l|}
\hline 
symbol & description & units \tabularnewline
\hline \hline
$N$, $N_a$&number of Kuhn segments in a linear chain, number of Kuhn segments in an arm& - \tabularnewline 
$b$&Kuhn length& nm \tabularnewline 
$t$ & time & ps\tabularnewline
	$T$ & temperature & K\tabularnewline
$\tau_0$&relaxation time of a Kuhn segment    & ps \tabularnewline
$\tau_p$&relaxation time of segment $p$ of the chain/arm, $p=1,2,...N$& ps  \tabularnewline 
	$p,q$ & mode indexes & -\tabularnewline
$\tau_R$, $\tau_{R_a}$&Rouse time of a linear chain, Rouse time of an arm ($\tau_1$)& ns \tabularnewline
	$\zeta_0$&friction of Kuhn segment& N.ps.nm$^{-1}$ \tabularnewline
$r_{\alpha,\ell,t}$ & position vector of $\ell$-th segment of the $\alpha$ arm at time $t$& -\tabularnewline
$C_{\infty}$& characteristic ratio of the chains& - \tabularnewline
n$_{\rm{max}}$ & number of bonds along the arm & - \tabularnewline
$\theta$& supplementary angle to the one between three atoms along the arm as defined in angle potential& degrees \tabularnewline
	MSD & monomer mean squared displacement & nm$^2$\tabularnewline
        $G(\bf{r}, t)$ & Van Hove function for the monomers in the region adjacent to the star kernel & -\tabularnewline
	$\alpha_2$ & non-Gaussian parameter for the monomers in the region adjacent to the star kernel & -\tabularnewline
	$D(t)$ & time-dependent monomeric diffusion coefficient & nm$^2$.ps$^{-1}$\tabularnewline
	$D_{\rm{CM}}$ & diffusion coefficient of the chain center of mass & nm$^2$.ps$^{-1}$\tabularnewline
\hline 
\end{tabular}
\end{center}
\caption{List of the variables used in the current study.}
\label{table:parameters}
\end{table*}

\section{\label{model}Simulation details}
\begin{figure*}[!ht]
{\centering{\includegraphics[width=0.8\textwidth]{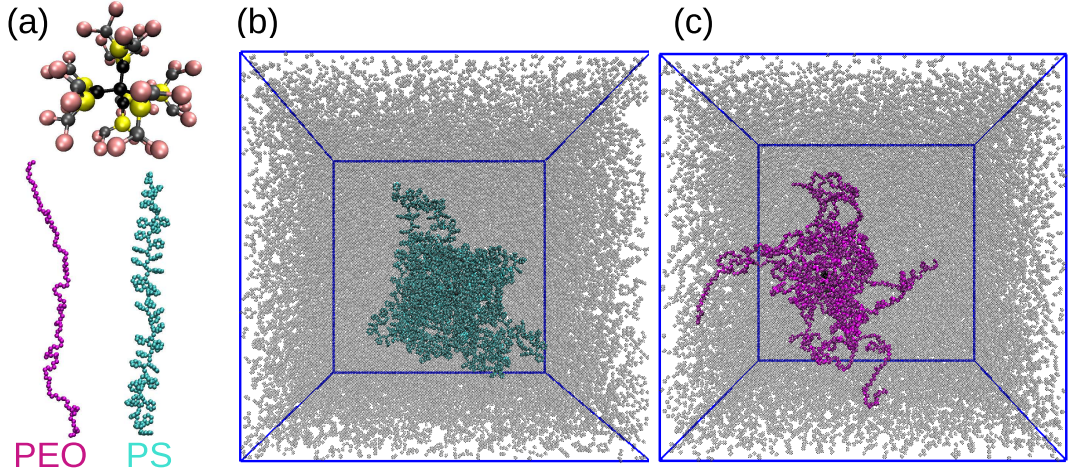}} \par}
	\caption{(a) Building blocks of a star molecule. The dendritic kernel with highlighted generations: black (1st), yellow (2nd), gray (3rd), pink (4th); together with snapshots of preequilibrated star arms. (b) Snapshot of a selected equilibrated (PS)$_{16}$ star (cyan) within the melt (grey color, transparent beads). (c) Snapshot of a selected equilibrated (PEO)$_{16}$ star (magenta) within the melt (grey color, transparent beads). For better visualization the color and the size of the beads were adjusted and do not correspond to actual characteristics given by the force field.}
\label{snaps}
\end{figure*}

The melts of non-entangled star-shaped polymers consist of stars with either atactic polystyrene (PS) or poly(ethylene oxide) (PEO) arms. 
The stars are symmetric, i.e., each arm has the same length of $m=40$ monomers.
The specific number of arms ($f$) is attached to a central kernel with a dendritic structure, mimicking the commonly used carbosilane dendrimers in experimental systems.~\cite{dendrimers,Glynos,core_PEO} 
The kernel is composed of C, CH and CH$_2$ carbon units and the number of generations in the kernel depends on the functionality $f$ (number of arms).
More specifically, the kernel of a star with $f=4$ consists of only 5 black units (1st generation) in Fig.~\ref{snaps}, a kernel of a star with $f=8$ includes 5 black and 8 yellow units (1st and 2nd generation) and so on. The total number of units that the kernel of a star is composed of is notated as $N_K$. 
As reference systems, we simulate additionally two melts, (PS)$_{\rm{lin}}$ and (PEO)$_{\rm{lin}}$, composed of linear chains of the same molecular weight as each arm of the above-mentioned stars.
The notation as well as the details about the composition of each simulated system are summarized in Table~\ref{table:systems}.

\begin{table}
\begin{center}
\begin{tabular}{|l|l|l|l|l|l|}
\hline 
	notation & f & m [mers]& $N_K$& $N_S$ & $M_{\textrm{w}}$ [g/mol]  \tabularnewline
\hline \hline
(PS)$_4$          &  4  & 40 &5 & 15 & 16768 \tabularnewline 
(PS)$_8$          &  8  & 40 &13& 15 & 33576 \tabularnewline 
(PS)$_{16}$       &  16 & 40 &29& 15 & 67192 \tabularnewline
(PS)$_{32}$       &  32 & 40 &61& 15 & 134424\tabularnewline
(PEO)$_{4}$       &  4  & 40 &5 & 30 & 7168  \tabularnewline 
(PEO)$_{8}$       &  8  & 40 &13& 30 & 14376 \tabularnewline
(PEO)$_{16}$      &  16 & 40 &29& 30 & 28792 \tabularnewline 
(PEO)$_{32}$      &  32 & 40 &61& 30 & 57624 \tabularnewline
(PS)$_{\rm{lin}}$ &  1  & 40 &0 &250 & 4176  \tabularnewline
(PEO)$_{\rm{lin}}$&  1  & 40 &0 &252 & 1775  \tabularnewline
\hline 
\end{tabular}
\end{center}
\caption{System composition: $f$ stands for functionality of each star, $m$ for number of monomers per arm/chain, $N_K$ represents the total number of united atoms constituting the kernel of each star, $N_S$ is the number of molecules in each system and $M_{\textrm{w}}$ the molecular weight of the star.}
\label{table:systems}
\end{table}

All the simulations were performed with the Gromacs~\cite{gromacs} simulation package employing the united-atom model of TRAPPE force field~\cite{trappe2, trappe1,PEOff}. 
In the united-atom model the hydrogens are not simulated explicitly and together with the carbon they form one unit, e.g., CH2. From now on, we refer to the united atoms as atoms.
The temperature was maintained at the value 600K for the PS and 450K for the PEO melts. Notice that the difference in the glass transition temperature of those two polymers is about 150K~\cite{Tg}, thus the two model systems are almost at equidistant temperatures from their glass transition.

We followed a multi-stage equilibration protocol for the preparation of the stars, which has been used in previous atomistic simulation studies of mikto-arm stars.~\cite{Petra2,Petra3,Petra4} 
In brief, first, we attached fully stretched arms to the atoms of the last generation of the kernel and run short runs to avoid overlaps and minimize the energy of the artificially created initial configuration.
Second, when the initial single-star configuration was ready, $N_S$ stars were inserted randomly into a simulation box and equilibrated with a sequence of heating and equilibration runs.
Due to the short, non-entangled arms and high simulation temperature, no advanced equilibration methods such as prepacking~\cite{petra_lau} reported previously for entangled systems were necessary. 
Then, production runs of about 100ns, are performed with the time step of 1 fs and with Nos\'{e}-Hoover thermostat in combination with Parrinello-Rahman barostat under a constant pressure of 1 atm.    
For more simulation details we refer the readers to our recent publication.~\cite{usMTS}

\begin{figure}[!ht]
{\centering{\includegraphics[width=0.45\textwidth]{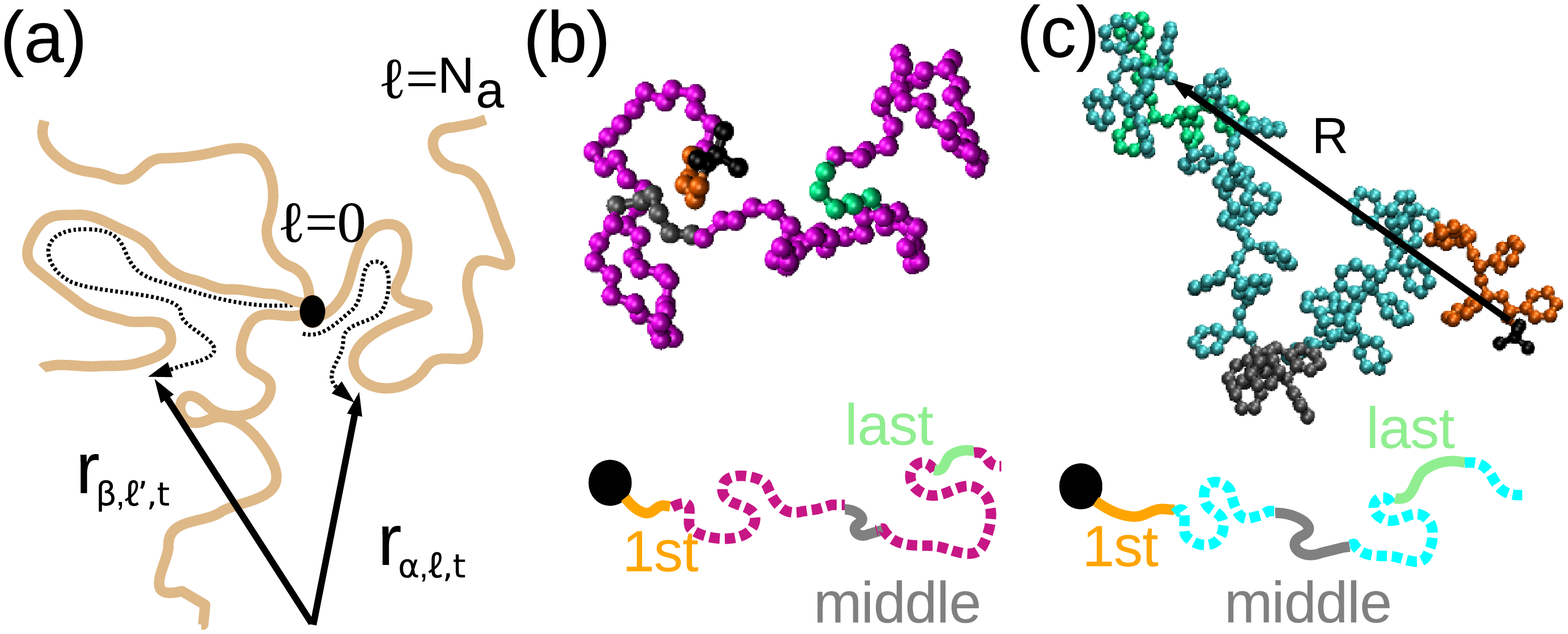}} \par}
	\caption{Schematic illusttration of the labelling of the position of the monomer on the arm in (a) the continuous Rouse model, (b) the PEO arm of (PEO)$_4$, (c) the PS arm of (PS)$_4$. The black dots represent the kernel, for (a) it is a point where $\ell=0$. Monomers in the middle region in (b) belong to the 8th Kuhn segments from the kernel, in (c) to the 4th Kuhn segments from the kernel (see the main text). The magenta and cyan monomers (dashed lines in scheme) in (b) and (c), respectively, were excluded from the analysis. The arrow in (c) illustrates an example of the center-to-end vector, pointing from the arm attachment to the end of the arm.} 
\label{label}
\end{figure}

\section{Rouse model for non-entangled stars}
\subsection{Description of the model}

In this section we recount the main features of the continuous Rouse model used further for the comparison with the simulation data. A thorough description
of the model as well as the full derivation of the theoretical equations for the MSDs of the star segments that are presented in Table~\ref{table:MSDPureRouse}
can be found in the original publication.~\cite{petra_lau} Irrespective of chain topology, the Langevin equation for an unentangled Rouse chain reads ~\cite{Rouse,Doi}
\begin{equation} \label{Rouse_Equation}
\zeta_0 \frac{{\partial} {\bf r}_{\alpha,\ell,t}}{\partial t} = \frac{3k_BT}{b^{2}} \frac{{\partial^2}  {\bf r}_{\alpha,\ell,t}}{\partial \ell^2 } +{\bf g}(\alpha,\ell,t)
\end{equation}
The left hand side of Eq.~\ref{Rouse_Equation} represents the drag force while the first and second terms on the right hand side represent the spring and random forces, respectively. Further, ${\bf r}_{\alpha,\ell,t}$ denotes the position vector of the $\ell th$ segment in the arm $\alpha$ at time $t$. 
Fig.~\ref{label}(a) offers a schematic illustration of ${\bf r}_{\alpha,\ell,t}$. The same figure also illustrates the labeling of the Rouse segments along an arm; $\ell=0$ corresponds to the branch point (kernel) while $\ell=N_a$ to the arm tip. It is important to note that the friction is uniformly distributed along the star arm, with the segmental friction being $\zeta_0$. The boundary conditions of Eq.~\ref{Rouse_Equation} are chain topology dependent. For linear chains, there is only one boundary condition, namely, the absence of tension (force) at both chain ends. Apart from this condition, there are two more boundary conditions for star polymer chains, namely, chain connectivity and force balance at the branch point.~\cite{petra_lau} As a result the expansion of ${\bf r}_{\alpha,\ell,t}$ to eigenfunctions is topology dependent too. As demonstrated in Ref.~\cite{petra_lau}, for symmetric stars of arbitrary functionality, the appropriate position vector expansion satisfying the aforementioned three boundary conditions is
\begin{multline}\label{Modes_Rouse_Star}
{\bf r}_{\alpha,\ell,t} = \sum_p {\bf X }_{p}^{c}(t) \Psi_{p}^{c}(\ell) +\\ +\sum_q \left( {\bf X}_{q}^{s_1}(t) \Psi_{q}^{s_1}(\alpha,\ell)+ \ldots +{\bf X}_{q}^{s_{f^{\prime}}}(t) \Psi_{q}^{s_{f^{\prime}}}(\alpha,\ell) \right).
\end{multline}
\begin{figure}[!ht]
{\centering{\includegraphics[width=0.43\textwidth]{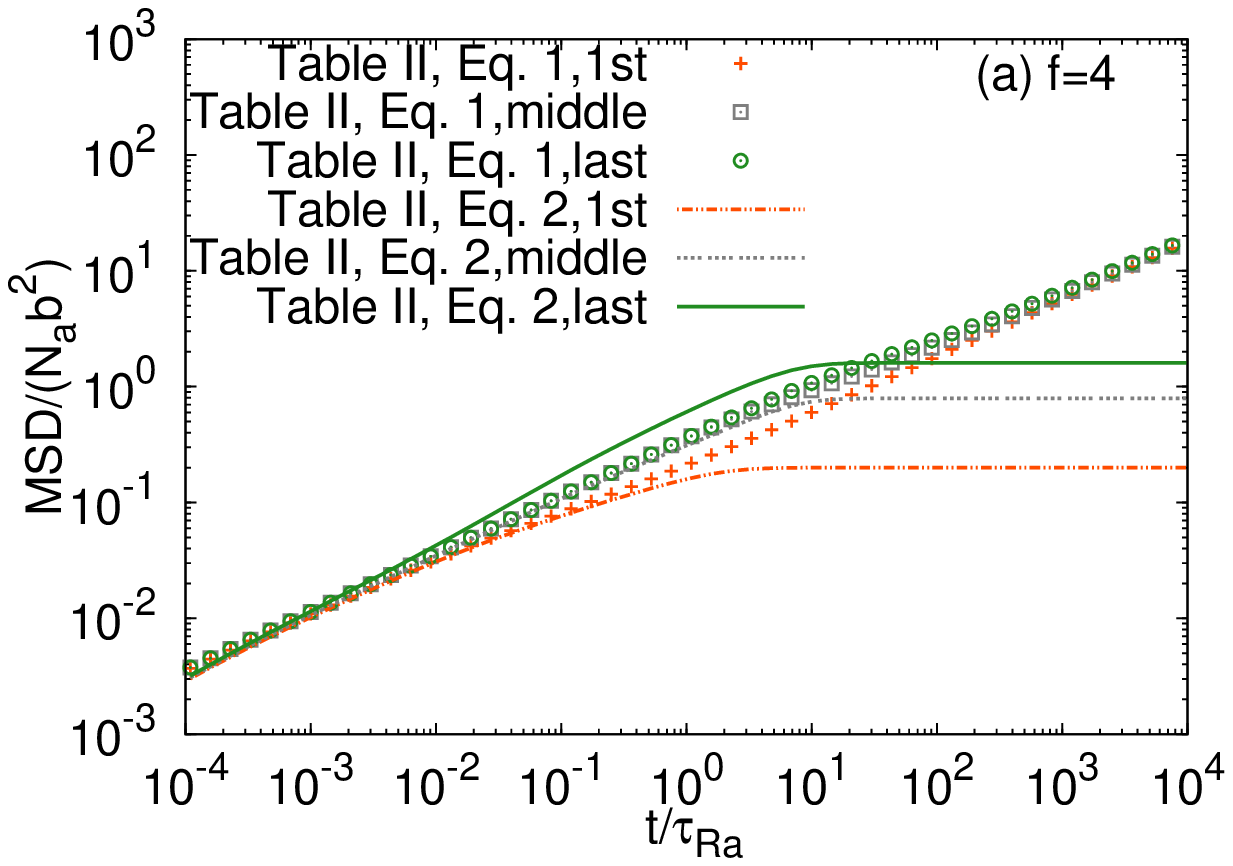}} \par}
{\centering{\includegraphics[width=0.43\textwidth]{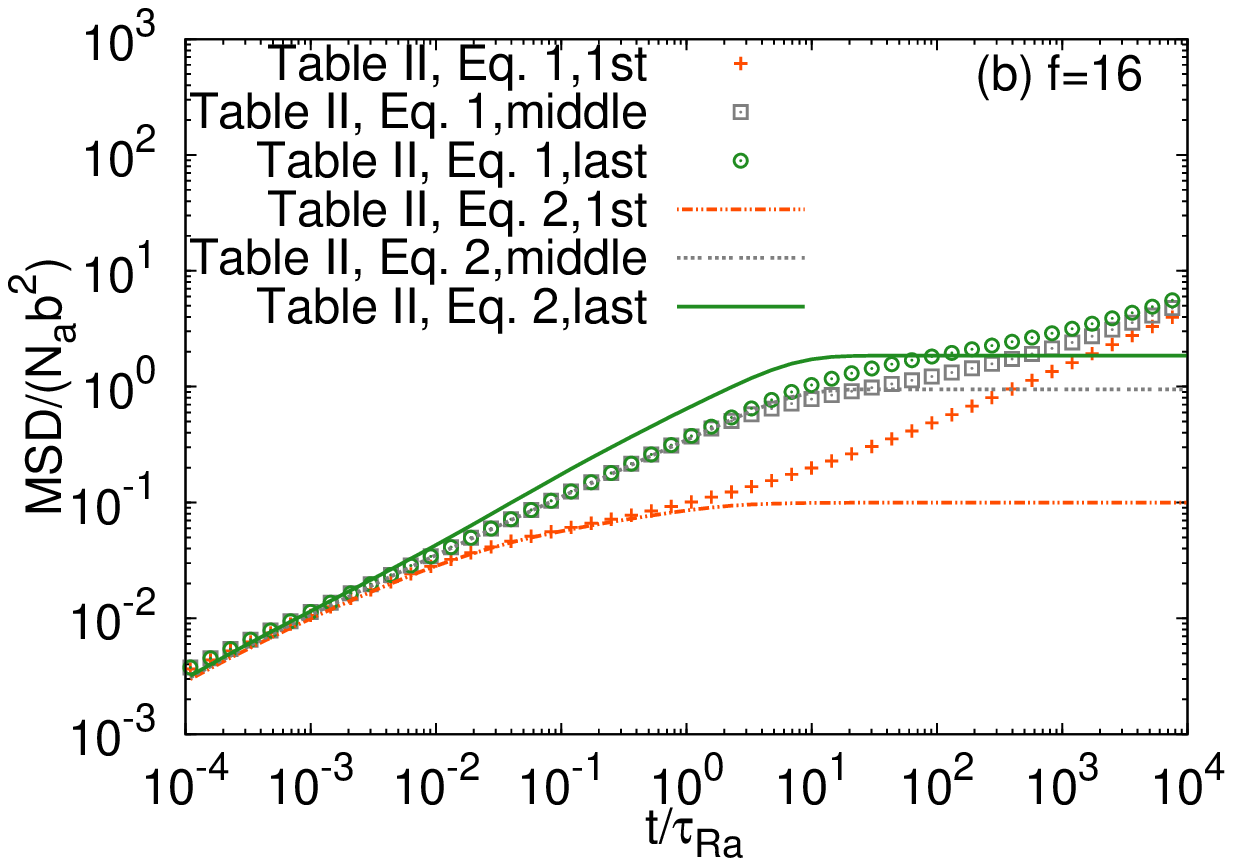}} \par}
	\caption{Comparison of the segmental MSD expressions of the Rouse star model for (a) $f=4$ and (b) $f=16$. Symbols correspond to the first equation of Table~\ref{table:MSDPureRouse} and lines to the second equation of Table~\ref{table:MSDPureRouse}. Labels ``1st,middle,last'' refer to the position of the segment, $\ell$. The parameters $N_a$ and $b$ for the PEO polymer type from Table.~\ref{table:par} were used.}
\label{MSDcoAnaSum}
\end{figure}
There are two types of eigenfunctions in Eq.~\ref{Modes_Rouse_Star}, namely cosine eigefunctions, $\Psi_{p}^{c}(\ell)$, with degeneracy of one, and sine eigenfunctions, $\Psi_{q}^{s_i}(\alpha,\ell)$, with degeneracy of $f^{\prime} = f-1$. For example, for a symmetric three arm star, there are two sine eigenfunctions and one cosine eigenfunction. The two types of eigenfunctions are indexed with the mode numbers $p$ and $q$, respectively. Notice that the expansion of Eq.~\ref{Modes_Rouse_Star} is similar to that of Refs.~\cite{Zimm_kilb,Watanabe_dist}. The explicit expressions of the cosine and sine eigenfunctions are respectively
%
\begin{subequations} \label{Eigenmodes}
\begin{align}
\Psi_{p}^{c}(\ell) =& \cos \left( \frac{p \pi \ell}{N_a} \right) \label{Eigenmodesa} \\
\Psi_{q}^{s_i}(\alpha,\ell) =& s_{i\alpha} \sin \left(\frac{ (2q-1) \pi \ell}{2N_a} \right). \label{Eigenmodeb}
\end{align}
\end{subequations}
The numerical coefficients $s_{i\alpha}$ of the sine eigenfunctions satisfy the following constraints
\begin{subequations}
\label{ConstrainsRouseStar}
\begin{align}
& \sum_{\alpha=1}^{f} s_{i\alpha}=0 \label{Constrainsa} \\
& \sum_{\alpha=1}^{f} s_{i\alpha}^2=f\label{Constrainsb}  \\
& \sum_{\alpha=1}^{f} s_{i\alpha} s_{j\alpha}=0, \label{Constrainsc}
\end{align}
\end{subequations}
where indices $i,j$ denote the $i$th and $j$th sine eigenfunction, respectively. Equation~\ref{Constrainsa} is a consequence of the force balance at the branch point. Equations~\ref{Constrainsb} and \ref{Constrainsc} arise from normalization and orthogonality, respectively.

In Eq.~\ref{Modes_Rouse_Star}, ${\bf X }_{p}^{c}(t)$ and ${\bf X }_{q}^{s_i}(t)$ denote eigenmode amplitudes. In deriving MSD correlation functions, only averages of the form $\langle  X_{p_{\mu}}^c (t)  X_{p^{\prime}_{\nu}}^c (t^{\prime}) \rangle$ and $\langle  X_{q_{\mu}}^{s_i} (t)  X_{q^{\prime}_{\nu}}^{s_i} (t^{\prime}) \rangle$ survive due to the constraints of Eqs.~\ref{ConstrainsRouseStar}. For a complete derivation of the MSD correlation functions the interested reader is referred to Appendix A of Ref.~\cite{petra_lau}. The final expression for segmental motion, i.e., $\langle ( {\bf r}_{\alpha,\ell,t}-{\bf r}_{\alpha,\ell,t^{\prime}} )^2 \rangle$, is quoted in Table~\ref{table:MSDPureRouse} as Eq. (1).
In this expression, $\Phi(x)$ is the error function given by $\Phi(x)=\tfrac{2}{\sqrt{\pi}} \int_{0}^{x} e^{-u^2} \, du $. Further, $\widetilde{t}_{R_a}=|t-t^{\prime}|\tau_{R_a}^{-1}$ is the time normalized by the arm Rouse time, $\tau_{R_a}$. The latter is related to the segmental relaxation time, $\tau_0$, and the number of arm segments, $N_a$, through $\tau_{R_a}=\tau_0 N_a^2$. For $f=2$, $\langle ( {\bf r}_{\alpha,\ell,t}-{\bf r}_{\alpha,\ell,t^{\prime}} )^2 \rangle$ provides the segmental MSD for a segment that belongs to a linear chain, i.e., $\tfrac{2N_ab^2}{\pi^{1.5}} \sqrt{\widetilde{t}_{R_a}}$. Concerning the branch point, the segmental MSD prediction reduces to $\frac{2}{f}\frac{2N_ab^2}{\pi^{1.5}} \sqrt{\widetilde{t}_{R_a}}$. Compared to the segmental motion for linear chains, the latter expression exhibits the same power law dependence of MSD on time, meaning that during local chain reorientation the branch point undergoes subdiffusive motion similar to a linear chain segment. Nevertheless, the branch point experiences stronger localization that is expressed by the $\frac{2}{f}$ prefactor. It should be stressed that in deriving the first equation of Table~\ref{table:MSDPureRouse}, sums over $p$ and $q$ modes are approximated by integrals. From a physics perspective, this approximation means that fast Rouse modes (i.e., large $p$ and $q$) dominate the dynamics. 
In this respect, as will be shown below the first formula of Table~\ref{table:MSDPureRouse} is unable to describe large scale reorientation of the star chains. Such reorientation is anticipated to occur at timescales of order $\tau_{R_a}$ and above.

To prescribe slow Rouse modes to the model, the MSD correlation functions should be evaluated by summation. Concerning the segmental MSD, the final expression is presented in the second equation of Table~\ref{table:MSDPureRouse}. Unlike the first expression, it is valid up to timescales of the order the Rouse relaxation time of the arms since it incorporates slow Rouse modes. The upper summation limit is $N_a$, i.e., the number of Rouse segments in an arm. 
For self-consistency, the predictions of the first expression in Table~\ref{table:MSDPureRouse} should be recovered in the limit of large $N_a$, provided that $t \ll \tau_{R_a}$. To verify this and, moreover, to identify the range of timescales over which the expression incorporating integrals is valid, we compare the predictions of the first two expressions of Table~\ref{table:MSDPureRouse}. Note that the first term of the second expression is neglected in the calculation as it reflects CM motion of the star chains. Such motion is obviously omitted in the expression (1) as well. The comparison can be seen in Fig.~\ref{MSDcoAnaSum}. There, lines refer to the second equation of Table~\ref{table:MSDPureRouse} while symbols represent the predictions of the first expression. 
The $x$ axis is normalized time, i.e., $\widetilde{t}_{R_a}$. The $y$ axis is rescaled MSD, i.e., MSD divided by $N_a b^2$. The $\tau_{R_a}$ and $N_a b^2$ parametrization is discussed later in the manuscript. Notice that, from the Rouse model perspective, $\tau_{R_a}$ and $N_a b^2$ values are functionality independent. With respect to the summation upper limit, $N_a=300$. Higher $N_a$ values do not alter the presented result.

\begin{table*}[t]
\begin{center}
	\caption{Segmental MSD of segments placed on the same arm $\alpha$ in an unentangled star. Expressions (1) with fast Rouse modes dominating the dynamics, (2) including slow Rouse modes, (3) with the branch point friction being $f\zeta_0$ are presented.}   
\begin{tabular} {|l|} \hline \hline
$\left\langle \left({\bf r}_{\alpha,\ell,t} - {\bf r}_{\alpha,\ell,t^{\prime}} \right)^2 \right\rangle$ expression \\ \hline \hline
(1.) $2b^2 \ell \left(\frac{f-2}{f} \right) \left[ 1- \Phi \left( \frac{\pi \ell}{\sqrt{\widetilde{t}_{R_a}}N_a}  \right)\right] +\frac{2 N_a b^2 }{\pi^{1.5}} \sqrt{\widetilde{t}_{R_a}} \left[ 1- \left(\frac{f-2}{f} \right) \exp \left({\frac{-\pi^2 \ell^2 }{\widetilde{t}_{R_a}N_a^2}} \right) \right] $ 
$$\\
(2.) $\left(\frac{2 N_a b^2}{\pi^2 \tau_{R_a} f} \frac{N_a}{N_a + f^{-1}}\right)t + \frac{4 N_a b^2}{\pi^2 f} \displaystyle \sum_{p=1}^{N_a} \cos^2 \left(\frac{p \pi \ell}{N_a} \right) \frac{ \left[ 1 - \exp \left(-\widetilde{t}_{R_a} p^2\right)  \right] }{p^2} + \frac{16 N_a b^2}{\pi^2} \frac{f-1}{f} \displaystyle \sum_{p=1}^{N_a} \sin^2 \left(\frac{ (2p-1) \pi \ell}{2 N_a} \right) \frac{ \left[ 1 - \exp \left(-\frac{\widetilde{t}_{R_a} (2p-1)^2 }{4} \right)  \right] }{(2p-1)^2}  $ \\
$$\\
(3.) $\left(\frac{2 N_a b^2}{\pi^2 \tau_{R_a} f} \frac{N_a}{N_a+1}\right)t  + \frac{4 N_a b^2}{\pi^2 f} \displaystyle \sum_{p=1}^{N_a} \cos^2 \left(\frac{ \pi (2 \ell +1) p}{2(N_a+1)} \right) \frac{ \left[ 1 - \exp \left(-\widetilde{t}_{R_a} p^2\right)  \right] }{p^2}  + \frac{16 N_a b^2}{\pi^2} \frac{f-1}{f}  \displaystyle \sum_{p=1}^{N_a} \cos^2 \left(\tfrac{(2p-1) \pi ( 2 N_a - 2 \ell + 1  )}{2 (2 N_a + 1)} \right)
\frac{ \left[ 1 - \exp \left(\frac{-\widetilde{t}_{R_a} (2p-1)^2 }{4} \right)  \right] }{(2p-1)^2} $ \\ \hline
\end{tabular}
\label{table:MSDPureRouse}
\end{center}
\end{table*}

From Fig.~\ref{MSDcoAnaSum}, it becomes obvious that, in the $t \ll \tau_{R_a}$ regime, the predictions of the expression (1) in Table~\ref{table:MSDPureRouse} are recovered. Concerning the validity range of the latter formula, it depends on the functionality and the position along the star arm. Irrespective of functionality, the expression incorporating integrals behaves poorly for the outer star segment (green curves) even at short timescales. For the middle monomer, the prediction using expression (1) in Table~\ref{table:MSDPureRouse} compares well with the summation result over the entire normalized time domain, irrespective of functionality (grey curves). As regards the section attached to the branch point (orange curves), the equation (1) performs reasonably up to $\widetilde{t}_{R_a} \simeq 0.5$, for low functionalities. For the highest two functionalities, it performs fairly up to the Rouse relaxation time of the arms. In view of the findings of Fig.~\ref{MSDcoAnaSum}, we disregard the integrated equation (1) in the remaining of the manuscript. Notice that the short time behavior of the summation expression is very sensitive to the upper limit value of the sum. This feature is discussed in more detail in Section I of the Supplementary Information (Fig. S1). Nevertheless, the behavior of the summation expression at intermediate and long times ($\widetilde{t}_{R_a} > 0.1$) is rather insensitive to the upper limit of the sum.

The third expression in Table~\ref{table:MSDPureRouse} corresponds to the segmental MSD prediction of Keesman {\it{et al.}}~\cite{Keesman}. It is derived from a Rouse model for symmetric star polymers also. From a conceptual point of view, the most salient difference between the two models is the branch point friction. Our model assumes that the branch point has the same friction as all other arm segments, i.e., $\zeta_0$. In contrast, the Rouse model variant of Keesman {\it{et al.}} assigns a larger friction coefficient to the branch point as compared to the arm segments. In particular, the branch point friction is $f\zeta_0$ rather than $\zeta_0$. Actually, the molecular picture of Ref.~\cite{Keesman} represents the actual chemistry as well as the kernel structure in our simulations more realistically than the $\zeta_0$ scenario. In this respect, the Keesman {\it{et al.}} predictions will also be discussed later in the manuscript.

\subsection{Linking atomistic details to Rouse model parameters}
\label{match}
To compare Rouse model predictions and simulation results, the number of Rouse segments, $N_a$, the segmental relaxation time, $\tau_0$, and the Kuhn length, $b$ are required.
From the outset, we stress that we aim to obtain reasonable estimates of these parameters in order to discuss (in a semi-quantitative manner) simulation results with respect to Rouse model results that serve as reference predictions reflecting a well-defined friction distribution along the arm and, moreover, absence of interactions between the Rouse beads (apart form chain connectivity). 

Utilizing concepts regarding conformations of ideal chains~\cite{rubinstein}, the number of Rouse (Kuhn) segments per arm can be obtained from the expression 
\begin{equation}\label{N_Kuhn} 
N_{a}=\frac{n_{\rm{max}}\cos^2{(\theta/2)}}{C_{\infty}} 
\end{equation}
where $n_{\rm{max}}$ is the number of bonds along the star arm, $\theta$ is the supplementary angle to the one used in the angle potential between atoms along the arm and $C_{\infty}$ is the characteristic ratio of the chains. More specifically, $\theta=\pi-\theta'$, where $\theta'$ is the backbone bond angle.
There are $n_{\rm{max}}=79$ bonds between the united atoms along the PS arm (the side aromatic rings are excluded) and $n_{\rm{max}}=119$ bonds between the united atoms along the PEO arm. 
The angle $\theta$ of the current atomistic model is the same for both polymers, i.e., $\theta=68^{\circ}$.~\cite{trappe2, trappe1}
Concerning the characteristic ratio, we consider $C_{\infty}\approx 7.56$ and $C_{\infty}=5.5$ for the PS and PEO chains, respectively.
The former value refers to $T=600$K and is calculated from the experimentally determined dependence of $C_{\infty}$ on temperature.~\cite{PS_cinf} The value for PEO is reported in Ref.~\cite{PEO_cinf}. 
Utilizing Eq.~\ref{N_Kuhn} we obtain $N_{a,\rm{PS}}=7.18$ for PS and $N_{a,\rm{PEO}}=14.87$ for PEO. 
Nevertheless, for the sake of simplicity, the values $N_{a,\rm{PS}}=7$ and $N_{a,\rm{PEO}}=15$ are used hereafter. 

\begin{figure}[!ht]
{\centering{\includegraphics[width=0.43\textwidth]{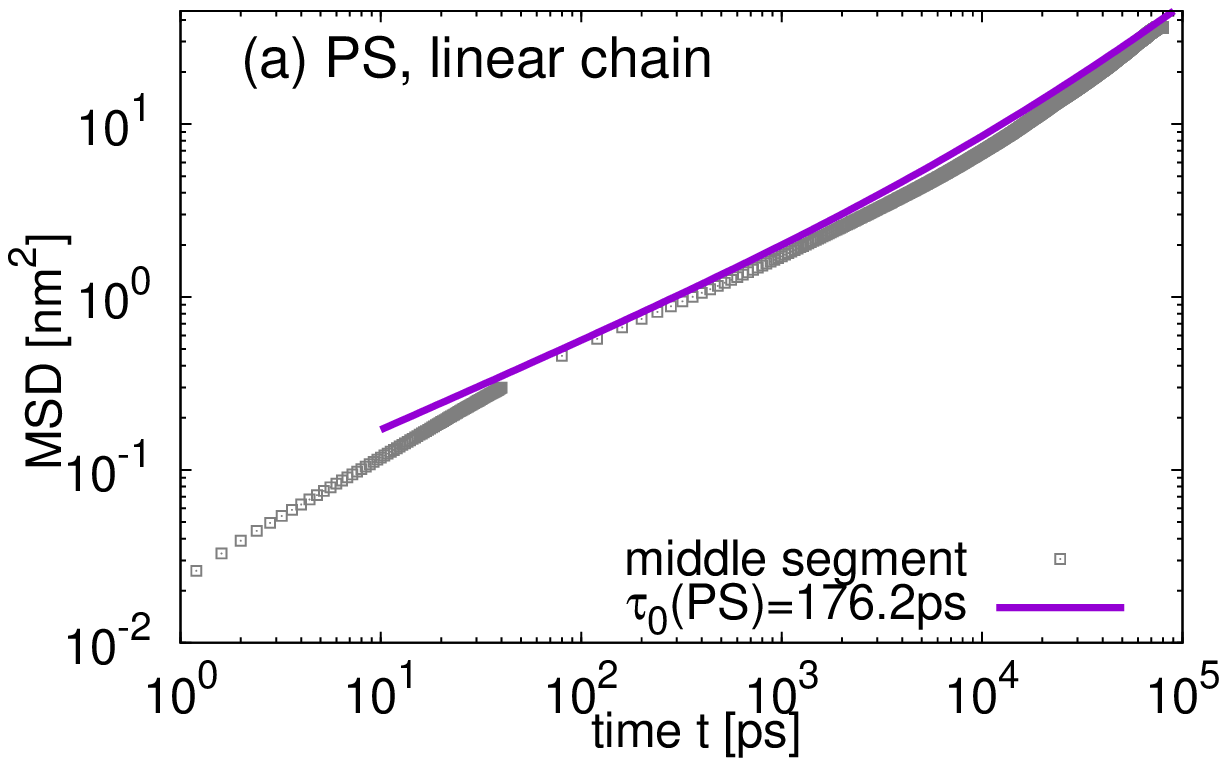}} \par}
{\centering{\includegraphics[width=0.43\textwidth]{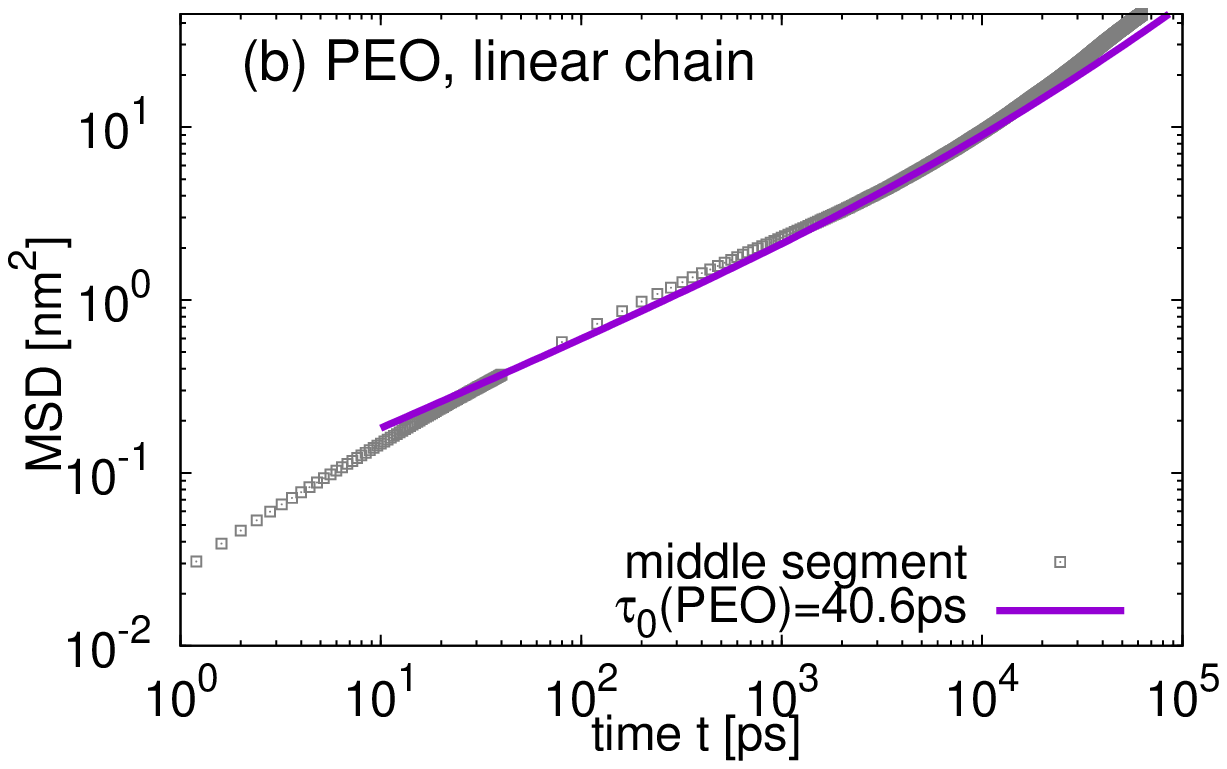}} \par}
	\caption{MSD of the middle segment of the (a) PS and (b) PEO linear chains (symbols) together with theoretical predictions (Eq.~\ref{RouseLinear}, lines) using the set of parameters given in the Table~\ref{table:par}.}
\label{rouse}
\end{figure}

The Kuhn lengths can be determined utilizing the same concepts as for the $N_{a}$ estimation. The Kuhn lengths are obtained from the following relation~\cite{rubinstein}:

\begin{equation}
b=\frac{C_{\infty}l_b}{\cos{(\theta/2)}}
\label{b_Kuhn}
\end{equation}
with $l_b$ being the average length of elementary bonds, which equals $0.154$nm and $0.147$nm for PS and PEO, respectively. According to Eq.~\ref{b_Kuhn}, $b_{\rm{PS}}=1.4$nm and $b_{\rm{PEO}}=0.98$nm. 

To estimate the remaining parameter $\tau_0$, we focus on the linear chains. 
We used three parametrization methods to obtain $\tau_0$, their description can be found in the Supplementary Information.

The method which gives the $\tau_0$ values which describe the simulation data satisfactorily utilizes the early MSD response of the middle chain monomer. 
In accordance with the approach adopted by Theodorou~\cite{Theodorou}, we consider $\tau_0$ to be the timescale at which the MSD simulation data equal to $b^2/3$. 
The so-obtained parameters are listed in Table~\ref{table:par}.
When this set of parameters is employed in the following equation for the Rouse model:
\begin{equation}\label{RouseLinear} 
\langle \big( {\bf r}_{\ell,t}-{\bf r}_{\ell,0} \big)^2 \big \rangle = \frac{2b^2}{\pi^2 N_a \tau_0}t + \frac{2N_ab^2}{\pi^{1.5}} \sqrt{\widetilde{t}_{R_a}}
\end{equation} 
where $\tau_{R_a}=\tau_{0}N_a^2$ and $\widetilde{t}_{R_a}=t\tau_{R_a}^{-1}$, the simulation data are described in good detail (compare the symbols and the solid lines in Fig.~\ref{rouse}).
It is to be noticed that Eq.~\ref{RouseLinear} actually represents an infinite long linear chain therefore it obeys sub-diffusive Rouse dynamics even at short timescales, unlike the simulation data.

Before focusing on the star polymers, we stress that, at early times (i.e., around $\tau_0$), the simulation MSD results for the middle linear chain monomer almost overlap
with the MSD data for the outer and middle arm monomers of the stars in both studied chemistries (see Figs.~\ref{msd_lk}(c,d)). 
Therefore, within the accuracy of the simulation data, essentially the same $\tau_0$ values can be obtained from the MSD of these star segments. 
Moreover, for a considerable amount of time, the outer star monomers behave (to a very good extent) as if they belonged to the central section of a linear chain - see the discussion related to Fig.~\ref{msd_lk} for more details about the time scales when the deviation in the behavior of the outer monomers occurs. 

\begin{table}
\begin{center}
\begin{tabular}{|l|l|l|}
\hline 
parameter & PS & PEO  \tabularnewline
\hline \hline
$b$   &  1.4 nm & 0.98 nm   \tabularnewline 
$N_a$ &  7  & 15  \tabularnewline 
$\tau_0$ &  176.2 ps & 40.6 ps  \tabularnewline
$\tau_{R_a}$ &  8.6 ns & 9.1 ns \tabularnewline
\hline 
\end{tabular}
\end{center}
\caption{The Rouse model parametrization that is used for the comparison with the simulation data.}
\label{table:par}
\end{table}

\section{Results}
\label{res}
In order to provide a consistent comparison between the two polymer types as well as between the simulation data and the theoretical model, we analyze the dynamics at the length scales of the order of the Kuhn length.  
We assume that a fully extended conformation of a star arm is identical to that of a fully extended linear chain. Thus, we consider 7 Kuhn segments per arm in the PS stars and more than twice as many, i.e., 15, Kuhn segments per arm in the PEO stars.
Hence, we proceed with the dynamical analysis as follows: firstly, we divide the monomers on each arm into 7 (PS stars) and 15 (PEO stars) regions, assigning in this way monomers to the particluar Kuhn segment. We start labelling from the arm attachment to the kernel, as it is schematically illustrated in Fig.~\ref{label}, thus the ``first'' region is the one adjacent to the kernel. We define as the ``middle'' region the 8th Kuhn segment in the case of (PEO)$_f$ and the 4th Kuhn segment in the case of (PS)$_f$. The ``last'' region corresponds to the last Kuhn segment from the arm tip, i.e., to the 7th one in the PS stars and 15th one on the PEO arms. 
The monomers which do not belong to the first, middle or last region as well as few remaining monomers located on the arm tip were excluded from the comparison with the theoretical model.

\subsection{Initial observations from simulation results}
\label{msd_sim}
\begin{figure*}[!ht]
\includegraphics[width=0.47\textwidth]{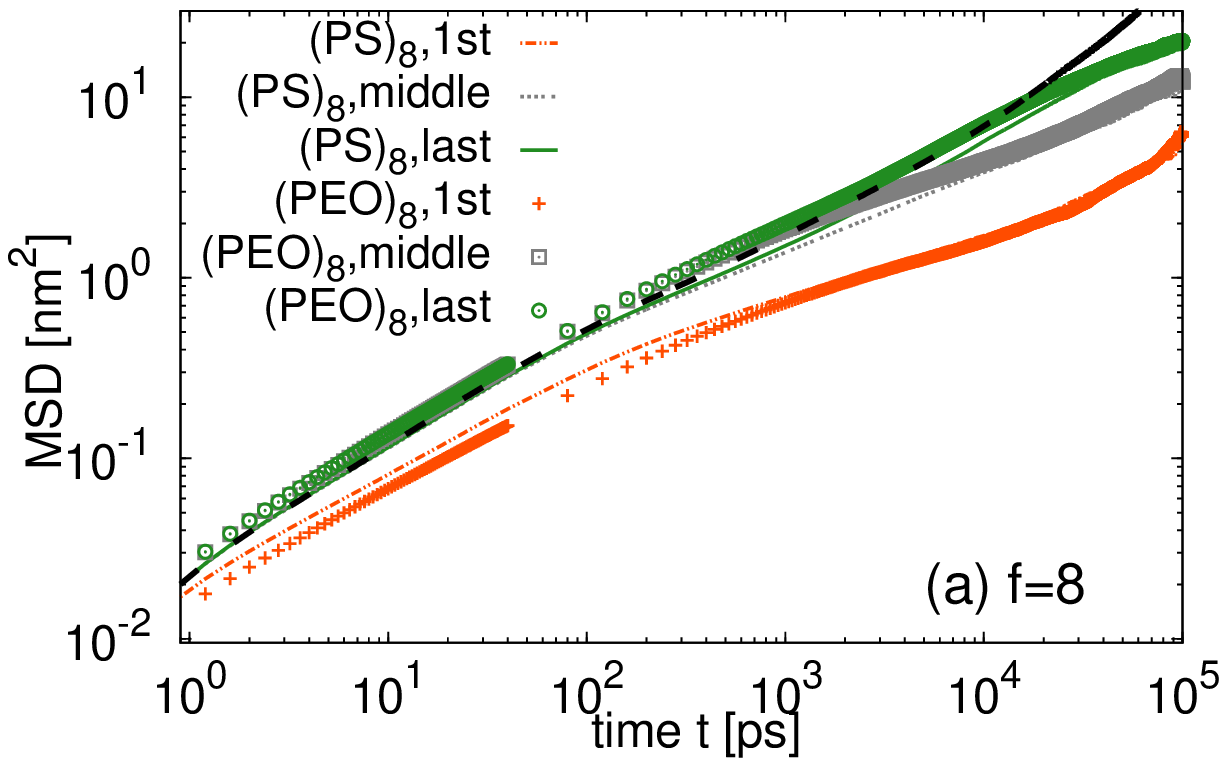} 
\includegraphics[width=0.47\textwidth]{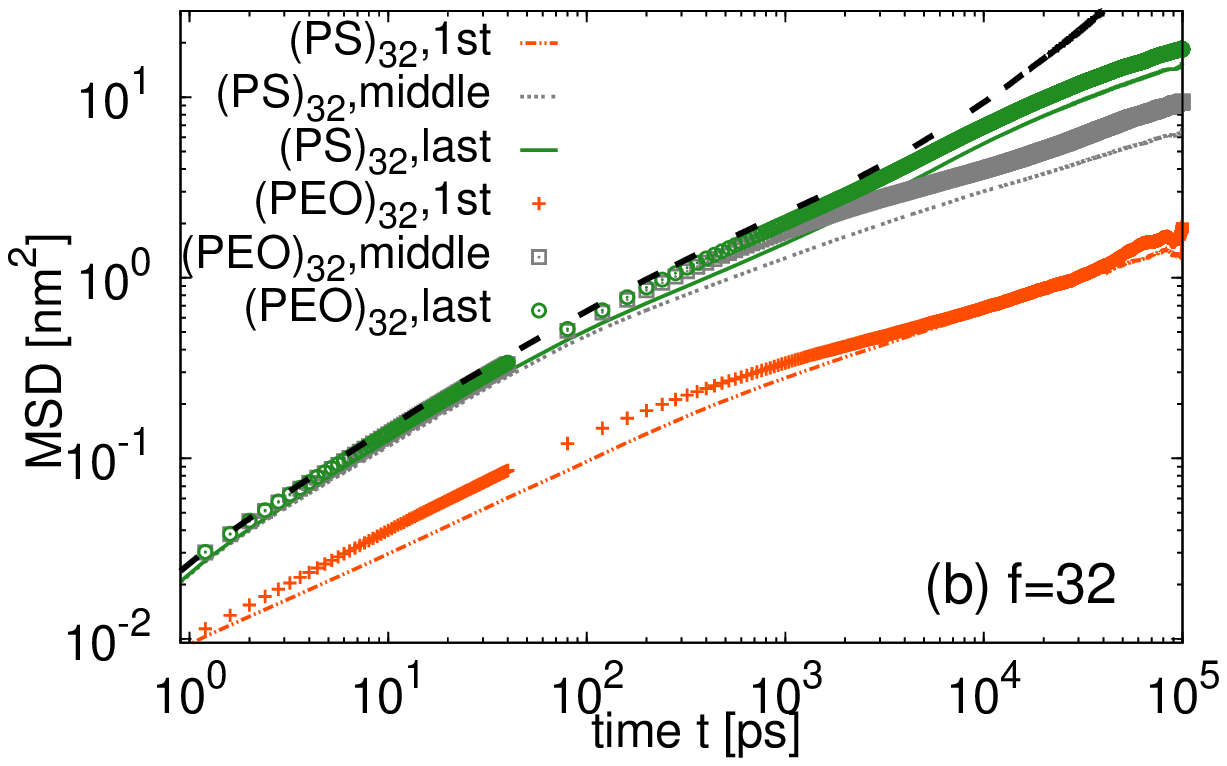}
\includegraphics[width=0.47\textwidth]{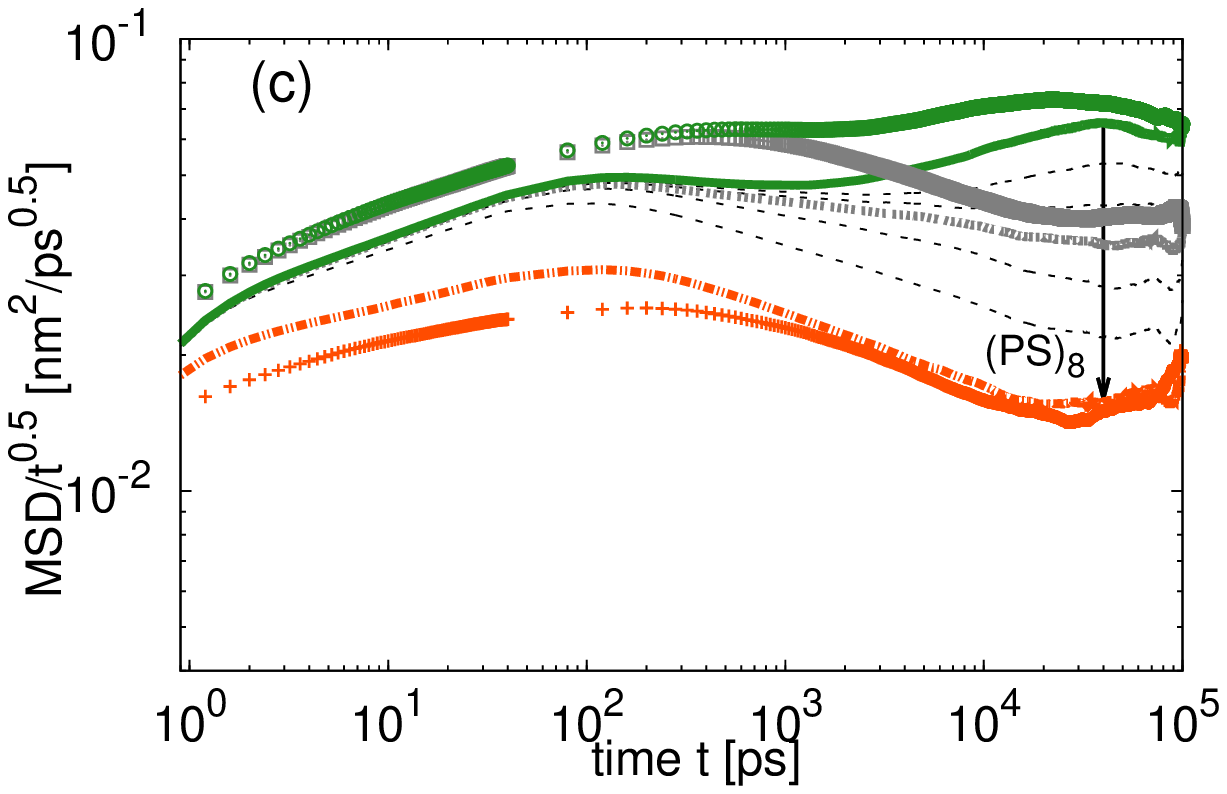} 
\includegraphics[width=0.47\textwidth]{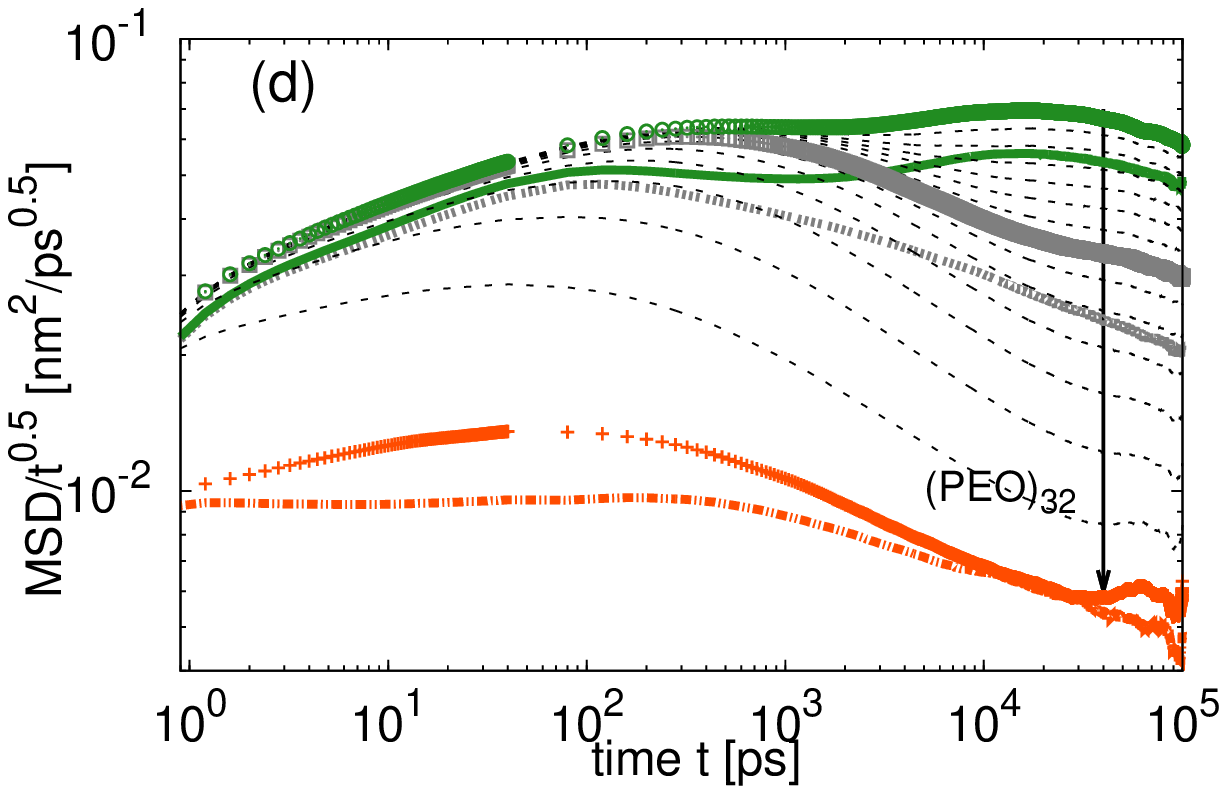}
	\caption{Monomer mean square displacement for stars with (a,c) $f=8$ and (b,d) $f=32$ averaged over monomers in the first, middle and end arm region. The colors are in agreement with the schematic division of arms in Fig.~\ref{label}. Figures (c,d) show the same set of data as in (a,b), normalized with $t^{0.5}$. The thick dashed lines in (a,b) are the data for PS and PEO linear chain, respectively. The thin dashed lines in (c,d) correspond to segments in the PS and PEO star, which are not labelled as ``first, middle'' or ``end''. The vertical arrows illustrate the dynamical gradient, pointing from the last to the first segment in the PS and PEO star, respectively.} 
\label{msd_lk}
\end{figure*}

The monomer mean squared displacement (MSD) averaged over the monomers in given region are plotted in Fig.~\ref{msd_lk} as a function of time, $t$, for two selected functionalities, $f=8$ and $f=32$. 
As discussed in Sec.~\ref{match}, the motion of the outer segments follows closely the behavior of the segments in linear chains (compare thick dashed lines with corresponding dark green data in Fig.~\ref{msd_lk}(a,b)) up to the Rouse time approximately.
 At times comparable to the Rouse times (see Table~\ref{table:par}), both linear chains enter the diffusive regime, with a characteristic slope MSD$\sim t^{1}$, while the outer segments are slowed down by the more localised inner sections of the arms, causing a deviation between the these two sets of data.   
The middle and the first region show significant deceleration with respect to the outer segments and thus to the dynamics of the linear chain as well. 
Concerning the comparison of the (PEO)$_f$ and (PS)$_f$ stars, at the first glance, the data for different chemistries but same functionalities overlap (compare symbols and lines in Fig.~\ref{msd_lk}(a,b)), confirming a general, chemistry-independent, dynamical behavior of the statistical segments of the order of the Kuhn length. 
However, a closer inspection reveals significant differences.
For a better visualization, we present the same data from Fig.~\ref{msd_lk}(a,b) normalized by $t^{0.5}$ in Fig.~\ref{msd_lk}(c,d). Note that in this representation the Rouse scaling (i.e., MSD$\sim t^{0.5}$) is easily detected.
Moreover, in this representation, the subtle differences between the stars made of two different polymer types are clearly visible. 
Except of the first region in (PS)$_8$, the data for the PS stars are systematically slower than those for the segments in the PEO stars. The origin of this systematic trend might be in higher molecular weight of the PS monomer. 
Interestingly, looking at the first region only (orange set of data in Fig.~\ref{msd_lk}(c,d)), the data for the two different chemistries deviate at the short time scales but merge at the times of the order of the theoretical Rouse time (see Table~\ref{table:par}).
 
We also plot the displacements of all 7 segments in the (PS)$_8$ star and all 15 segments in the (PEO)$_{32}$ star in Fig.~\ref{msd_lk}(c) and (d), respectively, to demonstrate the internal gradient in mobility in both types of star-like polymers. 
Note that a whole spectrum of scaling laws can be detected when following the time evolution of the motion of one particular segment.  
This dynamical gradient, i.e., heterogeneous mobility of the segments along the star arm, is due to the molecular architecture of the star molecule and has been previously reported for other star-like molecules, namely for bead-spring entangled stars~\cite{petra_lau,PetraStefan} and also atomistic nonentangled mikto-arm stars.~\cite{Petra2} 
The data reported here further emhpasise the role of specific chemistry (mainly flexibility and the molecular packing) on the gradient of the translational dynamics of stars.

\subsection{Comparison of the Rouse model and simulation data}\label{comp}
\begin{figure*}[!ht]
\includegraphics[width=0.47\textwidth]{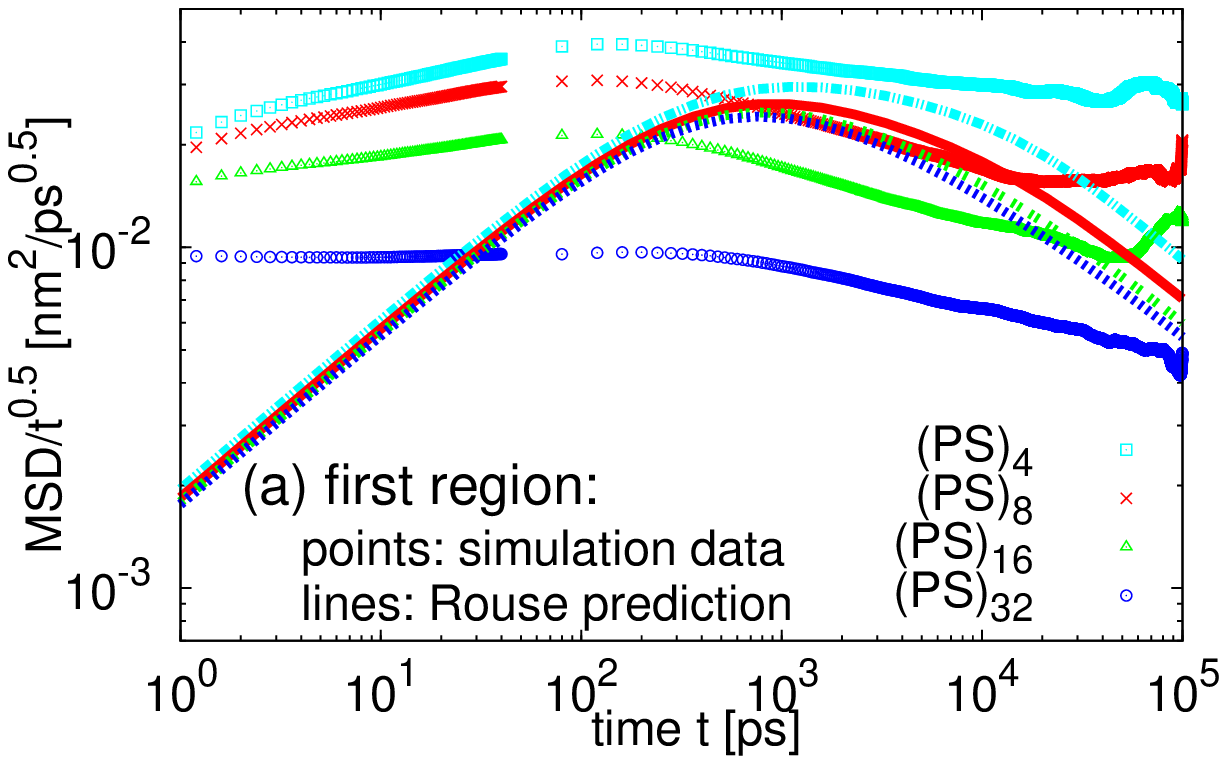} 
\includegraphics[width=0.47\textwidth]{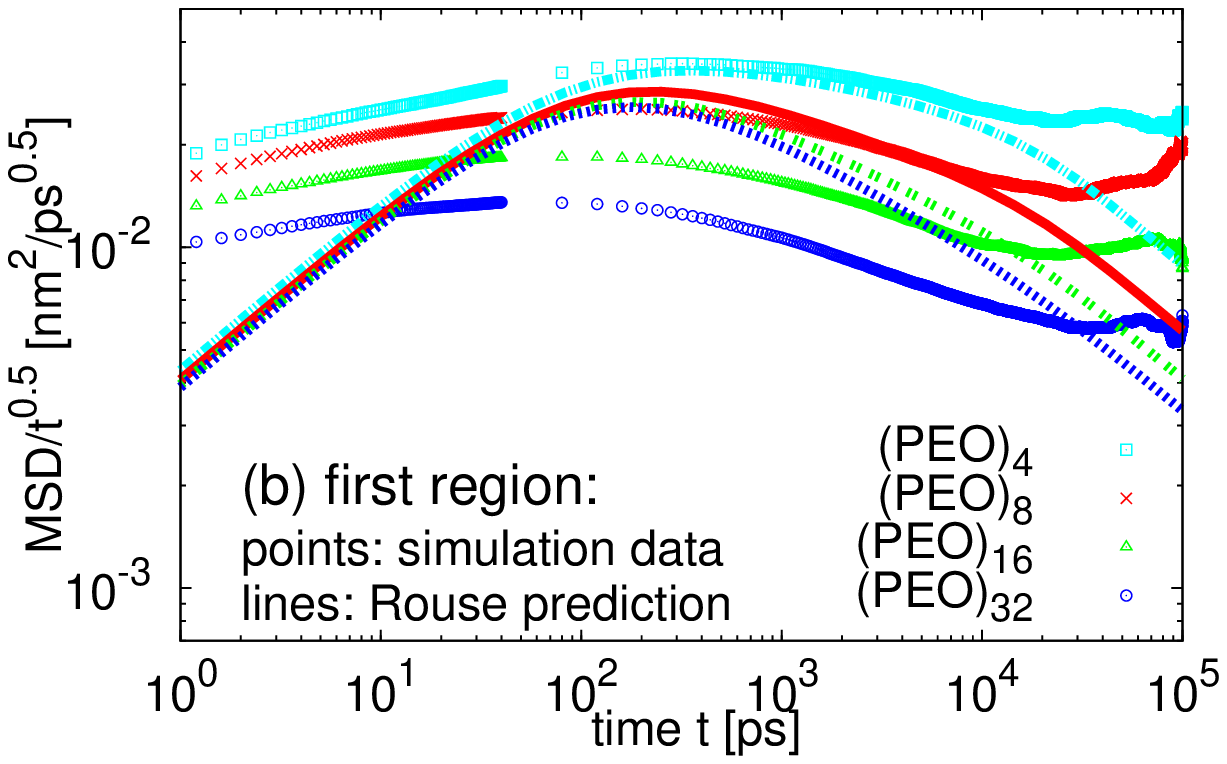} 
	\caption{Monomer mean square displacement averaged over monomers in the first region divided by $t^{0.5}$ as a function of time $t$ for (a) the PS and (b) the PEO stars: simulation data (symbols) together with the theoretical predictions (lines), namely expression (2) in Table~\ref{table:MSDPureRouse} discounting CM diffusion.}
\label{rouse_msd}
\end{figure*}

With the exception of some segments attached to the kernel, the remaining simulation segments do not appear to manifest diffusive modes. Hence, we initially choose to compare simulation data with Rouse model predictions obtained without the center of mass (CM) diffusion contribution. Fig.~\ref{rouse_msd} compares simulation data and theoretical predictions of the second equation of Table~\ref{table:MSDPureRouse}, excluding the CM motion, i.e., the first term of this equation. The comparison refers to the segments that are attached to the kernel (branch point in the Rouse model).
For completeness, we note that to better represent the continuous character of the model, we divide the segments of interest (e.g., 1, 4, and 7 for PS) into several points, averaging their MSD. In all Rouse results presented herein each simulation segment of interest is divided into 51 points. 
Nevertheless, the results are rather insensitive to the exact discretization; the same results are obtained with 11 and 101 points. 

Figure~\ref{rouse_msd} reveals several features. First, the model predictions deviate from the anticipated $t^{0.5}$ power behavior.
This is a direct consequence of the finite number of modes considered in the summation. As aforementioned, in the limit of large $N_a$ the predictions manifest the expected power law behavior (see Fig.~S1 in the Supplementary Information). Since at early timescales ($ t \lesssim \tau_0$) theoretical predictions are very sensitive to the number of modes used in the summation, we exclude this time regime from any subsequent discussion. Second, unlike the simulation data, model predictions do not exhibit a strong dispersion. Focusing on intermediate timescales, the model captures reasonably well the simulation data for the two lowest functionalities, with the comparison being better in the case of PEO. At the two highest functionalities, the model underperforms severely, except from (PEO)$_{16}$ for which the comparison is tolerable. Notice that the model performance is poorer for PS. Overall, the model overpredicts the mobility of this particular chain section. These findings suggest that at high functionalities, and especially for PS, the friction in the branch point vicinity is considerably larger than the one considered in the model. Likely, the severely stronger localization in the simulation data arises from (excluded volume) interactions between different arms.
Naturally, such interactions are expected to increase as the functionality increases. Unlike the simulations, the model disregards such interactions, meaning that segments can overlap even if the density near the star kernel is high.

According to the model, segmental motion saturates at long timescales. To examine whether CM diffusive modes can improve the model performance at such timescales, we compare simulation data and model predictions that include the CM contribution. The comparison can be seen in Fig. S3 of the Supplementary Information.
It reveals a good correspondence between theoretical outcomes and simulation data for the lowest two functionalities and irrespective of polymer chemistry.
In other words, the inclusion of CM contribution to the MSD appears to extend the time range over which the model exhibits a decent performance as regards low functionalities and segments in the vicinity of the branch point.

\begin{figure*}[!ht]
\includegraphics[width=0.47\textwidth]{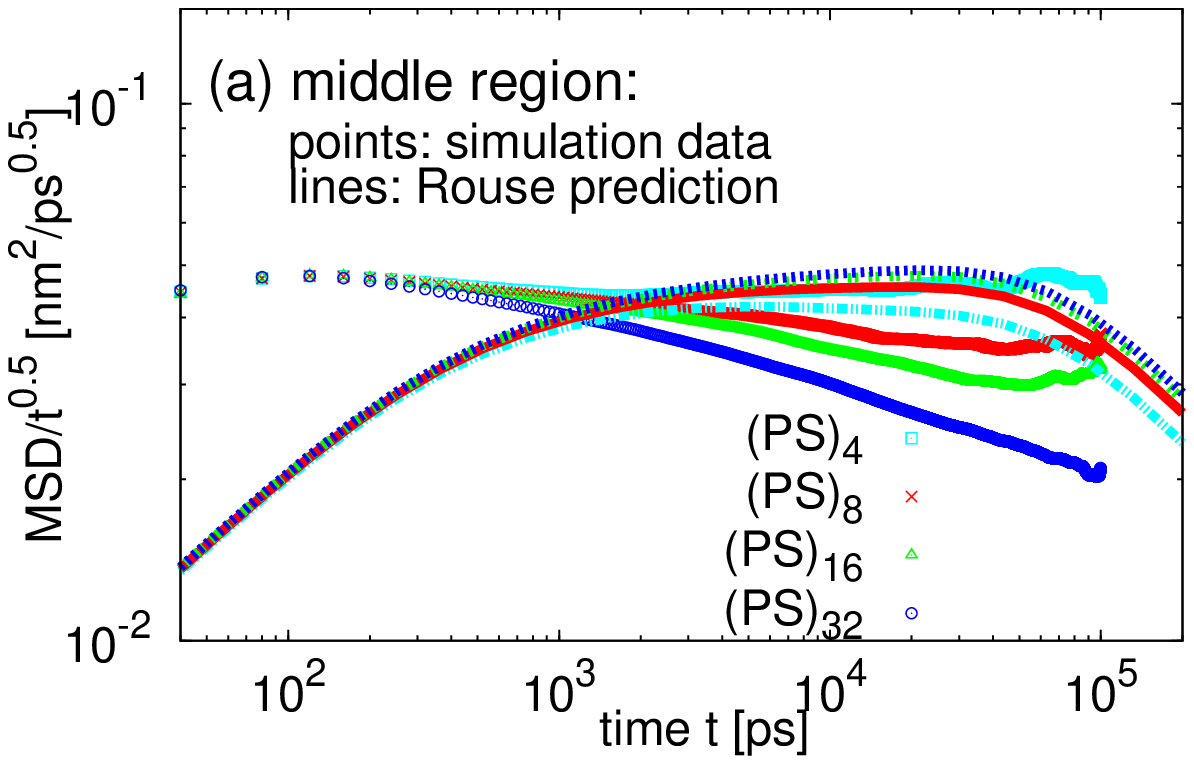}
\includegraphics[width=0.47\textwidth]{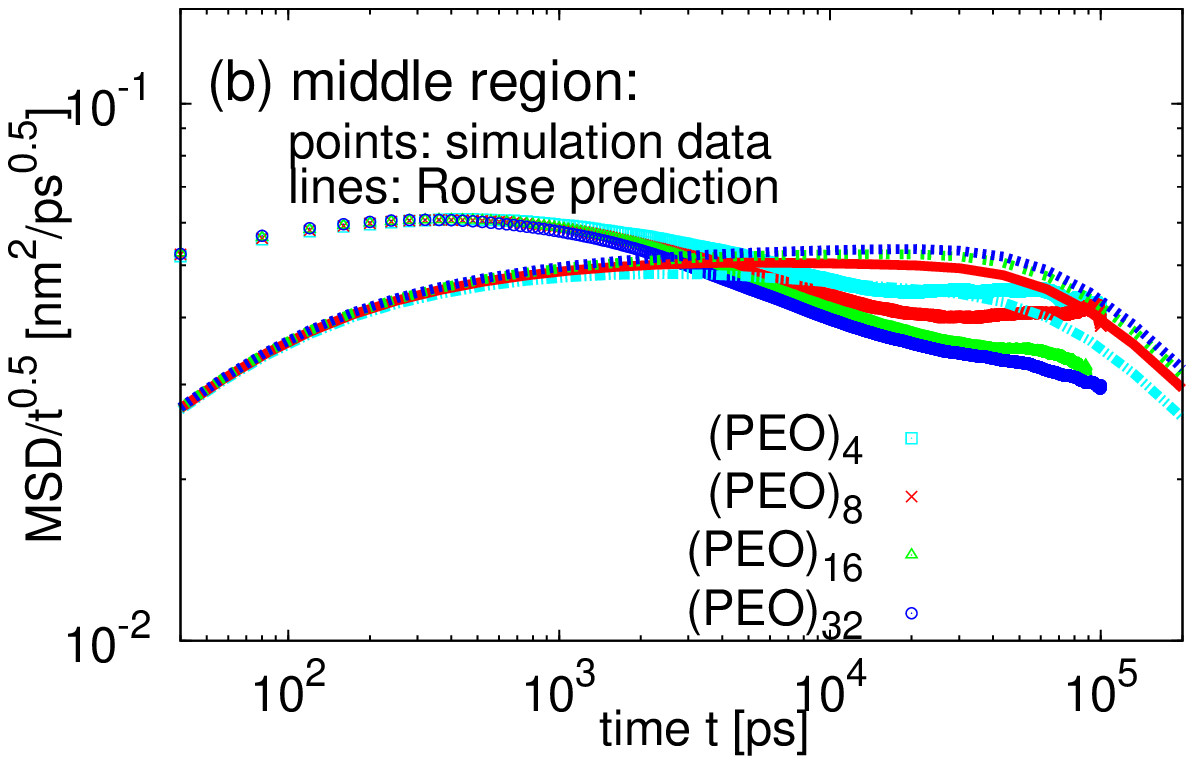}
	\caption{Monomer mean square displacement averaged over monomers in the middle region divided by $t^{0.5}$ as a function of time $t$ for (a) the PS and (b) the PEO stars: simulation data (symbols) together with the theoretical predictions (lines), namely expression (2) in Table~\ref{table:MSDPureRouse} discounting CM diffusion.}
\label{rouse_msd_mid}
\end{figure*}
Next, we turn our attention to the middle arm segments. Figure~\ref{rouse_msd_mid} compares simulation findings and model outcomes (the second equation of Table~\ref{table:MSDPureRouse}) which have been obtained in the absence of the CM contribution to the MSD. The comparison reveals a poor model performance overall. Concerning intermediate and long timescales, the model manifests a fair performance only at the lowest functionality, for both PS and PEO. The model performance deteriorates as the functionality increases. It is to be noticed, that the model fails even at a qualitative level. In particular, the theoretical MSD increases with increasing functionality whereas the simulation data exhibit the reverse trend. Such qualitative disagreement cannot be explained even if the adopted Rouse parametrization is inaccurate. In this respect, the findings of Fig.~\ref{rouse_msd_mid} strongly suggest that, between the branch point and the middle of the arm, the friction is not uniform. That is, the slow dynamics of the inner segments affect the dynamics of the consecutive segments placed along the star arm, which need to ``wait'' for the slowest component in the system to enter the terminal diffusive regime. From a conceptual standpoint, one could envisage Rouse beads of decaying size (friction) along the arm. To reflect the stronger interactions (correlations) near the kernel, the heaviest (largest) bead would be positioned at the branch point and the lightest (smallest) at the arm tip. The exact size decay profile of the beads would rather depend on the functionality. Such a molecular picture would be consistent with the decaying monomer density profile along the arms (see section VI of the Supplementary Information).

As readily seen in Fig. S4 of the Supplementary Information, the inclusion of CM diffusion worsens the model performance as it provides higher mobility at low functionalities for which the model performs reasonably without CM contribution. From a qualitative perspective, the findings of the simulation and Rouse comparison regarding the middle arm segment apply to the outer segments as well. From a quantitative standpoint, nevertheless, the model performs even worse as it significantly overestimates the mobility at all functionalities. These features can be readily appreciated in Figs.~S5 and S6 of the Supplementary Information. The former (latter) figure presents theoretical prediction with (without) CM diffusion contribution to the MSD. In summary, the MSD comparison indicates that the large-scale reorientation of the simulated stars is more complex than the model predicts.

\begin{figure}[h!]
\includegraphics[width=0.45\textwidth]{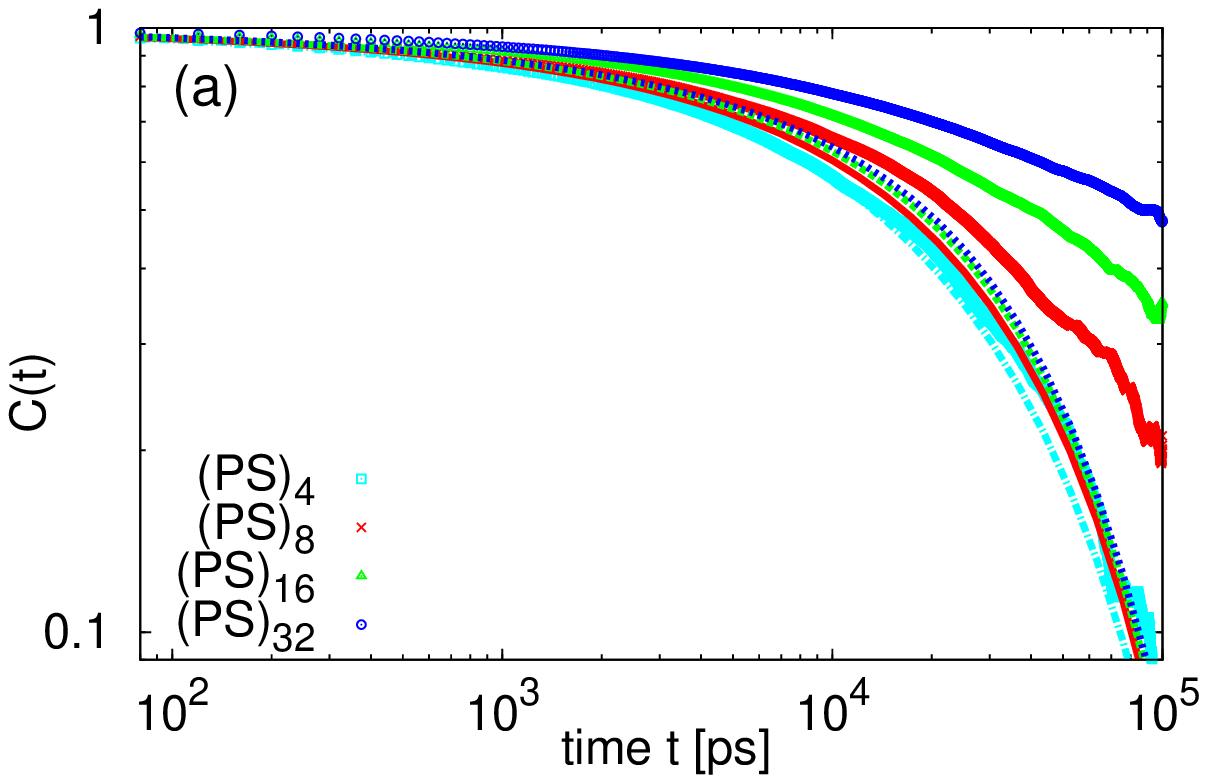}
\includegraphics[width=0.45\textwidth]{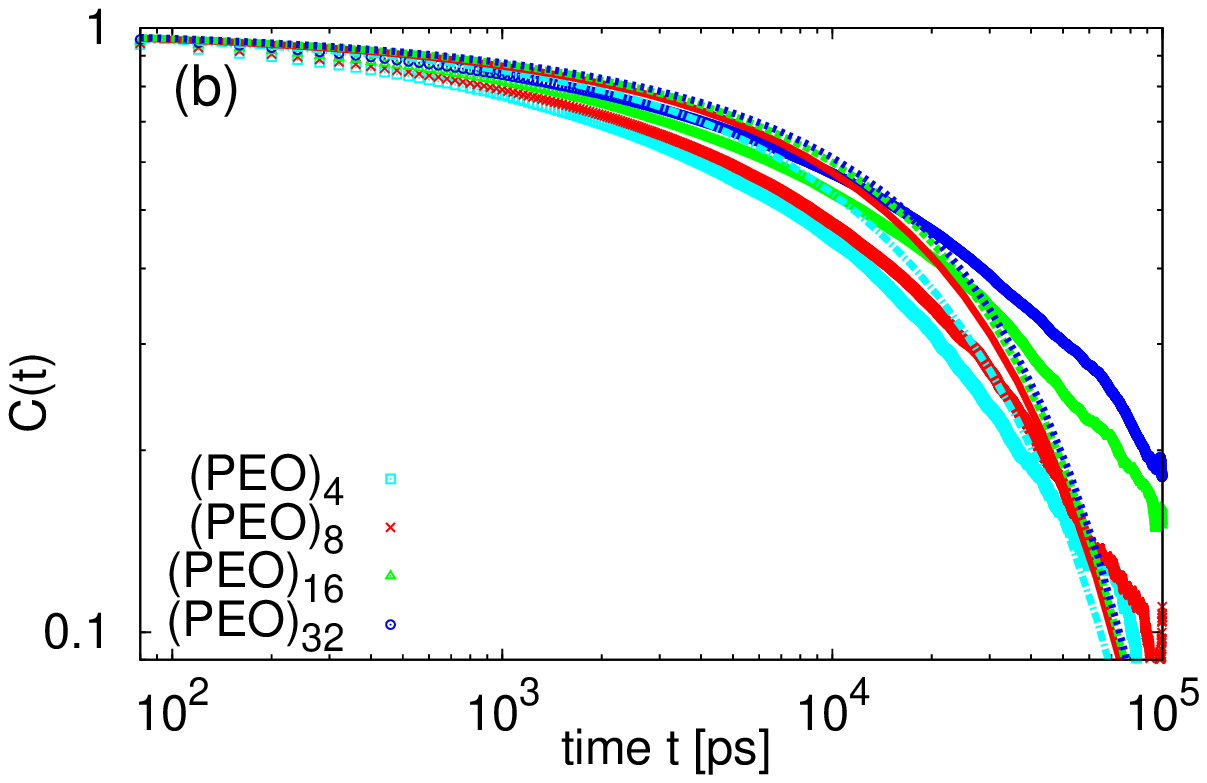}
	\caption{Autocorrelation function of the center-to-end vector $\bf{R}$ as a function of time for (a) the PS and (b) the PEO stars. The lines are theoretical predictions of the Rouse model.}
\label{end}
\end{figure}

A good indicator of the large-scale reorientation of the stars is the correlation function of the center-to-end vector, $\bf{R}=\bf{R_{\alpha}}$ (see Fig.~\ref{label}). The theoretical expression for the correlator can be found in section IV of the Supplementary Information. Similar to the second MSD expression of Table~\ref{table:MSDPureRouse}, this equation retains information for all modes. Figure~\ref{end} compares Rouse outcomes and the corresponding simulation results. Notice that theoretical predictions are obtained using the same parametrization (Table~\ref{table:par}) as for the calculation of the MSD. Focusing on the simulation data first, we observe that the decays of the correlation functions for all studied functionalities and for both chemistries obey a stretched exponential decay rather than a single exponential decay (an example of the fitting procedure is shown in section IV of the Supplementary Information). More importantly, the deviation from the single exponential decay increases as the functionality increases. Attention now shifts to the Rouse outcomes. As the correlation function is sensitive to the large-scale reorientation of the chains, the comparison for the correlator should yield similar outcomes as the corresponding comparison for the MSD results referring to the segment that is attached to the kernel (first). As readily seen from Figure~\ref{end}, this is indeed the case. In more detail, for both PS and PEO stars, the Rouse predictions provide either a good or tolerable description of the simulation data for the lowest two functionalities. Moreover, similar to the MSD description of the (PEO)$_{16}$ first segment, the comparison of the correlation function is tolerable for the (PEO)$_{16}$ stars as well. Overall, the model has the tendency to underpredict the center-to-end vector relaxation occurring in the simulations. It should be emphasized that the latter exhibits a strong $f$ dependence, unlike the model.

At a first glance the better model performance for the inner segment than the middle and last segments might seem surprising. As shown in the next section, the simulation kernel region mainly comprises slow components rendering, for a given functionality, this section of the chain rather homogeneous in terms of friction distribution, similar to the Rouse model assumption. In contrast, the middle and outer segments experience a more inhomogeneous environment. 
For example, outer segments of a given star arm are excluded from the impenetrable central regions of surrounding stars, lowering their mobility compared to the predictions anticipated by the model. Although this explanation is speculative, it complies with the simulation findings of the following section.

\section{Discussion}
In this section we address the internal dynamical heterogeneity observed in Sec.~\ref{msd_sim} and possible sources of discrepancies between the theoretical model and the data from the simulations observed in Sec.~\ref{comp}.
We performed additional analysis and tested three factors. Namely, 1) non-uniform friction distribution in the Rouse model, 2) non-Gaussian character of segmental motion, and 3) heterogeneous environment along the star arm induced by the star-like architecture. 

\begin{figure}[h!]
\includegraphics[width=0.45\textwidth]{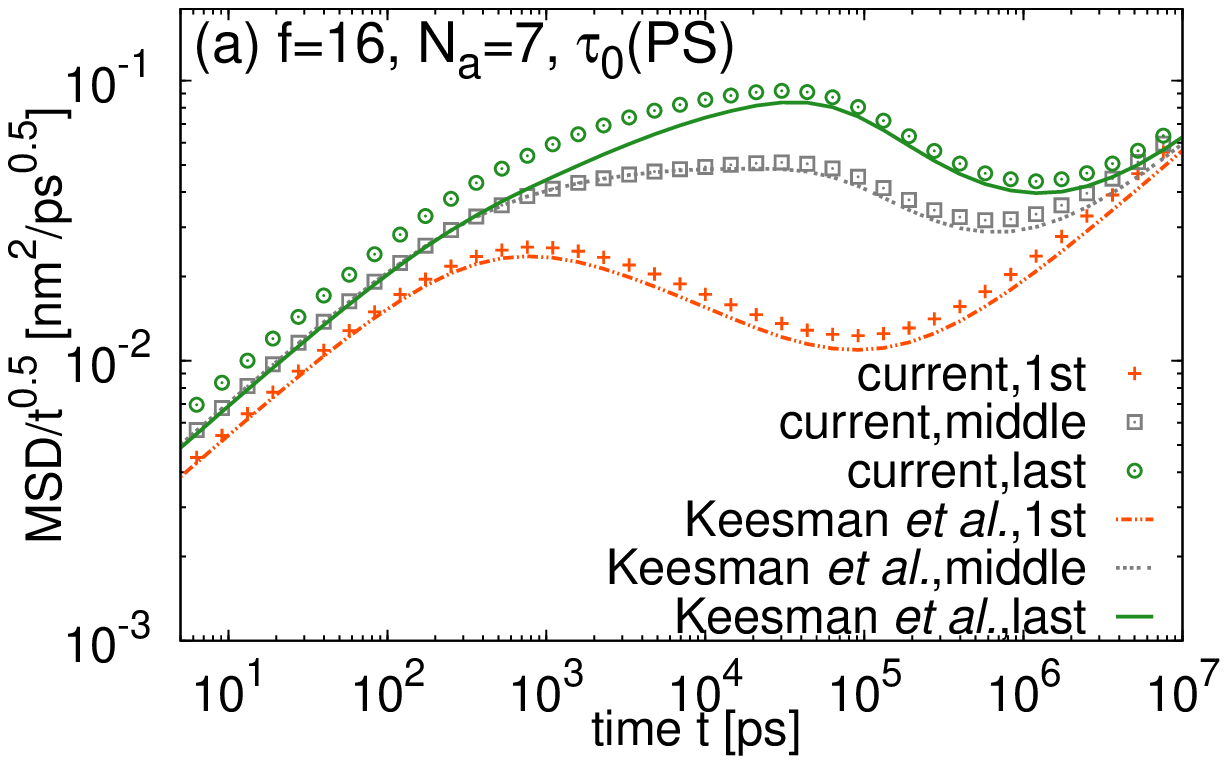}
\includegraphics[width=0.45\textwidth]{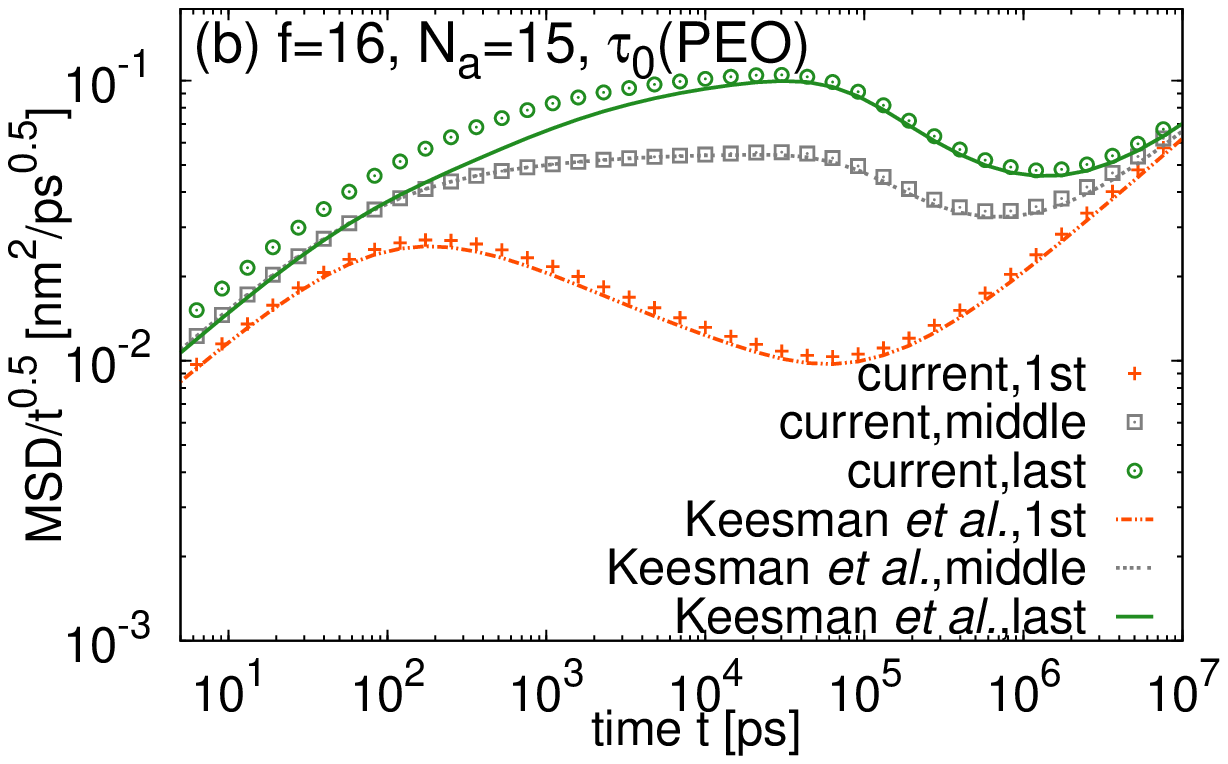}
\caption{Comparison of the two theoretical models, the one presented here (symbols) and the one reported by Keesman {\it{et al.}} in Ref.~\cite{Keesman} (lines) for functionality $f=16$ and parameters corresponding to the (a) PS and (b) PEO stars.}
\label{2model}
\end{figure}

Regarding the first factor, we compare our segmental MSD equation with that of Keesman {\it{et al.}} ~\cite{Keesman}, i.e., the third expression of Table~\ref{table:MSDPureRouse}. 
Recall that this expression is derived from a Rouse model that assumes a larger branch point friction ($f\zeta_0$) than our model. 
The comparison between the two expressions can be seen in Fig.~\ref{2model} for $f=16$. All results include CM motion and are obtained using the Rouse parametrization of Table~\ref{table:par}. Figure~\ref{2model} indicates that the assignment of a larger friction to the kernel than the other beads does not have a substantial effect on the Rouse model predictions. 
Although Fig.~\ref{2model} refers to $f=16$ the same outcome is found for all other functionalities (not shown). The localization of the segments increases slightly with the most noticeable effect being on the outermost segment at short and intermediate timescales. This behavior seems peculiar and it requires a more detailed investigation, which lies beyond the scope of the current work. Notice that our numerical implementation of the Keesman {\it{et al.}} expression was verified by reproducing the results of Fig.~2 of Ref.~\cite{Keesman}.

\begin{figure}[!ht]
{\centering{\includegraphics[width=0.43\textwidth]{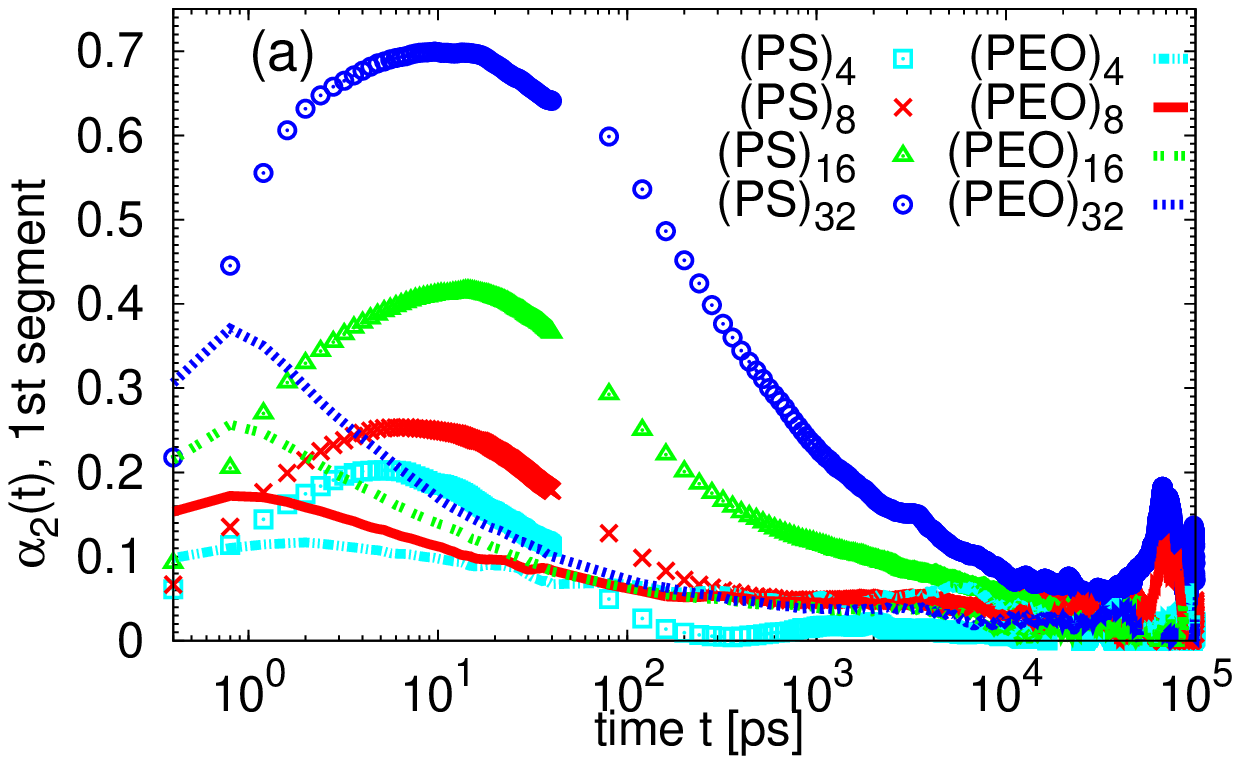}} \par}
{\centering{\includegraphics[width=0.43\textwidth]{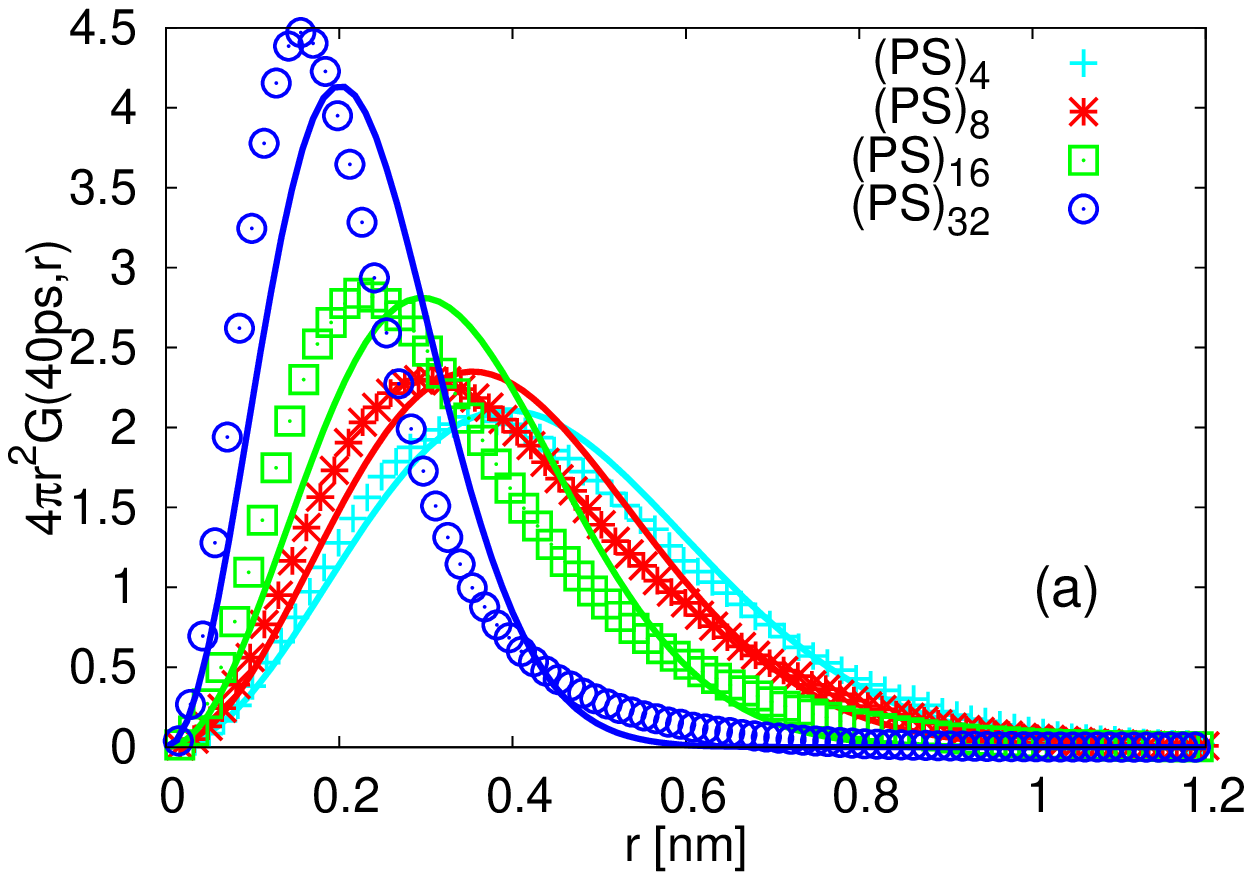}} \par}
	\caption{(a) Non-gaussian parameter $\alpha_2$ and (b) Van Hove function for the monomers in the (first) region adjacent to the star kernel. The solid lines in (b) are Gaussian distributions Eq.~\ref{VH_gaus} with the corresponding MSD taken from Fig.~\ref{msd_lk} and with $t=40$ ps.} 
\label{nonga}
\end{figure}

With respect to factor 2) above, we focus on the first region only, i.e., the one in the close vicinity of the kernel. In this region the factor 2) is expected to play a significant role. 
In order to fist examine the non-Gaussian features of the monomeric motions, which are presumably the main reasons for the discrepancies of the Rouse model~\cite{Krushev,Smith_nongaus}, we present in Fig.~\ref{nonga} the time-dependence of the non-Gaussian parameter, $\alpha_2 (t) = 3\langle \Delta r^4 (t)\rangle/5\langle \Delta r^2 (t)\rangle^2 -1$ where $\Delta r(t)$ denotes the displacement of the monomers in the first region at time $t$.
In both types of polymers, the non-Gaussian character of the translational motion is more evident in systems with high $f$. While the actual values of $\alpha_2$ are low for all functionalities of (PEO)$_f$ stars, the data for the (PS)$_f$ systems show a well-pronounced maxima, shifted to longer times in comparison to their PEO analogues.
Next, we calculate the Van Hove function $G({\bf{r}},t)$ for selected times $t$ in order to detect slow or fast components causing the deviation from the Gaussianity.  
The $G({\bf{r}},t)$ was calculated as follows:
\begin{equation}
	G({\bf{r}},t)=\frac{1}{M_i} \left< \sum_{i=1}^{M_i} \sum_{j=1}^{M_i} \delta \left({\bf{r}}+{\bf{r}}_j(0)-{\bf{r}}_i(t) \right) \right>
\end{equation}
with $M_i$ being the total number of monomers in the given region (i.e., first) in the system. The $G({\bf{r}},t)$ data were fitted by:
\begin{equation}
	G({\bf{r}},t)=(4 \pi D(t)t)^{-\frac{3}{2}} \exp \left( - \frac{{\bf{r}}^2}{4D(t)t} \right)
	\label{VH_gaus}
\end{equation}
where $D(t)$ denotes a time-dependent diffusion coefficient obtained as $D(t)=\rm{MSD}(t)/6t$ at the given time $t$ and MSD$(t)$ is the mean square displacement of the monomers in the first region at this time.
The Van Hove functions for $t=40$ ps and (PS)$_f$ stars are plotted in Fig.~\ref{nonga}(b) together with the corresponding Gaussian functions. 
In accordance with the observations in Fig.~\ref{nonga}(a), there is a certain heterogeneity in the distribution of the displacements of the monomers of the (PS)$_f$ stars in the first region at short time scales, however, this heterogeneity disappears at longer times, more specifically, already at $t=400$ps the $G({\bf{r}},t)$ for the PS stars are fairly Gaussian (see Fig.~S8 in the Supplementary Information).  
The $G({\bf{r}},t)$ functions for the PEO stars show only negligible deviations from the Gaussian description in all time frames, selected randomly in logarithmic scale: $t=40,400,4000,40000$ps (see Fig.S9 in the Supplementary Information).
  
\begin{figure}[!ht]
\includegraphics[width=0.43\textwidth]{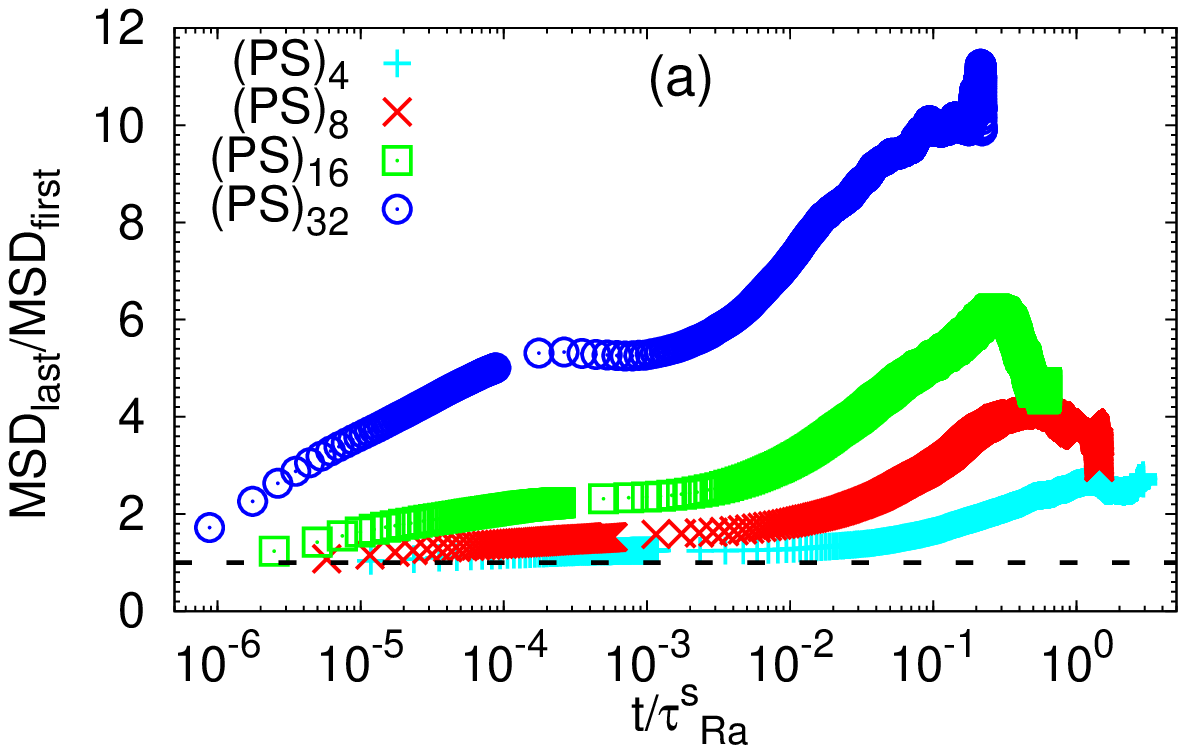}
\includegraphics[width=0.43\textwidth]{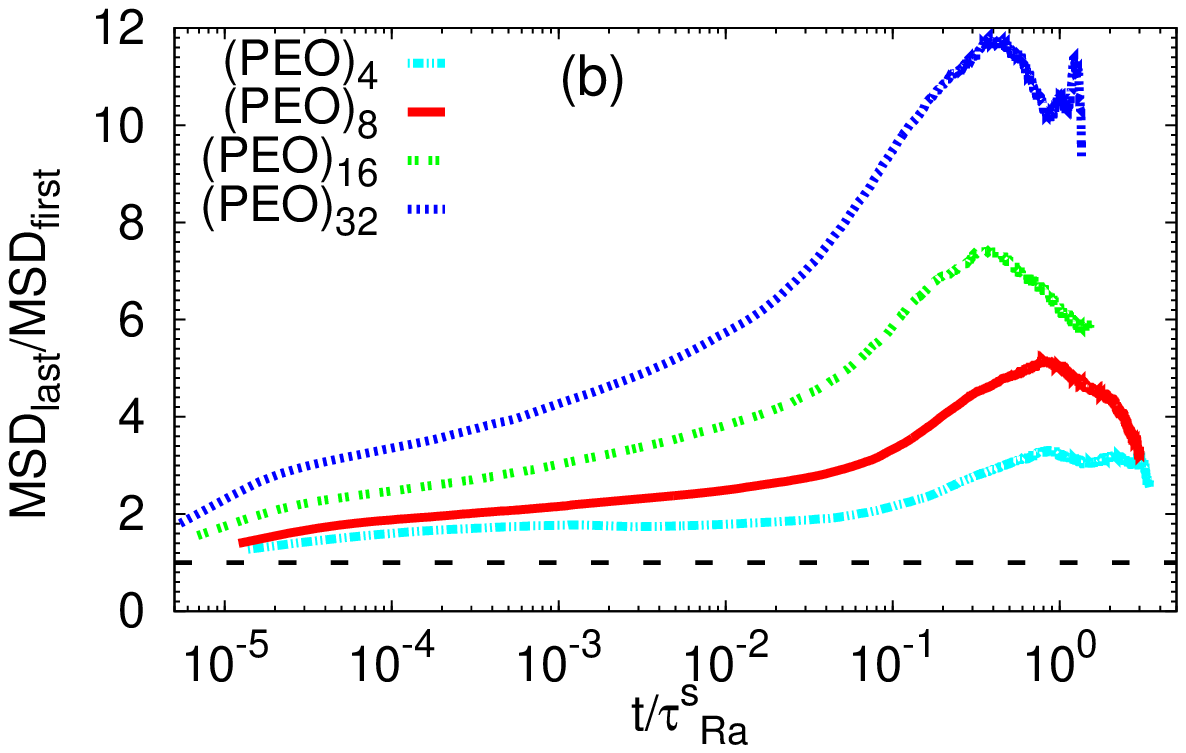}
	\caption{Ratio of the MSD in the last and in the first segment for (a) (PS)$_f$ and (b) (PEO)$_f$ stars as a function of rescaled time, $t/\tau_{R_a}^s$, where $\tau_{R_a}^s$ is the Rouse time obtained from the simulation (Table~S1 in the Supplementary Information). The black horizontal dashed line depicts the ratio equal to 1.} 
\label{frac}
\end{figure}

Having in mind the relatively low values of non-Gaussianity parameter in the time window where the Rouse model is expected to be valid, the restricted motion in a high-density, inpenetrable central star region turns out to be the most probable source of the discrepancy between the theoretical and simulation model. 
With increasing functionality the arms become closely packed around the kernel, which leads to a high density region in the vicinity of the kernel (see the monomer density profiles in Fig.~S10 in the Supplementary Information) and to the mutual interactions of the arms, generally not accounted in the theoretical models. 
This region is not penetrable by other stars~\cite{Daoud} and thus, the local environment consists exclusively of the slow segments attached to the kernel. 
This is particularly the case of (PS)$_{32}$ and (PS)$_{16}$ stars, due to the bulky side groups (aromatic rings) of the PS monomers. 
On the other hand, stars with low functionality (in our case $f=4$ and $f=8$) are much more penetrable and therefore the local environment of the segments along the arm is more homogeneous, resembling a melt of linear chains, as any segment of any neighboring star can come into contact with the star under investigation. 

In order to quantify the above observations, we calculate the ratio of the MSDs of the last over the first segment; results are shown for all stars in Fig.~\ref{frac}.
This quantity gives us a hint about the ``degree'' of the internal dynamical heterogeneity within the star, in other words, about the magnitude of the dynamical gradient of the segments along the arm (see the arrows in Fig.~\ref{msd_lk}(c,d)). 
In both types of polymer stars the ratio reaches the maximum slighly before the simulation Rouse time (see Table~S1 in the Supplementary Information), then the functions decrease, as the segmental dynamics tends to become more homogeneous before entering the final diffusive regime. Note that for the free diffusion of the molecule this ratio should equal to 1.  
The difference in the dynamics of the last and the first segments seems to be slighly bigger in the case of (PEO)$_f$ stars. 
For the highest functionalities, $f=32$, the dynamics in the first region is up to a factor 10 slower than the dynamics of the outer segments. 
Because the overall dynamics is mostly ruled by the slowest component in the material, this significant difference in the motion of the segments along the arm may be reflected in the properties such as dynamical structure factor or dielectric spectrum.

In summary, the segmental motion in (PEO)$_f$ stars shows only small deviations from the Gaussian behavior. In the case of the (PS)$_f$ stars more significant deviations from the ideal behavior have been found at times comparable to the segmental time, especially for the segments adjacent to the star kernel and for the stars with the highest studied functionality $f=32$. 
In those cases, the dense packing of the arms close to the kernel seems to be the main source of heterogeneous dynamical behavior.
As a consequence of the dynamical heterogeneities within the star molecule, a complex response of the material is expected, showing a colloidal-like behavior at relatively low functionalities, as reported recently for the PS stars with the same functionalities of those studied here.~\cite{Glynos} 

\section{Conclusions}
We present a detailed study of the translational motion of the non-entangled star-shaped polymer melts consisted of varying number of poly(ethylene oxide) or polystyrene arms. 
We used the atomistic molecular dynamics simulation to capture the chemical details and mimic the behavior of the analogous materials prepared by polymer synthesis. The simulation data reveal a presence of gradient in the mobility of the segments along the arm, with the slowest segments placed in the vicinity of the star kernel. The internal dynamical heterogeneity becomes more significant in the stars with high functionality, where the close packing of the arms close to the kernel seems to be the main source of deviation from the ideal behavior predicted by the theoretical model.

Overall the comparison between the simulation data and Rouse predictions for both the segmental MSD and the center-to-end vector correlator suggests that the simulated systems exhibit elements of Rouse motions. Nevertheless, the large-scale reorientation modes of the systems appear to be heavily perturbed from those anticipated from the Rouse model. The perturbation increases as the functionality increases. For identical functionality, it is stronger in PS than PEO. Without inclusion of excluded volume interactions (EVI), the Rouse star model appears to oversimplify the actual dynamics of the star molecules.

For a quantitative account of the simulation results, incorporation of such interactions appears to be essential. We anticipate that a Rouse model comprising beads the size (friction) of which decays from the branch point to the arm tip could provide an alternative to a detailed EVI model. The precise form of such a friction gradient could be guided by atomistic molecular dynamics simulations; this will be the subject of a future work.

\section*{Data Availability Statement}
The data that support the findings of this study are available within the article and its supplementary material.

\section*{Supplementary material}
See supplementary material for the effect of the number of modes on the MSD Rouse expressions with sums, estimation of $\tau_0$ parameter, further comparison of the Rouse model and simulation MSD data, description and fitting procedure of the autocorrelation function of the center-to-end vector, additional Van Hove functions and monomer density profiles.  

\begin{acknowledgments}
This research has been co-financed by the General Secretariat for Research and Technology (Action KRIPIS, project AENAO, MIS: 5002556). The work was supported by computational time granted from the Greek Research \& Technology Network (GRNET) in the National HPC facility ARIS under project named AMDStar.
L.G.D.H thanks the Fonds National de la Recherche Scientifique - FNRS for financial support.
V.H. acknowledges funding from the European Union’s Horizon 2020 research and innovation programme under grant agreement no. 810660.
\end{acknowledgments}

\nocite{*}
\bibliography{ref}

\end{document}


\title{SUPPLEMENTARY INFORMATION: Dynamical heterogeneities in non-entangled polystyrene and poly(ethylene oxide) star melts}
\author{Petra Ba\v{c}ov\'a}
 \email{pbacova@iacm.forth.gr}
 \affiliation{Institute of Applied and Computational Mathematics (IACM), Foundation for Research and Technology Hellas (FORTH), GR-70013 Heraklion, Crete, Greece}
\affiliation{Computation-based Science and Technology Research Center, The Cyprus Institute, 20 Constantinou Kavafi Str., Nicosia 2121, Cyprus}
\author{Eirini Gkolfi}%
\affiliation{Institute of Applied and Computational Mathematics (IACM), Foundation for Research and Technology Hellas (FORTH), GR-70013 Heraklion, Crete, Greece}
\affiliation{Department of Mathematics and Applied Mathematics, University of Crete, GR-71409 Heraklion, Crete, Greece}

\author{Laurence G. D. Hawke}
\affiliation{
Institute of Condensed Matter and Nanosciences (IMCN), Bio and Soft Matter Division (BSMA), Universit\'e catholique de Louvain, Croix du Sud 1 \& Place L. Pasteur 1, B-1348 Louvain-la-Neuve, Belgium
}
\author{Vagelis Harmandaris}
\affiliation{Institute of Applied and Computational Mathematics (IACM), Foundation for Research and Technology Hellas (FORTH), GR-70013 Heraklion, Crete, Greece}
\affiliation{Department of Mathematics and Applied Mathematics, University of Crete, GR-71409 Heraklion, Crete, Greece}
\affiliation{Computation-based Science and Technology Research Center, The Cyprus Institute, 20 Constantinou Kavafi Str., Nicosia 2121, Cyprus}
\maketitle

The following article has been accepted by Physics of Fluids. After it is published, it will be found at: \url{https://publishing.aip.org/resources/librarians/products/journals/}.

\section{\label{ModesEffect}The effect of the number of modes on the MSD Rouse expressions with sums}
\begin{figure}[h!]
\includegraphics[width=0.45\textwidth]{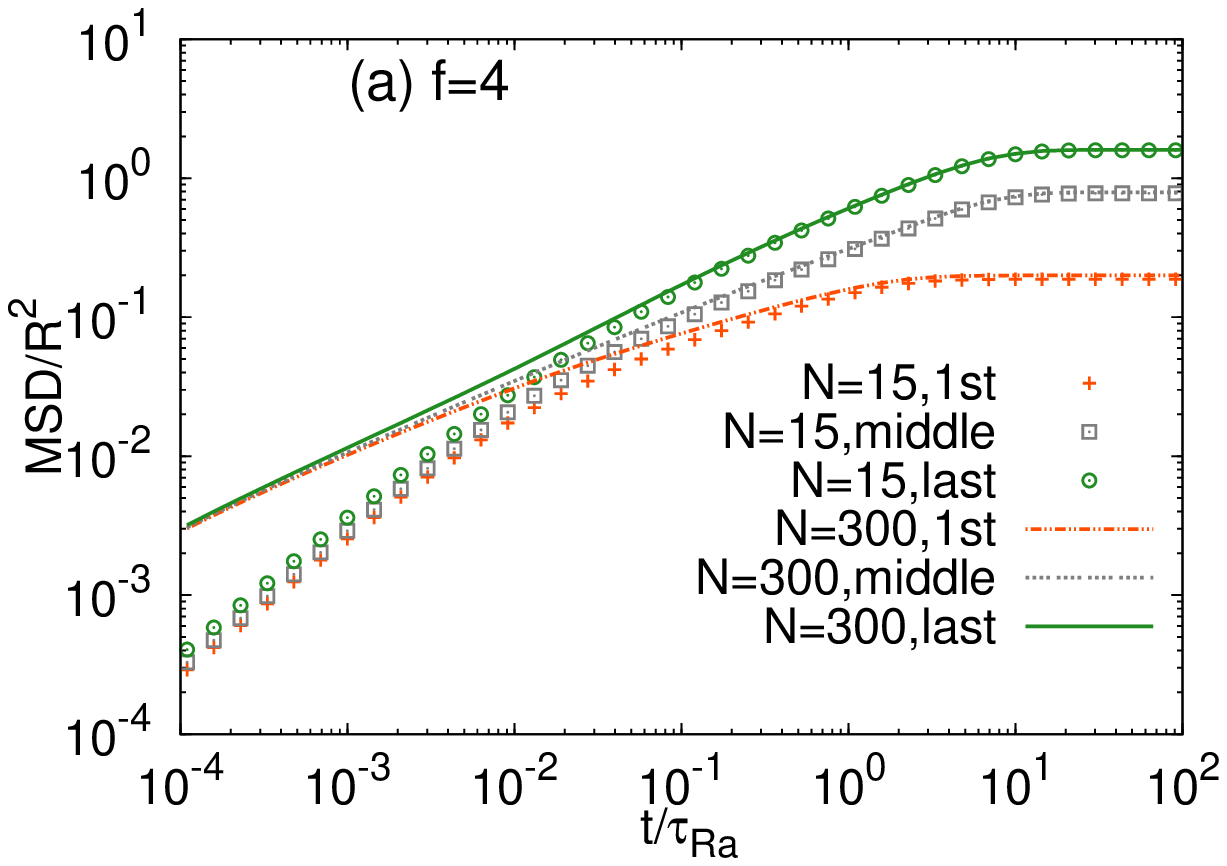}
\includegraphics[width=0.45\textwidth]{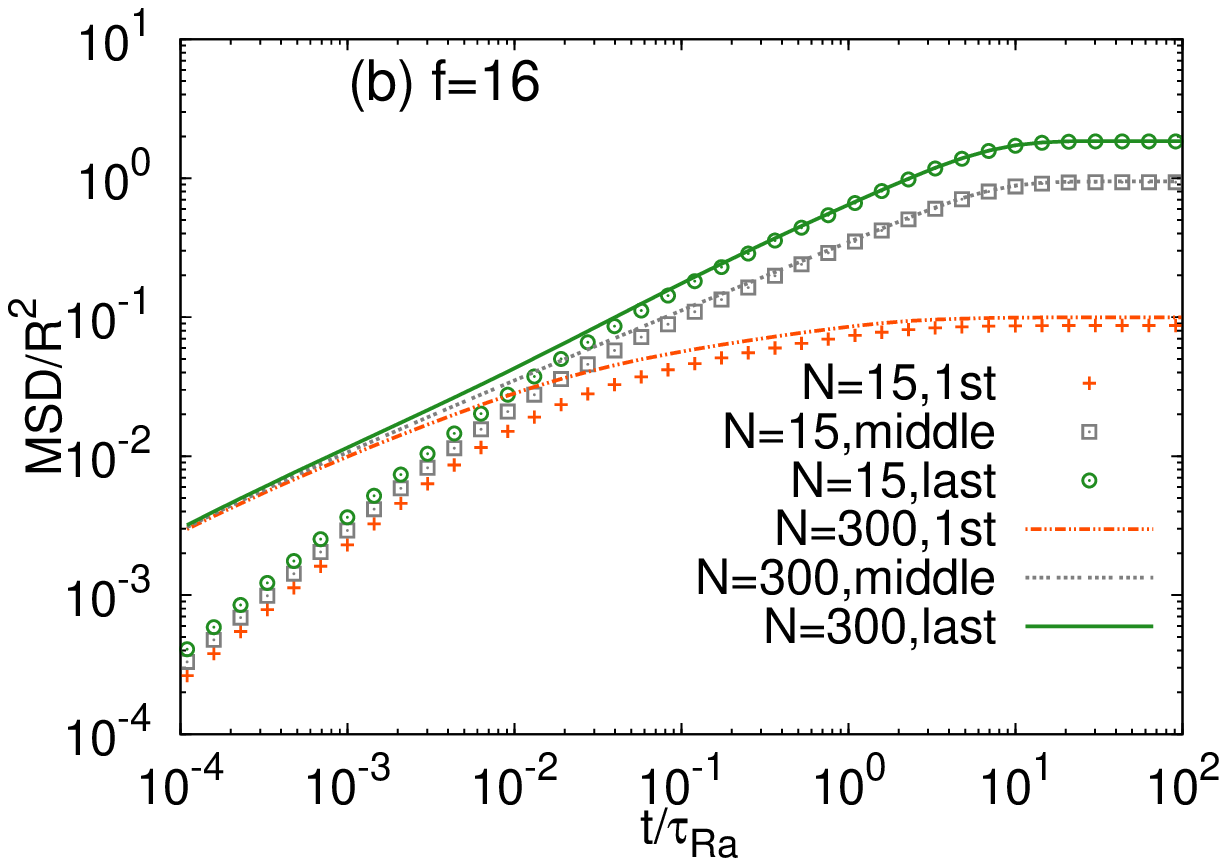}
	\caption{Rouse model predictions of the MSDs: The effect of the number of modes (N) on the summation outcomes at short timescales.}
\label{ModesEffectFig}
\end{figure}

This section examines the effect of the number of modes that is considered as the upper sum limit in the second expression of Table II of the main manuscript.
Figure~\ref{ModesEffectFig} compares predictions for two different upper limits, namely $N_a=N=15$ and $N_a=N=300$. We use the notation $N$ in order to stress that the prefactors of the equations including $N_a$ and the $N_a$ in the equation $\tau_{R_a}=\tau_0 N_a^2$ are unchanged (i.e., in the case of the PEO stars $N_a=15$), only the upper limit of the sums is altered. 
The comparison is made for two functionalities, i.e., $f=4$ and $f=16$. In the comparison, $\tau_0=\tau_0$(PEO), $R^2=N_ab^2$, $b=0.98$ nm and $N_a=15$. Figure~\ref{ModesEffectFig} reveals that, irrespective of $f=4$ and the position along the arm the early response, i.e., $\widetilde{t}_{R_a} < 0.1$ strongly depends on the number of modes considered. For low $N$ values, the early response differ form the typical subdiffusive power law bahavior of $t^{0.5}$. This behavior is recovered as the number of modes in which the chain is split increases. As readily seen from Figure~\ref{ModesEffectFig} this is the case when $N=300$. Nevertheless, summation outcomes at intermediate and long times ($\widetilde{t}_{R_a} > 0.1$) are rather insensitive to $N$.

\section{\label{tau0Esti}Estimation of $\tau_0$ parameter}
To estimate the $\tau_0$ parameter, we focus on the linear chains.
The fact that the diffusion limit is reached in these simulations allows the estimation of $D_{\rm{CM}}$. In more detail, the latter is obtained from the long time plateau of the MSD of the chain center of mass, MSD$_{cm}$, divided by $6t$, namely: $D_{\rm{CM}}=\tfrac{MSD_{cm}}{6t}$ for $t\to\infty$.
The corresponding values for PS and PEO are $7,7.10^{-5} {\rm{nm}^2}/{\rm{ps}}$ and $1,15.10^{-4}{\rm{nm}^2}/{\rm{ps}}$, respectively.
Having determined $D_{\rm{CM}}$, the segmental relaxation time of a Rouse bead can be extracted by combining the following two equations:
\begin{subequations} \label{Dcm-tau}
\begin{align}
D_{\rm{CM}} =& \frac{k_BT}{N_a\zeta_0} \label{Dcm} \\
\tau_0 =& \frac{\zeta_0b^2}{3\pi^2k_BT}. \label{tau}
\end{align}
\end{subequations}
That is, $\tau_0=b^2/(3\pi^2N_aD_{\rm{CM}})$, where $N_a=7$ for the PS and $N_a=15$ for the PEO linear chains. The corresponding $\tau_0$ values are
$\tau_{0}(1)=120$ps for the PS chains and $\tau_{0}(1)=20$ps for the PEO ones. So-obtained values are in turn utilized to obtain the Rouse prediction:
\begin{equation}\label{RouseLinear}
\langle \big( {\bf r}_{\ell,t}-{\bf r}_{\ell,0} \big)^2 \big \rangle = \frac{2b^2}{\pi^2 N_a \tau_0}t + \frac{2N_ab^2}{\pi^{1.5}} \sqrt{\widetilde{t}_{R_a}}
\end{equation}
where $\tau_{R_a}=\tau_{0}N_a^2$ and $\widetilde{t}_{R_a}=|t-t^{\prime}|\tau_{R_a}^{-1}$. According to the Rouse prediction all segments of a linear chain exhibit identical response.
In reality this is not entirely correct as chain ends have always higher mobility than the central sections of the chain. Nevertheless, the Rouse prediction is accurate for the central section of the chain.

\begin{figure}[!ht]
\includegraphics[width=0.47\textwidth]{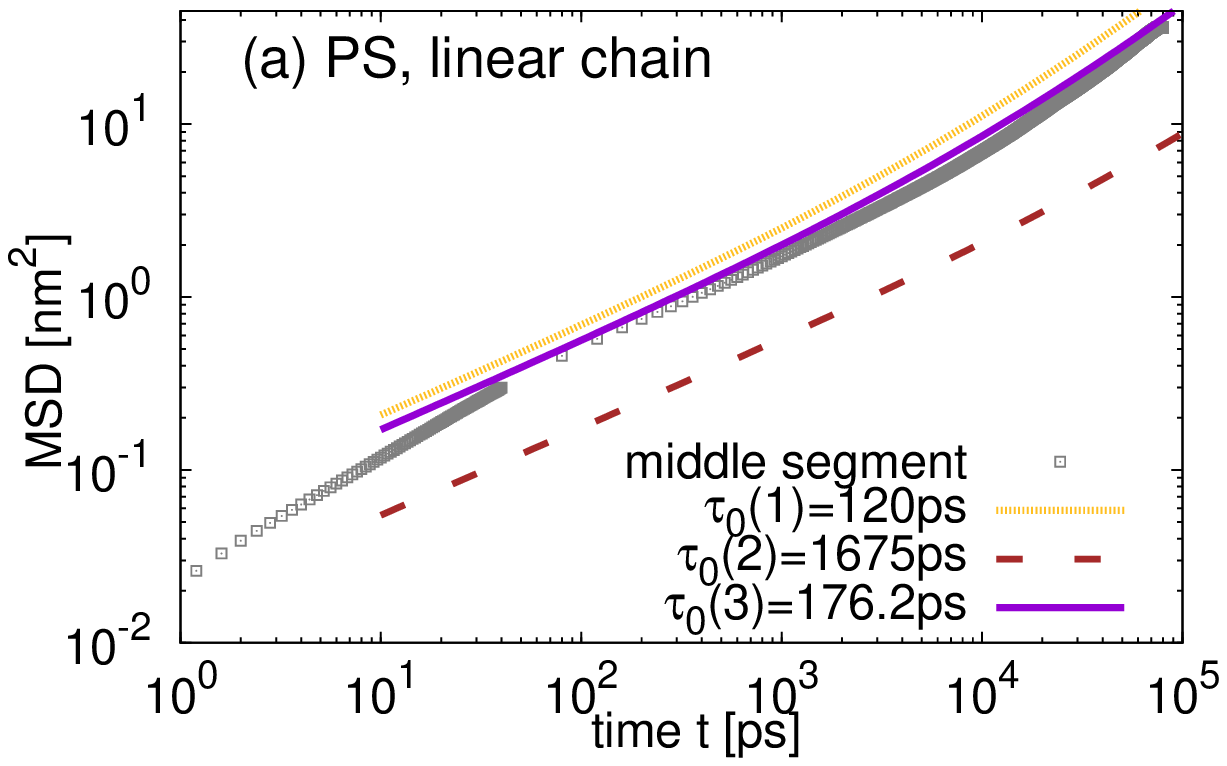}
\includegraphics[width=0.47\textwidth]{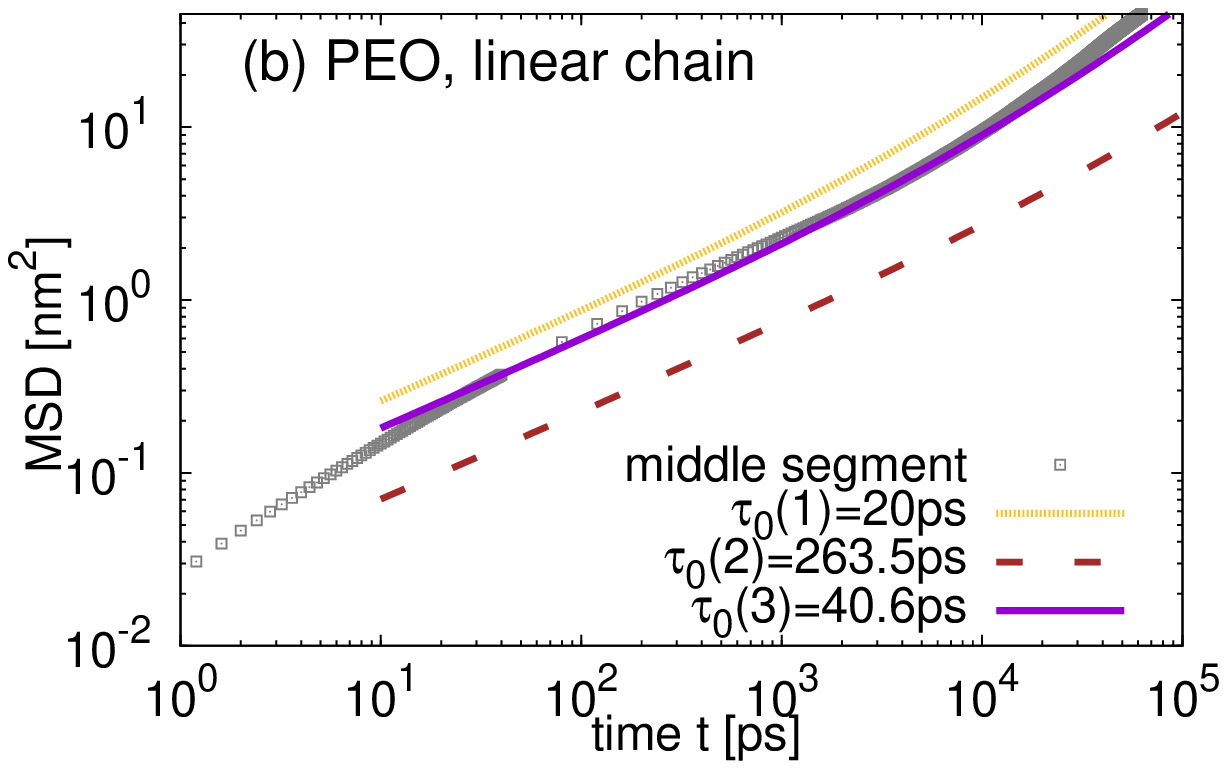}
\caption{MSD of the middle segment of the (a) PS and (b) PEO linear chains (symbols) together with theoretical predictions (lines) using the $\tau_0$ values estimated from three different methods (indices 1,2,3). For further comparison, the set of data represented by solid lines was used.}
\label{rouse}
\end{figure}

In this respect, the predictions of Eq.~\ref{RouseLinear} are only compared to MSD simulation data for the middle monomer. The comparison can be seen in Fig.~\ref{rouse}. There, symbols refer to simulation data while golden dotted lines represent Rouse predictions. The comparison reveals that, for both PS and PEO, the theoretical curves express the data in reasonable detail. That is, apart from very early timescales, the shape of the simulation curves is reproduced faithfully by the model. However, compared to the simulation results, the predicted curves are horizontally shifted to faster timescales. As from a quantitative perspective the Rouse predictions agree satisfactorily with the simulation data, we do not attribute this discrepancy to non-Gaussian effects (i.e. the simulated linear PS and PEO chains being semi-flexible). We rather attribute the discrepancy to $\tau_{0}$ values being slightly longer than the $\tau_{0}(1)$ values utilized. We can only speculate on the origin of this discrepancy. Since Eq.~\ref{tau} is based on scaling arguments and not on a microscopic derivation, there could be prefactors of the order of unity omitted in this equation when applied to short chains. Another possibility is that the considered values for $N_a$ and $b$ are somewhat different from the actual values. Since the discrepancy between simulation data and Rouse predictions is not dramatic, we have chosen to maintain the values of $N_a$ and $b$ constant, but adjust the $\tau_0$ values. Hence, uncertainties regarding omitted prefactors in Eq.~\ref{Dcm} and the exact $N_a$ and $b$ values will be absorbed in the $\tau_0$ values.

Therefore, alternative methods of $\tau_0$ estimation are considered. One such method utilizes the early response of the MSD data of the linear chains. In particular, we consider $\tau_0$ to be the timescale at which the MSD simulation data equal to $b^2$. So-obtained $\tau_0$ values from this method are $\tau_{0}(2)=1675$ps and $\tau_{0}(2)=263.5$ps for PS and PEO, respectively. The corresponding theoretical predictions using Eq.~\ref{RouseLinear} are shown in Fig.~\ref{rouse} by dashed brown lines. As readily seen, the model fails dramatically in predicting the simulation response of the linear chains. Clearly, the $\tau_{0}(2)$ values are considerably longer than the actual $\tau_{0}$ values. Actually, they are more than an order of magnitude longer than the $\tau_{0}(1)$ values suggested by the chain diffusivity approach of the first method above. Such a big discrepancy between $\tau_0(1)$ and $\tau_0(2)$ can not be easily attributed to uncertainties in the $N_a$ and $b$ values or omitted prefactors in Eqs.~\ref{Dcm} and ~\ref{tau}. We also note that the
corresponding Rouse relaxation times $\tau_{0}(2)N_a^2$ ($8.10^4$ps and $6.10^4$ps for PS and PEO, respectively) correlate with timescales that correspond to ordinary diffusion of the linear chains' center of mass. On all aforementioned grounds, the $\tau_{0}(2)$ values are disregarded.

Similar to the second approach above, the third examined method of $\tau_0$ estimation utilizes the early MSD response of the middle chain monomer. Nevertheless, in this scenario, we consider $\tau_0$ to be the timescale at which the MSD simulation data equal to $b^2/3$. This approach was adopted by Theodorou~\cite{Theodorou} who included a factor of three in the MSD when estimating the tube diameter of entangled linear chains from MSD simulation data. The values of relaxation times obtained from this method are $\tau_{0}(3)=176.2$ps for the PS and $\tau_{0}(3)=40.6$ps for the PEO chains. With the third parametrization method, the Rouse model, i.e., Eq.~\ref{RouseLinear}, describes the simulation data in great detail (compare the symbols and the solid lines in Fig.~\ref{rouse}). Moreover, the obtained $\tau_0$ values are in reasonable agreement with those extracted from the diffusion coefficient of the chains ($\tau_{0}(1)$ values). In this respect, only the $\tau_0(3)$ values will be used in all subsequent comparisons between the simulation data and Rouse
predictions referring to star polymer chains.

\newpage
\section{\label{Rouse_MD-comparisons}Comparison of the Rouse model and simulation MSD data}
\begin{figure*}[!ht]
\includegraphics[width=0.47\textwidth]{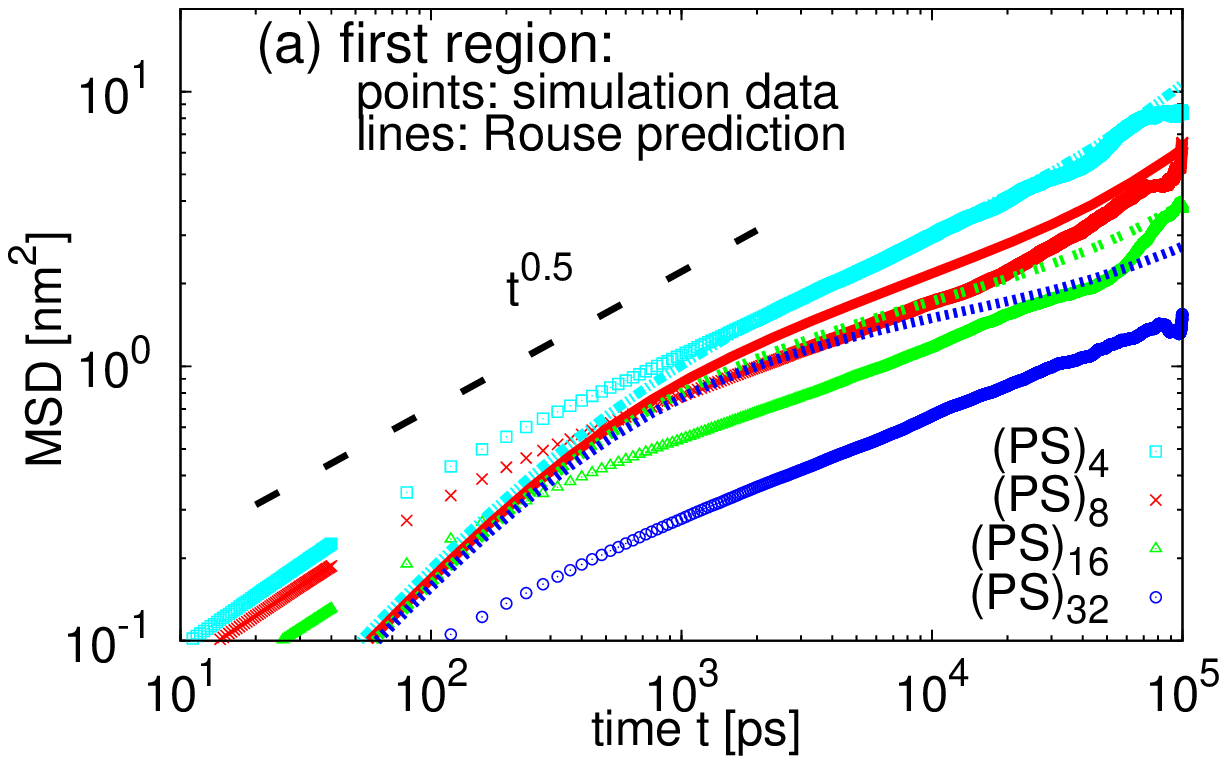}
\includegraphics[width=0.47\textwidth]{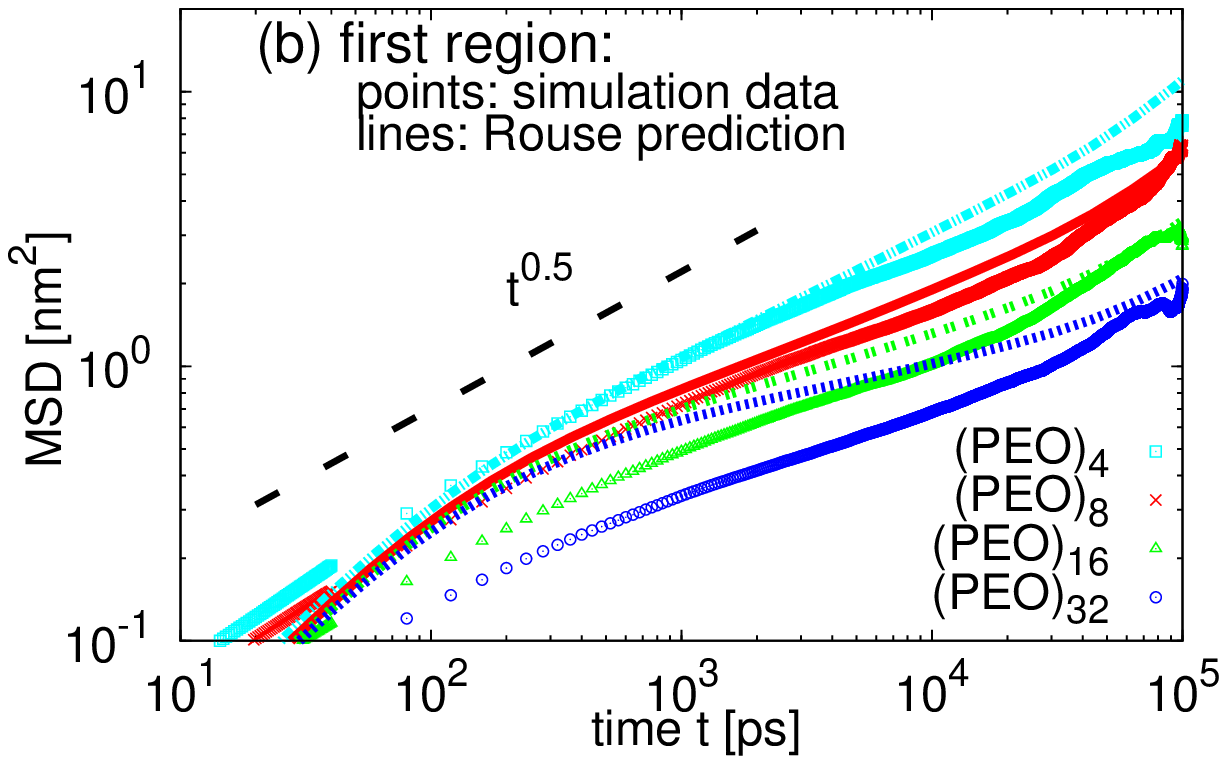}
\caption{Segmental MSD averaged over monomers in the first region: simulation data (symbols) and model predictions (lines) that include CM diffusion. The thick black dashed line is plotted to guide the eye and represents the scaling MSD$\sim t^{0.5}$.}
\label{rouse_msd_1st_with_CM}
\end{figure*}

\begin{figure*}[!ht]
\includegraphics[width=0.47\textwidth]{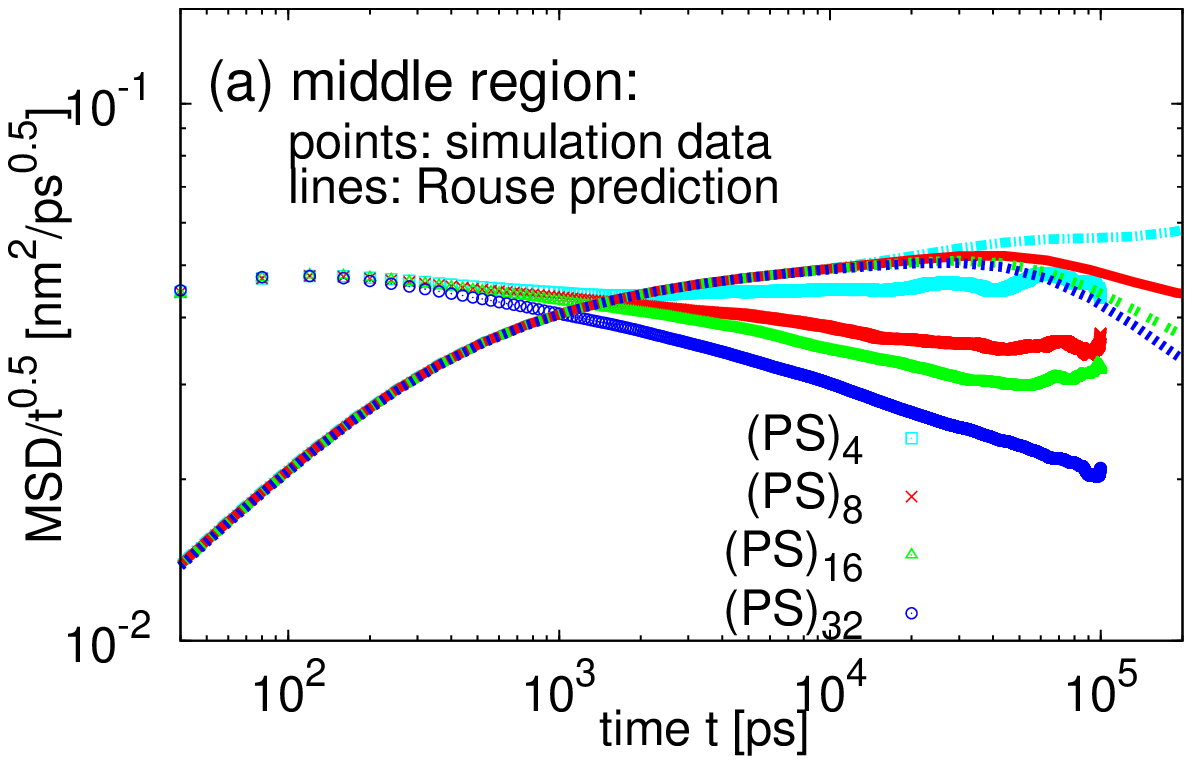}
\includegraphics[width=0.47\textwidth]{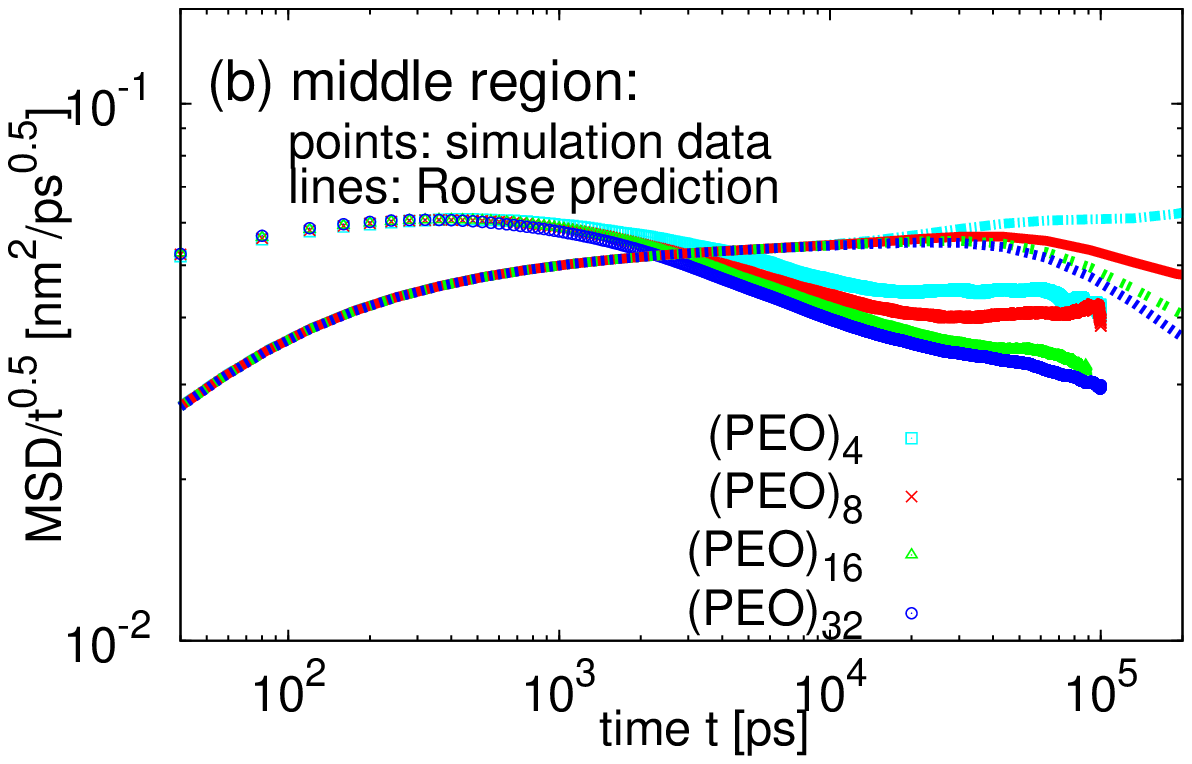}
\caption{Segmental MSD averaged over monomers in the middle region divided by $t^{0.5}$: simulation data (symbols) and model predictions (lines) that include CM diffusion.}
\label{rouse_msd_mid_with_CM}
\end{figure*}

\begin{figure*}[!ht]
\includegraphics[width=0.47\textwidth]{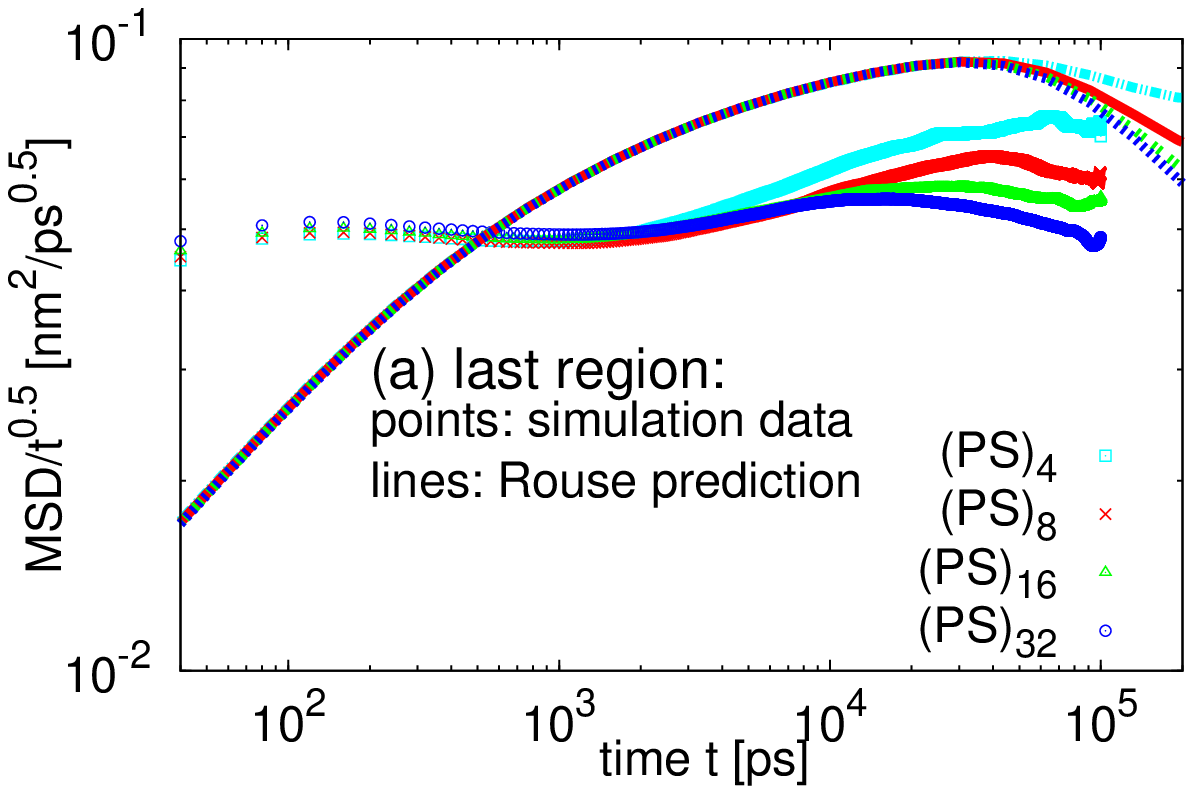}
\includegraphics[width=0.47\textwidth]{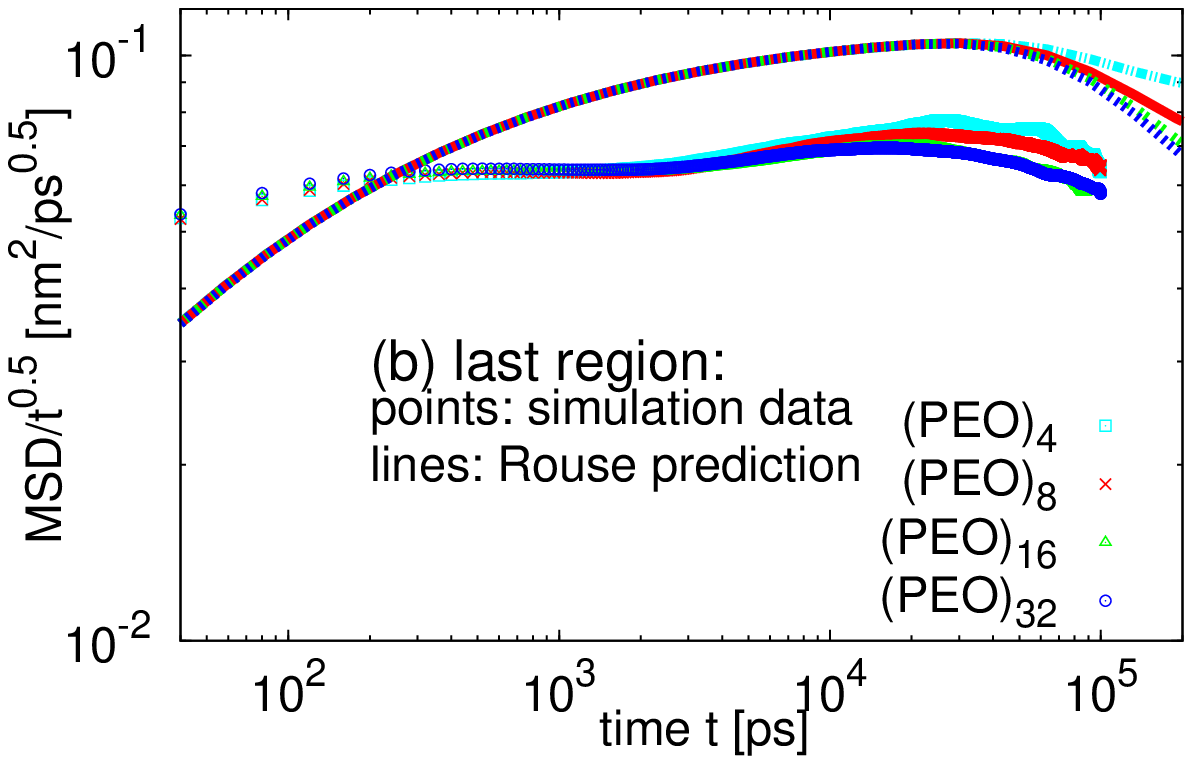}
\caption{Segmental MSD averaged over monomers in the last region divided by $t^{0.5}$: simulation data (symbols) and model predictions (lines) that include CM diffusion. }
\label{rouse_msd_end_with_CM}
\end{figure*}

\begin{figure*}[!ht]
\includegraphics[width=0.47\textwidth]{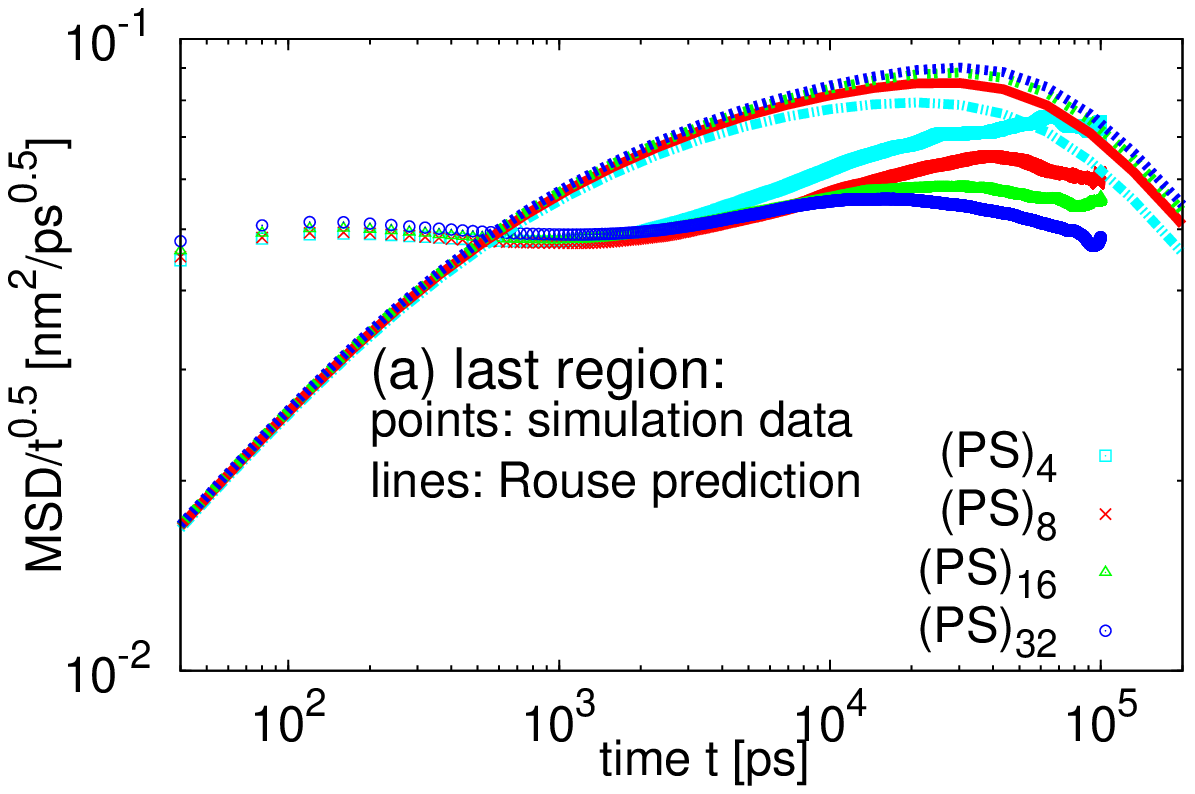}
\includegraphics[width=0.47\textwidth]{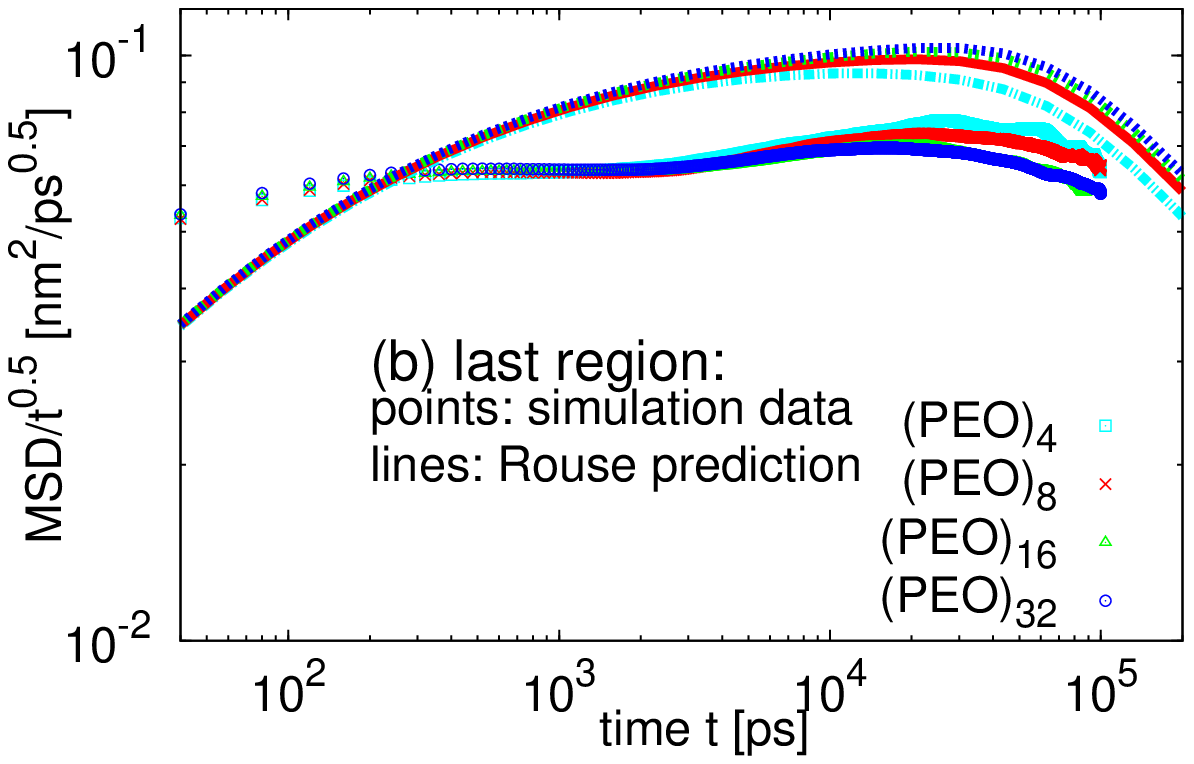}
\caption{Segmental MSD averaged over monomers in the last region divided by $t^{0.5}$: simulation data (symbols) and model predictions (lines) that discount CM diffusion. }
\label{rouse_msd_end_without_CM}
\end{figure*}

\newpage
\section{\label{Correlator}Autocorrelation function of the center-to-end vector}
\begin{figure}[h!]
\includegraphics[width=0.45\textwidth]{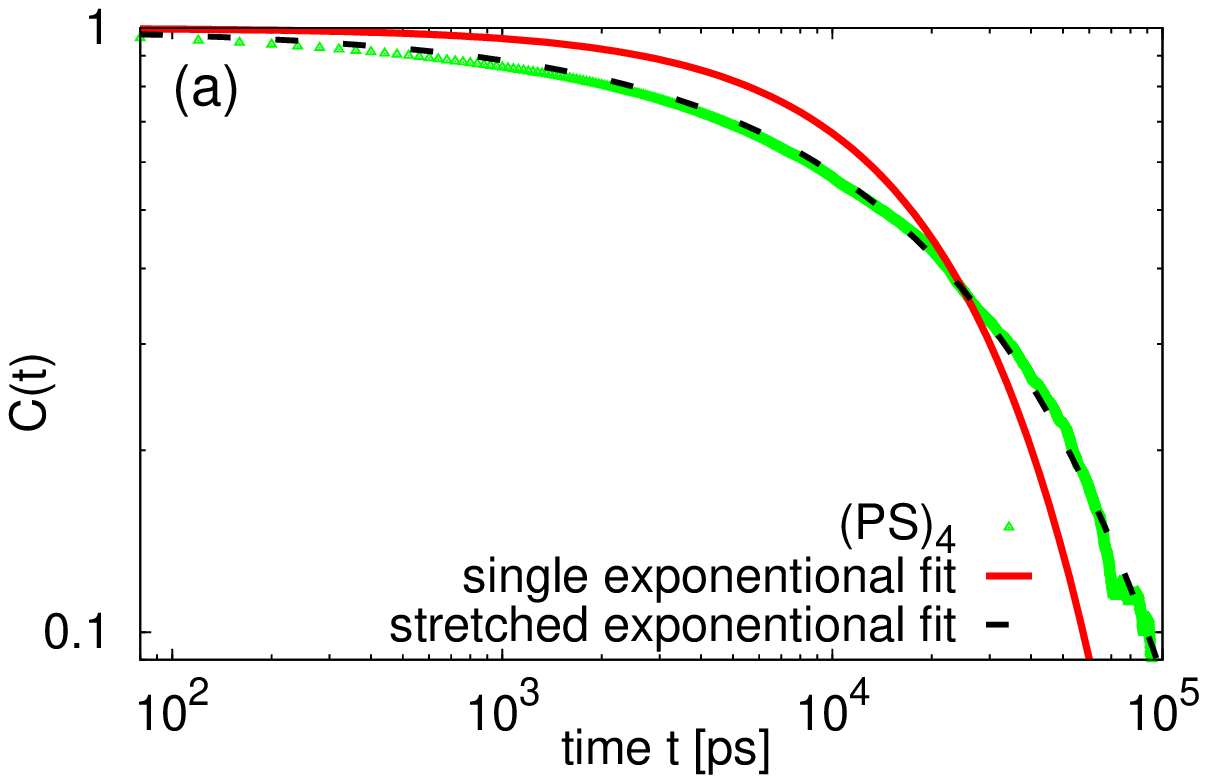}
\includegraphics[width=0.45\textwidth]{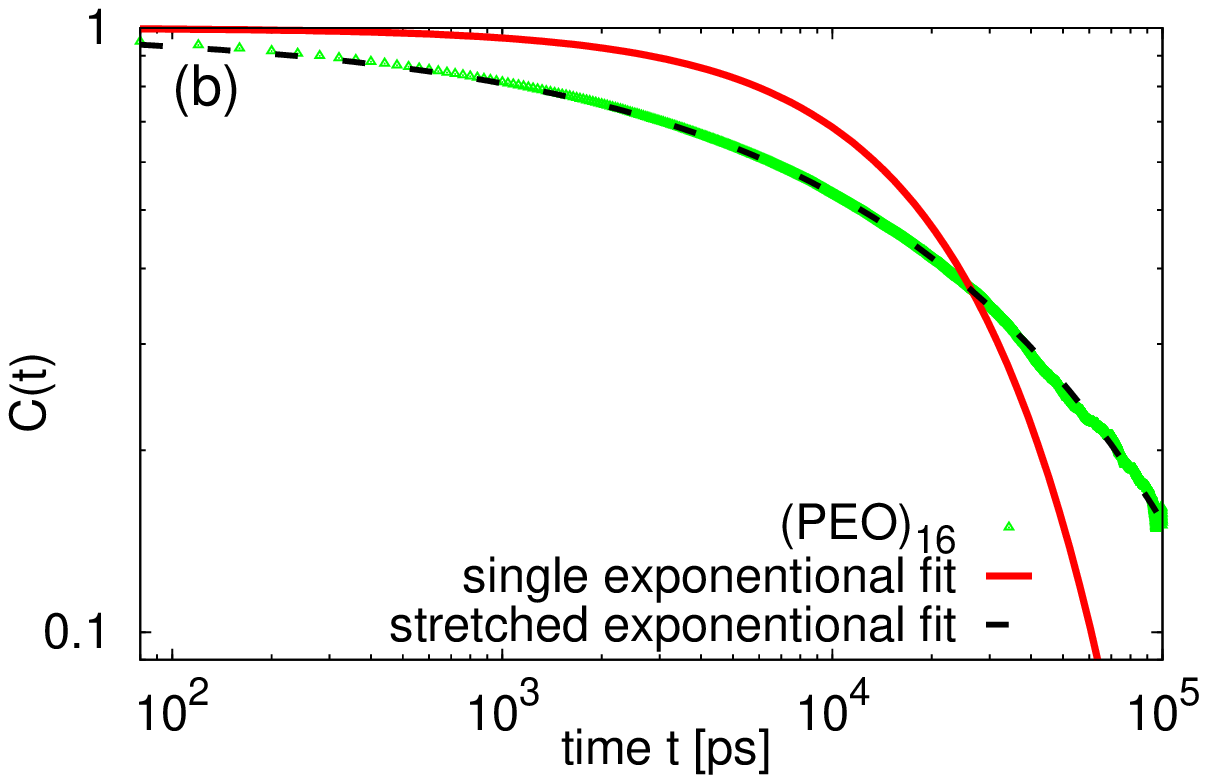}
	\caption{Examples of fitting the autocorrelation function of the center-to-end vector $R$ for (a) (PS)$_4$ and (b) the (PEO)$_{16}$ stars. The lines are single and stretched exponential fits.}
\label{end}
\end{figure}
The autocorrelation function $C(t)$ of the center-to-end vector is defined as follows:
\begin{equation}
C(t) = \frac{\left< {\bf{R}}_{\alpha}(t_0+t)\cdot  {\bf{R}}_{\alpha}(t_0) \right>}{||{\bf{R}}_{\alpha}(t_0+t) ||\cdot|| {\bf{R}}_{\alpha}(t_0)||}
\end{equation}
where
${\bf{R}}_{\alpha}(t_0+t)$ is the end-to-end vector of an arm $\alpha$ at time $t_0+t$ and correspondingly, ${\bf{R}}_{\alpha}(t_0)$ is the end-to-end vector of arm $\alpha$ at the time origin.
The brackets denote an average over different time origins $t_0$ and over all possible vectors, i.e., over all the arms $f$.

The theoretical expression for $C(t)$ can be expressed as:
\begin{align*}
	C(t)=\frac{\left< {\bf{R}}_{\alpha}(t_0+t)\cdot  {\bf{R}}_{\alpha}(t_0) \right>}{N_{a}b^2}&=\frac{8}{f\pi^2} \left( \sum_{p,odd}{\frac{1}{p^2}e^{-\frac{tp^2}{\tau_{R_a}}}} + [ (s_{1\alpha})^2+...+(s_{f\alpha})^2] \sum_{q}{\frac{1}{(2q-1)^2} \sin^2 {\left( \frac{(2q-1)\pi}{2}  \right)}  e^{-\frac{t(2q-1)^2}{4 \tau_{R_a}}}}\right) \\
	&=  \frac{8}{f\pi^2} \left(  \sum_{p,odd}{\frac{1}{p^2}e^{-\frac{tp^2}{\tau_{R_a}}}} +(f-1) \sum_{q,odd}{\frac{1}{q^2}   e^{-\frac{tq^2}{4\tau_{R_a}}}} \right)	
\end{align*}

We fit the correlation functions for each functionality by Kohlrausch-Williams-Watts (KWW) stretch exponential function $f(t)=A\exp\left(-(t/\tau_{\textrm{KWW}})^{\beta_{\textrm{KWW}}}\right)$, where $A$ is a prefactor, $\beta_{\textrm{KWW}}$ a stretch exponent and $\tau_{\textrm{KWW}}$ a relaxation time. One example of the fitting procedure of $C(t)$ functions for each chemistry is shown in Fig~\ref{end}. We also show a fit with a simple exponential function, i.e., a function with $\beta_{\textrm{KWW}}=1$ to demonstrate the stretched character of the $C(t)$ decay.
The average relaxation time  $\tau_{R_a}^s$ is then calculated as $\tau_{R_a}^s=\frac{\tau_{\textrm{KWW}}}{{\beta_{\textrm{KWW}}}}\Gamma\left(\frac{1}{{\beta_{\textrm{KWW}}}}\right)$ where $\Gamma$ stands for the gamma function. We keep the superscript $^s$ to distinguish $\tau_{R_a}^s$ obtained from the fit of the simulation data from the theoretically predicted Rouse time, $\tau_{R_a}$. The so-obtained $\tau_{R_a}^s$ are listed in Tab.~\ref{tab}.
\begin{table}
\begin{center}
\begin{tabular}{|l|l|l|}
\hline
notation & $\tau_{R_a}^s$ [ns] & $\beta_{\textrm{KWW}}$ \tabularnewline
\hline \hline
(PS)$_4$          & $34$  & $0.65$\tabularnewline 
(PS)$_8$          & $69$  & $0.58$\tabularnewline 
(PS)$_{16}$       & $161$  & $0.51$\tabularnewline
(PS)$_{32}$       & $455$  & $0.46$\tabularnewline
(PEO)$_{4}$       & $29$  & $0.51$\tabularnewline 
(PEO)$_{8}$       & $33$  & $0.52$\tabularnewline
(PEO)$_{16}$      & $58$  & $0.48$\tabularnewline 
(PEO)$_{32}$      & $74$  & $0.48$\tabularnewline
\hline
\end{tabular}
\end{center}
\caption{The Rouse time $\tau_{R_a}^s$ obtained by fitting the simulation data with KWW function. $\beta_{\textrm{KWW}}$ denotes the stretch exponent of the given function.}
\label{tab}
\end{table}

\newpage
\section{\label{vanhovefits}Van Hove functions}

\begin{figure}[h!]
\includegraphics[width=0.45\linewidth]{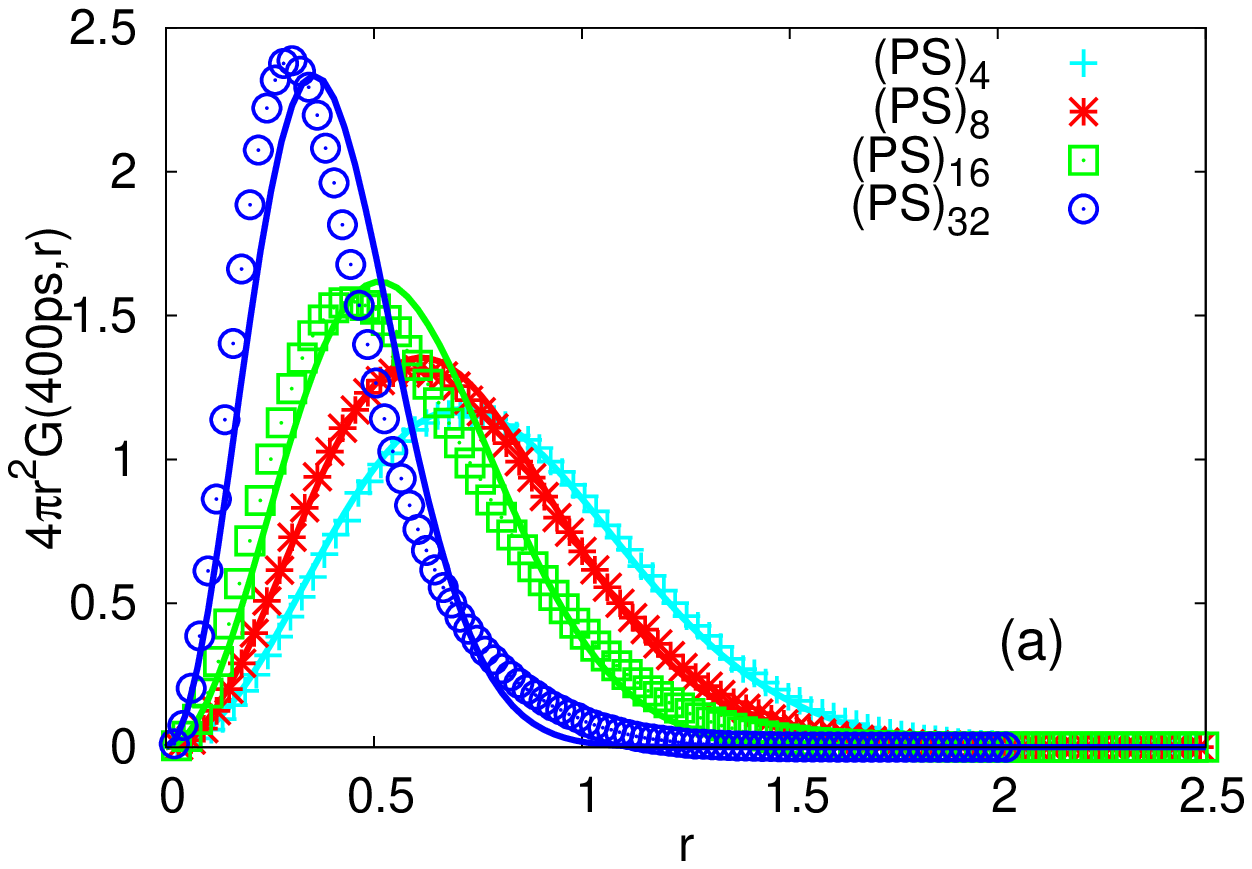}
\includegraphics[width=0.45\linewidth]{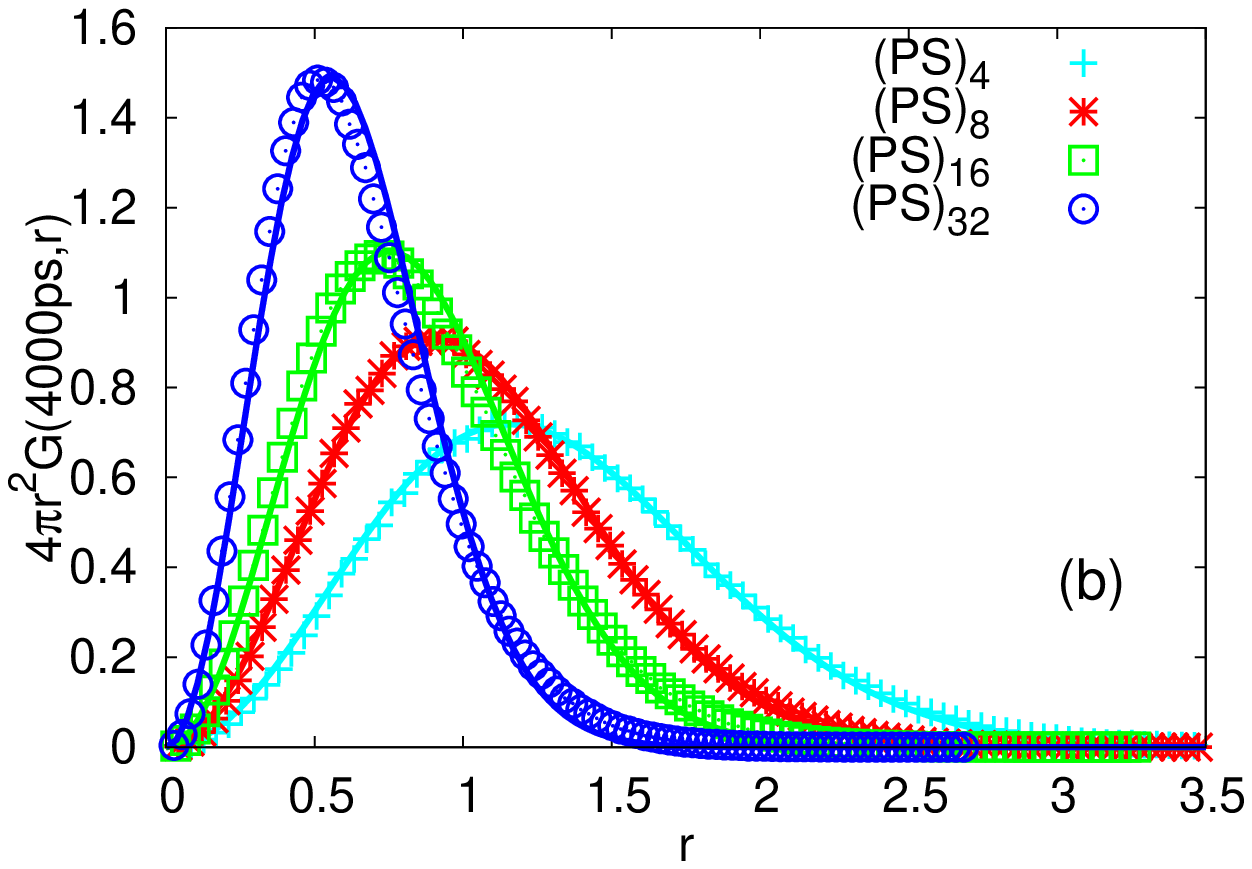}
\includegraphics[width=0.45\linewidth]{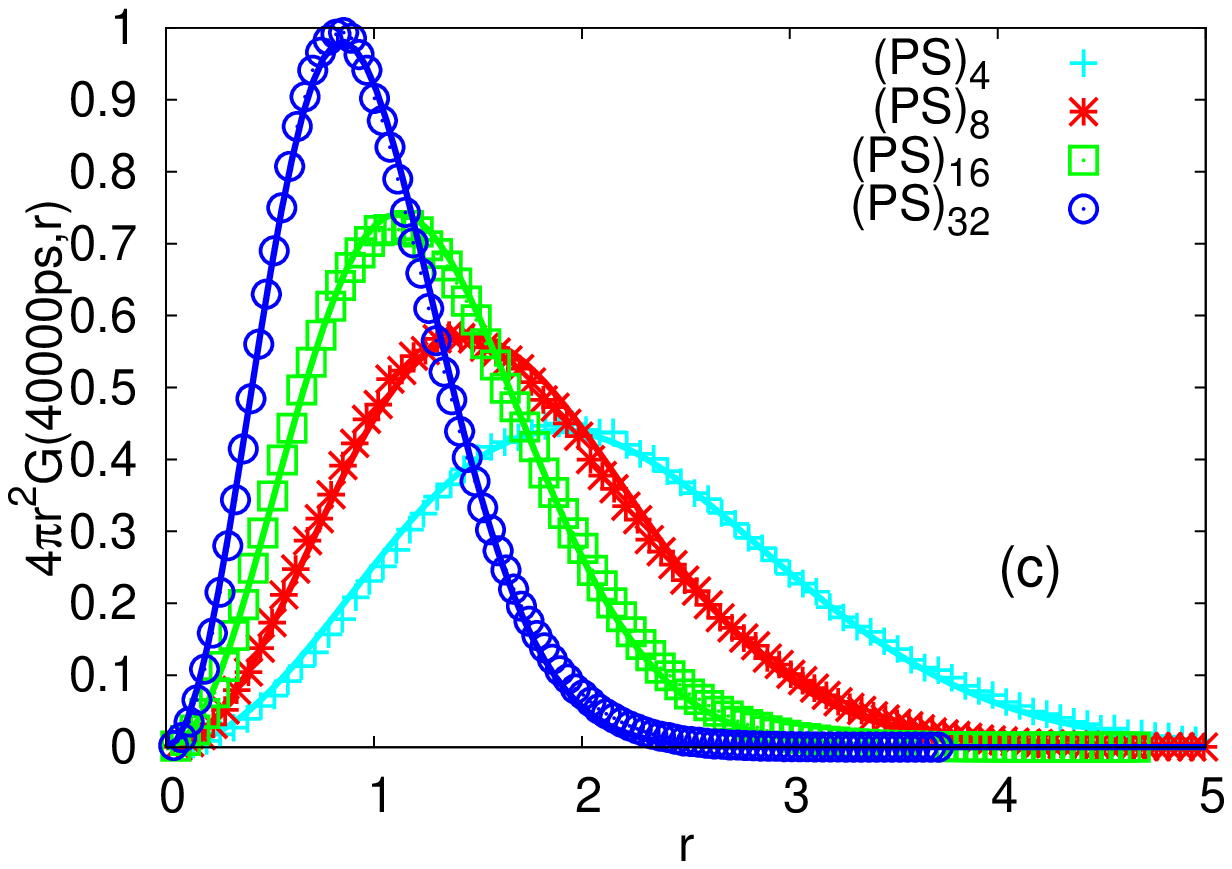}
	\caption{Van Hove function for the monomers in the (first) region adjacent to the star kernel in (PS)$_f$ stars for randomly selected time frames: (a)400, (b)4000 and (c)40000ps. The solid lines are Gaussian distributions with the corresponding MSD taken from the simulation.}
\label{van_PS}
\end{figure}

\begin{figure}[h!]
\includegraphics[width=0.45\linewidth]{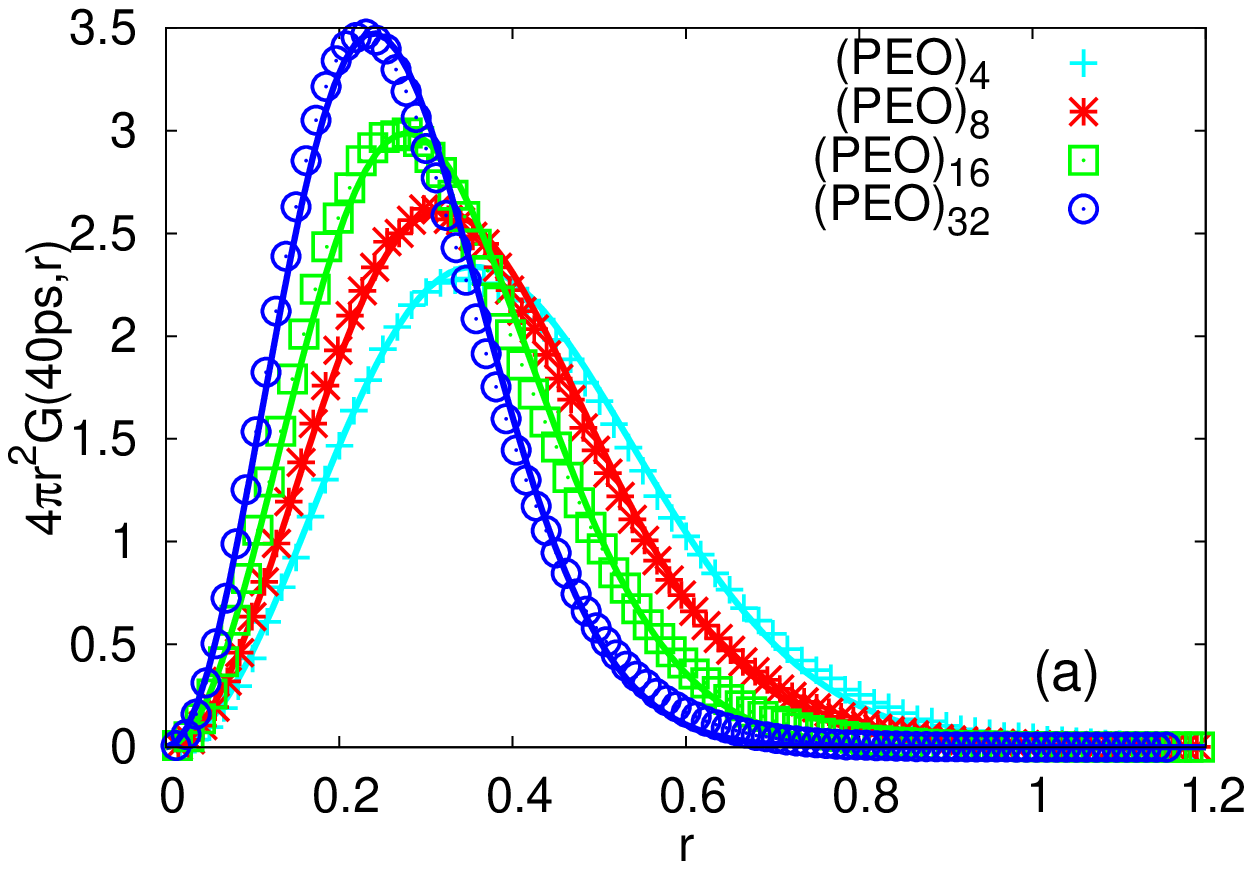}
\includegraphics[width=0.45\linewidth]{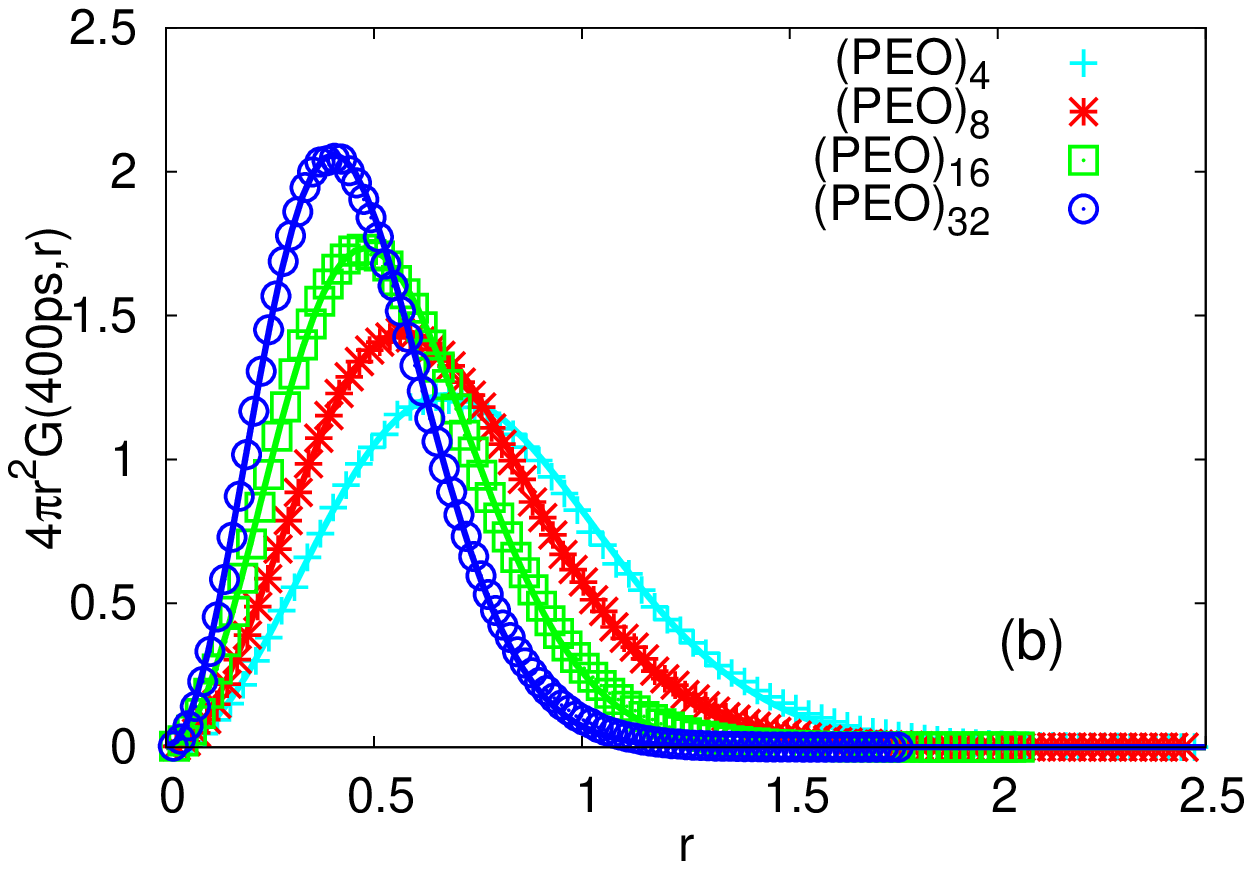}
\includegraphics[width=0.45\linewidth]{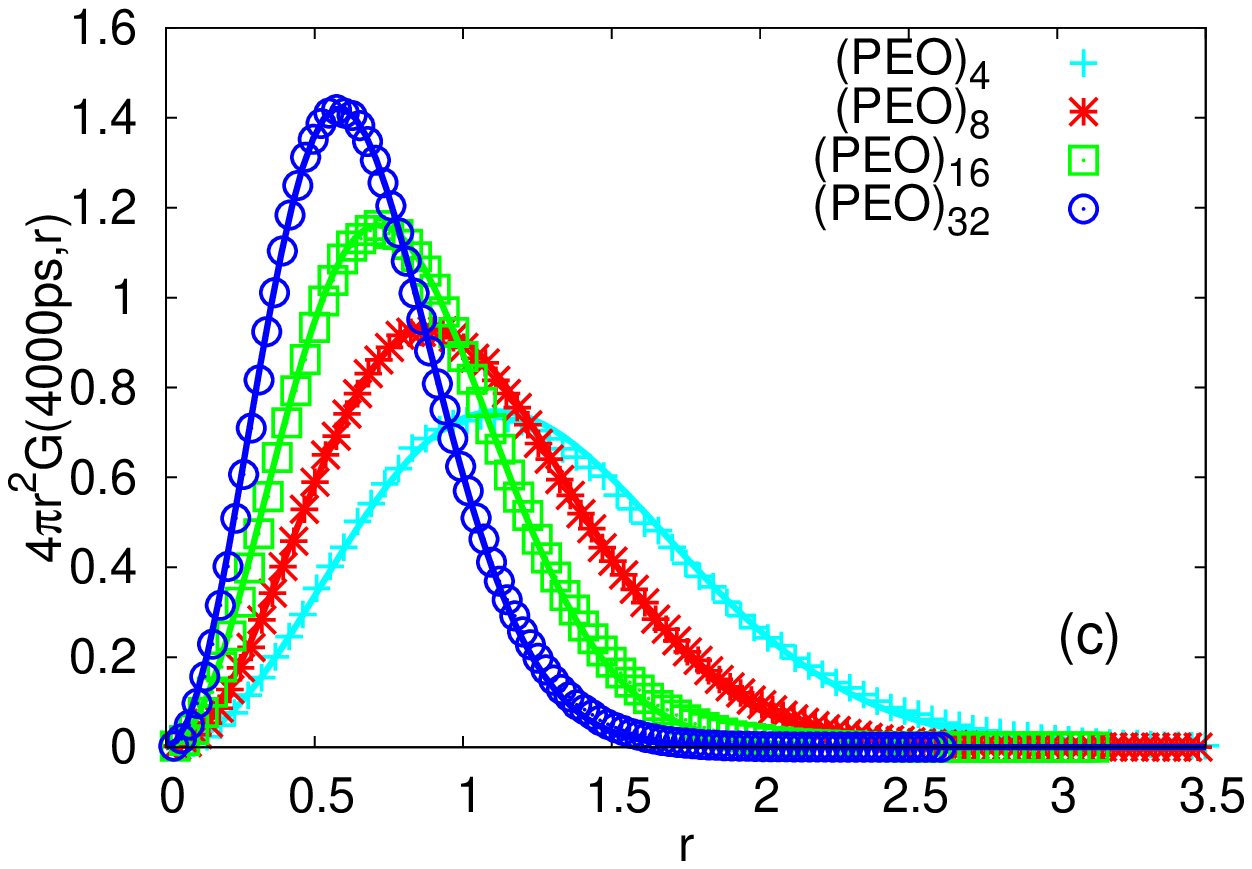}
\includegraphics[width=0.45\linewidth]{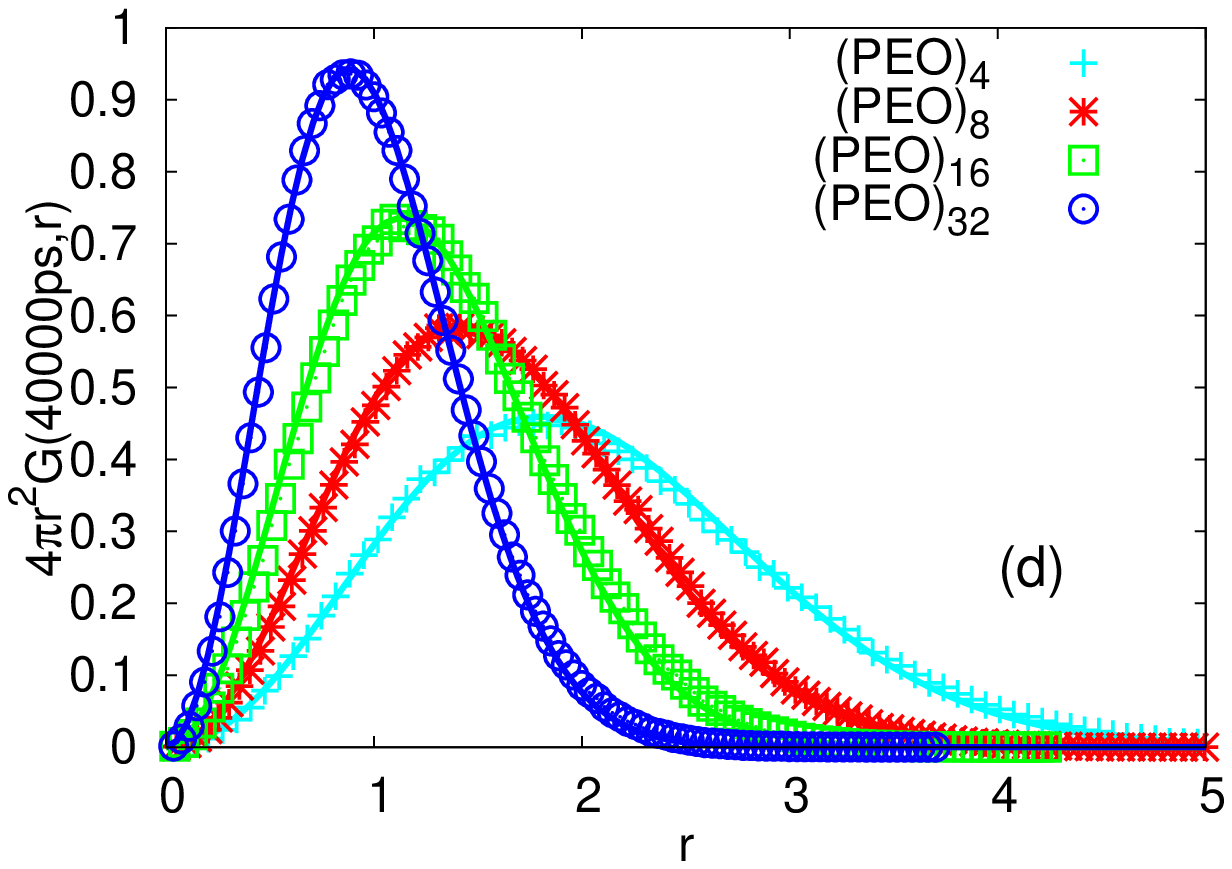}
	\caption{Van Hove function for the monomers in the (first) region adjacent to the star kernel in (PEO)$_f$ stars for randomly selected time frames: (a)40, (b)400, (c)4000 and (d)40000ps. The solid lines are Gaussian distributions with the corresponding MSD taken from the simulation.}
\label{van}
\end{figure}

\newpage
\section{\label{density}Monomer density profiles from the simulations}

\begin{figure}[!ht]
\includegraphics[width=0.43\textwidth]{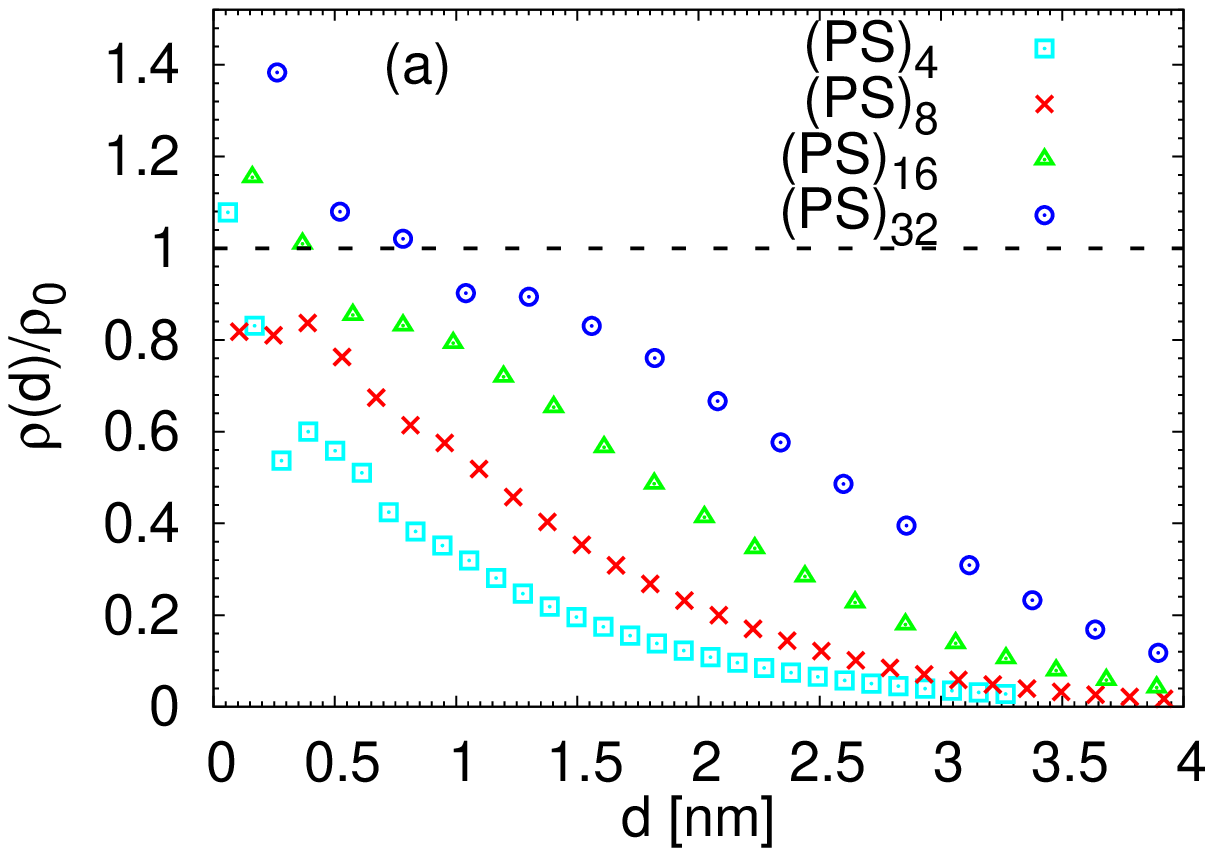}
\includegraphics[width=0.43\textwidth]{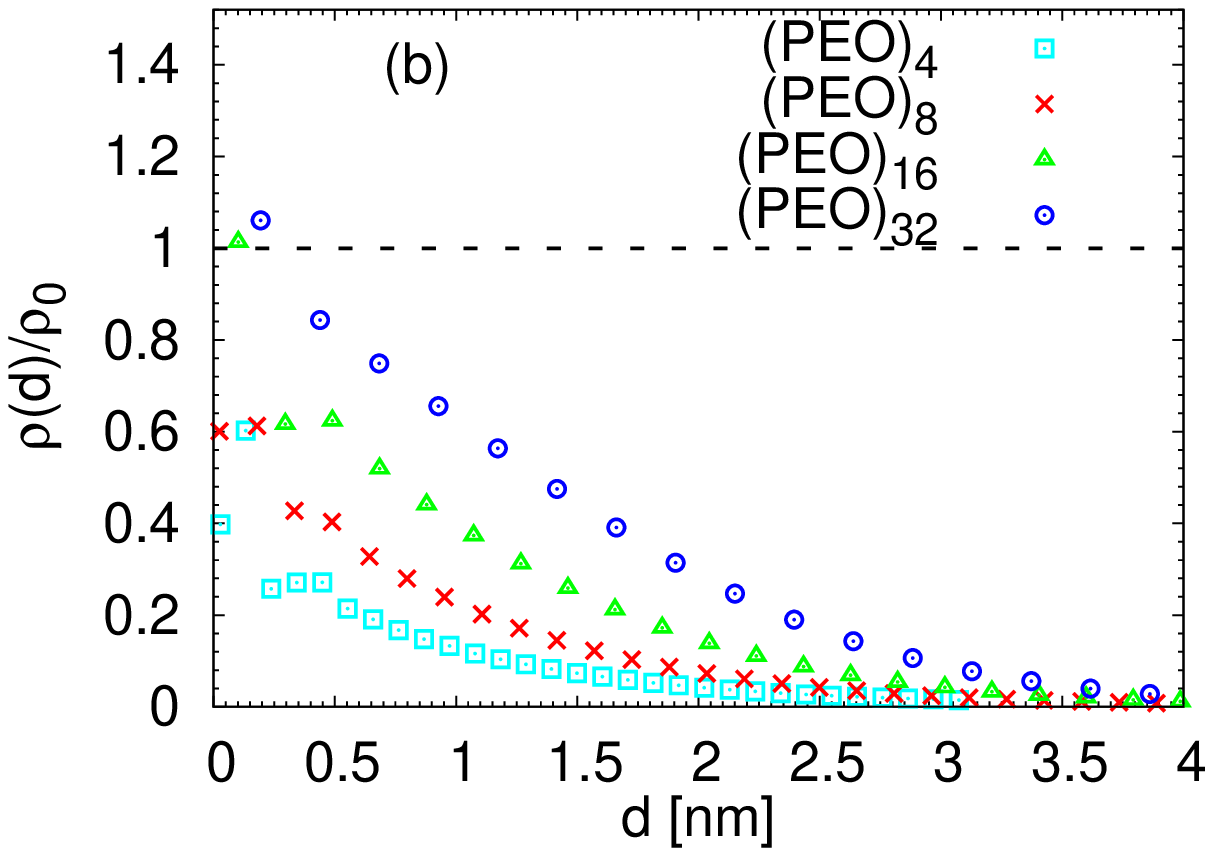}
	\caption{Monomer density profile for (a) (PS)$_f$ and (b) (PEO)$_f$ as a distance from the kernel, $d$. The local density $\rho(d)$ is normalized by the density of the system, $\rho_0$.}
\label{dens}
\end{figure}

\FloatBarrier
\nocite{*}
\bibliography{ref_SI}